\documentclass[longauth]{aa}

\usepackage{graphicx,color}
\usepackage[T1]{fontenc}
\usepackage[utf8]{inputenc}
\usepackage{lmodern}

\usepackage{txfonts}

\usepackage[colorlinks,bookmarks]{hyperref}
\definecolor{linkblue}{rgb}{0,0,0.8}
\definecolor{linkgreen}{rgb}{0,0.5,0}
\hypersetup{linkcolor=linkblue, citecolor=linkgreen, urlcolor=linkblue}

\usepackage{booktabs}
\usepackage{supertabular}

\graphicspath{{./figs/}}

\DeclareUnicodeCharacter{0303}{~}

\definecolor{valecol}{rgb}{0,0.5, 1.}

\definecolor{lcol}{rgb}{1,0,1}

\begin{document}

\title{The miniJPAS survey: star-galaxy classification using machine learning}

\author{
P.~O.~Baqui\inst{\ref{inst1}}\thanks{These authors contributed equally to this work.}
\and
V.~Marra\inst{\ref{inst1},\ref{inst2},\ref{inst2x},\ref{inst2y}}\footnotemark[1]
\and
L.~Casarini\inst{\ref{inst3}}
\and
R.~Angulo\inst{\ref{inst4},\ref{inst15}}
\and
L.~A.~Díaz-García\inst{\ref{inst5}}
\and
C.~Hernández-Monteagudo\inst{\ref{inst22},\ref{inst9a},\ref{inst9b}}
\and
P.~A.~A.~Lopes\inst{\ref{inst10}}
\and
C.~López-Sanjuan\inst{\ref{inst22}}
\and
D.~Muniesa\inst{\ref{inst6}}
\and
V.~M.~Placco\inst{\ref{inst7}}
\and
M.~Quartin\inst{\ref{inst10},\ref{inst21}}
\and
C.~Queiroz\inst{\ref{inst13}}
\and
D.~Sobral\inst{\ref{inst12}}
\and
E.~Solano\inst{\ref{inst11}}
\and
E.~Tempel\inst{\ref{inst8}}
\and
J.~Varela\inst{\ref{inst22}}
\and
J.~M.~V\'ilchez\inst{\ref{inst9}}
\and
R.~Abramo\inst{\ref{inst13}}
\and
J.~Alcaniz\inst{\ref{inst14}}
\and
N.~Benitez\inst{\ref{inst9}}
\and
S.~Bonoli\inst{\ref{inst6},\ref{inst4},\ref{inst15}}
\and
S.~Carneiro\inst{\ref{inst16}}
\and
A.~J.~Cenarro\inst{\ref{inst22}}
\and
D.~Cristóbal-Hornillos\inst{\ref{inst6}}
\and
A.~L.~de Amorim\inst{\ref{inst23}}
\and
C.~M.~de Oliveira\inst{\ref{inst17}}
\and
R.~Dupke\inst{\ref{inst14},\ref{inst18},\ref{inst19}}
\and
A.~Ederoclite\inst{\ref{inst17}}
\and
R.~M.~González Delgado\inst{\ref{inst9}}
\and
A.~Marín-Franch\inst{\ref{inst22}}
\and
M.~Moles\inst{\ref{inst6}}
\and
H.~Vázquez Ramió\inst{\ref{inst22}}
\and
L.~Sodré\inst{\ref{inst17}}
\and
K.~Taylor\inst{\ref{inst20}}
}

\institute{\small
PPGFis \& Núcleo de Astrofísica e Cosmologia (Cosmo-ufes), Universidade Federal do Espírito Santo, 29075-910, Vitória, ES, Brazil\label{inst1}
\and
PPGCosmo \& Departamento de Física, Universidade Federal do Espírito Santo, 29075-910, Vitória, ES, Brazil\\
\email{marra@cosmo-ufes.org}\label{inst2}
\and
INAF -- Osservatorio Astronomico di Trieste, via Tiepolo 11, 34131 Trieste, Italy \label{inst2x}
\and
IFPU -- Institute for Fundamental Physics of the Universe, via Beirut 2, 34151, Trieste, Italy \label{inst2y}
\and
Departamento de Física, Universidade Federal de Sergipe, 49100-000, Aracaju, SE, Brazil \label{inst3}
\and
Donostia International Physics Center (DIPC),  Manuel Lardizabal Ibilbidea, 4, San Sebasti\'an, Spain\label{inst4}
\and
Ikerbasque, Basque Foundation for Science, E-48013 Bilbao\label{inst15}
\and
Academia Sinica Institute of Astronomy \& Astrophysics (ASIAA), 11F of Astronomy-Mathematics Building, AS/NTU, No.~1, Section 4, Roosevelt Road, Taipei 10617, Taiwan\label{inst5}
\and
Centro de Estudios de F\'\i sica del Cosmos de Arag\'on (CEFCA), Unidad Asociada al CSIC,  Plaza San Juan, 1, 44001, Teruel, Spain\label{inst22}
\and
Instituto de Astrof\'isica de Canarias, C/ V\'ia L\'actea, s/n, E-38205, La Laguna, Tenerife, Spain\label{inst9a}
\and
Departamento de Astrof\'isica, Universidad de La Laguna, E-38206, La Laguna, Tenerife, Spain\label{inst9b}
\and
Observat\'orio do Valongo, Universidade Federal do Rio de Janeiro, 20080-090, Rio de Janeiro, RJ, Brazil\label{inst10}
\and
Centro de Estudios de Física del Cosmos de Aragón (CEFCA), Plaza San Juan 1, 44001 Teruel, Spain\label{inst6}
\and
NSF's Optical-Infrared Astronomy Research Laboratory, Tucson, AZ 85719, USA\label{inst7}
\and
Instituto de Física, Universidade Federal do Rio de Janeiro, 21941-972, Rio de Janeiro, RJ, Brazil\label{inst21}
\and
Instituto de F\'isica, Universidade de S\~ao Paulo, 05508-090, São Paulo, SP, Brazil\label{inst13}
\and
Physics Department, Lancaster University, United Kingdom\label{inst12}
\and
Departamento de Astrofísica, Centro de Astrobiología (CSIC-INTA), ESAC Campus, Camino Bajo del Castillo s/n, E-28692 Villanueva de la Cañada, Madrid, Spain\label{inst11}
\and
Tartu Observatory, University of Tartu, Observatooriumi~1, 61602 T\~oravere, Estonia\label{inst8}
\and
Instituto de Astrof\'isica de Andaluc\'a - CSIC, Apdo 3004, E-18080, Granada, Spain\label{inst9}
\and
Observat\'orio Nacional, Minist\'erio da Ciencia, Tecnologia, Inovaç\~ao e Comunicaç\~oes, 20921-400, Rio de Janeiro, RJ, Brazil\label{inst14}
\and
Instituto de F\'isica, Universidade Federal da Bahia, 40210-340, Salvador, BA, Brazil\label{inst16}
\and
Departamento de Física-CFM, Universidade Federal de Santa Catarina, 88040-900, Florianópolis, SC, Brazil\label{inst23}
\and
Departamento de Astronomia, Instituto de Astronomia, Geofísica e Ciências Atmosféricas, Universidade de São Paulo, 05508-090, São Paulo, SP, Brazil\label{inst17}
\and
Department of Astronomy, University of Michigan, 311West Hall, 1085 South University Ave., Ann Arbor, USA\label{inst18}
\and
Department of Physics and Astronomy, University of Alabama, Box 870324, Tuscaloosa, AL, USA\label{inst19}
\and
Instruments4, 4121 Pembury Place, La Cañada Flintridge, CA 91011, USA\label{inst20}
}

\titlerunning{star-galaxy classification using machine learning in miniJPAS}
%
\authorrunning{Baqui, Marra et al.}



\abstract
{Future astrophysical surveys such as J-PAS will produce very large datasets, the so-called ``big data'', which will require the deployment of accurate and efficient Machine Learning (ML) methods.
In this work, we  analyze the miniJPAS survey, which observed about $\sim$$1 \text{deg}^2$ of the AEGIS field with  56 narrow-band filters and  4 $ugri$ broad-band filters.
The miniJPAS primary catalogue contains approximately 64000 objects in the $r$ detection band (mag$_{AB} \lesssim 24$), with forced-photometry in all other filters.}
{We discuss the classification of miniJPAS sources into extended (galaxies) and point-like (e.g.~stars) objects, a necessary step for the subsequent scientific analyses. We aim at developing an ML classifier that is complementary to traditional tools based on explicit modeling. In particular, our goal is to release a value added catalog with our best  classification.
}
{In order to train and test our classifiers, we crossmatched the miniJPAS dataset with SDSS and HSC-SSP data, whose classification  is trustworthy within the intervals $15\le r \le 20$ and $18.5 \le r \le 23.5$, respectively.
We trained and tested 6 different ML algorithms on the two crossmatched catalogs:
K-Nearest Neighbors (KNN), Decision Trees (DT), Random Forest (RF), Artificial Neural Networks (ANN), Extremely Randomized Trees (ERT) and  Ensemble Classifier (EC). EC is a hybrid algorithm that combines ANN and RF with J-PAS's stellar/galaxy loci classifier (SGLC).
As input for the ML algorithms we use the magnitudes from the 60 filters together with their errors, with and without the morphological parameters. We also use the mean PSF in the $r$ detection band for each pointing.}
{We find that the RF and ERT algorithms perform best in all scenarios.
When analyzing the full magnitude range of $15\le r \le 23.5$ we find $AUC=0.957$ (area under the curve) with RF when using only photometric information, and $AUC=0.986$ with ERT when using photometric and morphological information.
Regarding feature importance, when using morphological parameters, FWHM is the most important feature.
When using photometric information only, we observe that broad bands are not necessarily more important than narrow bands, and errors (the width of the distribution) are as important as the measurements (central value of the distribution).
In other words, the full characterization of the measurement seems to be important.
}
{ML algorithms can compete with traditional star/galaxy classifiers, outperforming the latter at fainter magnitudes ($r \gtrsim 21$).
We use our best classifiers, with and without morphology, in order to produce a value added catalog available at \href{https://j-pas.org/datareleases}{j-pas.org/datareleases} via the ADQL table \texttt{minijpas.StarGalClass}.
The ML models are available at \href{https://github.com/J-PAS-collaboration/StarGalClass-MachineLearning}{github.com/J-PAS-collaboration/StarGalClass-MachineLearning}.
}


\maketitle
%

\section{Introduction}

An important step in the analysis of data from wide-field surveys is the classification of  sources into stars and galaxies.
Although challenging, this separation is crucial for many areas of cosmology and astrophysics.
Different classification methods have been proposed in the literature,  each having their respective advantages and disadvantages.  One of the most used methods is based on morphological separation, where parameters related to the object structure and photometry are used \citep{Bertin:1996fj,2011MNRAS.412.2286H,Molino:2013oia,2019A&A...631A.156D,2019A&A...622A.177L}.
In these methods one assumes that stars appear as point sources while galaxies as extended sources.
This has been shown to be consistent with previous spectroscopic observations \citep{LeFevre:1995pq,Dawson:2012va,2013ApJS..208....5N}.  However, at fainter magnitudes, the differences between these point-like and extended structures decrease and this method becomes unreliable.
In what follows, by ``stars'' we mean point-like objects that are not galaxies, that is, both stars and quasars.%
\footnote{Also very compact galaxies such as Green Peas fall into the category of  point-like objects \citep{2009MNRAS.399.1191C, 2010ApJ...715L.128A}.
}

Future photometric surveys such as the Javalambre-Physics of the Accelerating Universe Astrophysical Survey \citep[J-PAS,][]{2014arXiv1403.5237B}\footnote{\href{http://www.j-pas.org}{www.j-pas.org}}
and the Vera Rubin Observatory Legacy Survey of Space and Time \citep[LSST,][]{2017arXiv170804058L}\footnote{\href{https://www.lsst.org}{www.lsst.org}} will detect a large number of objects and are facing the management of data produced at an unprecedented rate. The LSST, in particular, will reach a rate of petabytes of data per year \citep{garofalo_botta_ventre_2016}.
This wealth of data demands very efficient numerical methods but also gives us the opportunity to deploy Machine Learning (ML) algorithms, which, trained on big astronomical data, have the potential to outperform traditional methods based on explicit programming, if biases due to potentially unrepresentative training sets are kept under control.

ML has been widely applied in the context of cosmology and astrophysics, see \citet{2017ConPh..58...99I}. A non-exhaustive list of applications is photometric classification of supernovae \citep{2016ApJS..225...31L,2017ascl.soft05017C,VargasdosSantos:2019ovq}, gravitational wave analysis \citep{2013PhRvD..88f2003B,2015JPhCS.654a2001C}, photometric redshift \citep{2018A&A...616A..69B,2015MNRAS.452.3100C}, morphology of galaxies \citep{2010arXiv1005.0390G,2010MNRAS.406..342B}, and determination of atmospheric parameters for stellar sources \citep{2019A&A...622A.182W}.

ML applications to star-galaxy separation have been successfully performed on many surveys. \citet{2011AJ....141..189V}, for example, used various tree methods to classify SDSS sources. \citet{2015MNRAS.453..507K} used classifiers that mix supervised and unsupervised ML methods with CFHTLenS data. Recently, Convolutional Neural Networks (CNN) have been adopted: using images as input, they achieve an Area Under the Curve (AUC) > 0.99 for CFHTLenS and SDSS data \citep{2017MNRAS.464.4463K}. For more ML applications in the context of star/galaxy classification see \citet{2019arXiv190908626C,2018MNRAS.481.5451S,2019MNRAS.483..529C,2012ApJ...760...15F,2004AJ....128.3092O}.

Our goal here is to classify the objects detected by  Pathfinder miniJPAS  \citep{Bonoli:2020ciz}, which observed  $\sim$$1 \text{deg}^2$ of the AEGIS field
with the 56 narrow-band J-PAS filters and the 4 $ugri$ broad-band filters, for a total of approximately 64000 objects (mag$_{AB} \lesssim 24$).
The ML algorithms that we consider in this work are supervised and, for the learning process, need an external trustworthy classification.
We adopt Sloan Digital Sky Survey \citep[SDSS,][]{2015ApJS..219...12A} and Hyper Suprime-Cam Subaru Strategic Program \citep[HSC-SSP,][]{2019PASJ...71..114A} data.
We compare different ML models to each other and to the two classifiers adopted by the JPAS survey: the \texttt{CLASS\_STAR} provided by SExtractor \citep{Bertin:1996fj} and the stellar/galaxy loci classifier (SGLC) introduced in \citet{2019A&A...622A.177L}.

This paper is organized as follows.
In Section~\ref{jpas}, we briefly describe J-PAS and miniJPAS and we review the classifiers adopted in miniJPAS.
In Section~\ref{mlsec} we present the ML algorithms used in this work, and in Section~\ref{metrics} we define the metrics that we use to assess the performance of the classifiers.
Our results are presented in Sections~\ref{results} and~\ref{vac}, and our conclusions in Section~\ref{conclusions}.

\section{J-PAS and miniJPAS} \label{jpas}

J-PAS is a ground-based imaging survey that will observe 8500 deg$^2$ of the sky via the technique of quasi-spectroscopy: by observing with 56 narrow-band filters and 4 $ugr(i)$ broad-band filters%
\footnote{miniJPAS features also the $i$ band, while J-PAS is not expected to have it.}
it will produce a pseudo-spectrum ($R\sim 50$) for every pixel \citep[for the filters' specifications see][]{Bonoli:2020ciz}.
It features a dedicated 2.5m telescope with an excellent étendue, equipped with a 1.2 Gigapixel camera with a very large field of view of $4.2 \deg^2$. The observatory is on the mountain range ``Sierra de Javalambre'' (Spain), at an altitude of approximately 2000 meters, an especially dark region with the very good median seeing of $0.7''$ \citep{2010SPIE.7738E..0VC}.
Therefore, J-PAS  sits between photometric  and spectroscopic surveys, fruitfully combining the advantages of the former (speed and low cost) with the ones of the latter (spectra).
In particular, thanks to its excellent photo-$z$ performance, it will be possible to accurately study the large scale structure of the universe using the galaxy and quasar catalogs produced by J-PAS \citep{Bonoli:2020ciz}.

Between May and September 2018, the 2.5m J-PAS telescope with its filter set was equipped with the Pathfinder camera, used to test the telescope performance and execute the first scientific operations. The camera features a 9k $\times$ 9k CCD, with a 0.3 deg$^2$ field-of-view and 0.225 arcsec pixel size.
This led to the miniJPAS survey which covered a total of $\sim 1 \text{deg}^2$ of the AEGIS field,%
\footnote{See \citet{Davis:2006tn} for informations on the All-wavelength Extended Groth strip International Survey (AEGIS).}
reaching the target depth planned for J-PAS (mag$_{AB}$, 5$\sigma$ in a 3'' aperture, between 21.5 and 22.5 for the narrow-band filters and up to 24 for the broad-band filters).
miniJPAS  consists of the 4 fields/pointings AEGIS1-4, each of approximately 0.25 deg$^2$ field-of-view.
The miniJPAS primary catalogue contains 64293 objects in the $r$ detection band, with forced-photometry in all other filters. See \citet{Bonoli:2020ciz} for the presentation paper.  The miniJPAS Public Data Release was presented to the public in December 2019.\footnote{\href{https://j-pas.org/datareleases/minijpas_public_data_release_pdr201912}{j-pas.org/datareleases/minijpas\_public\_data\_release\_pdr201912}}



\subsection{Crossmatched catalogs} \label{xmatch}

\begin{figure}
\centering
 \includegraphics[width=.98 \columnwidth]{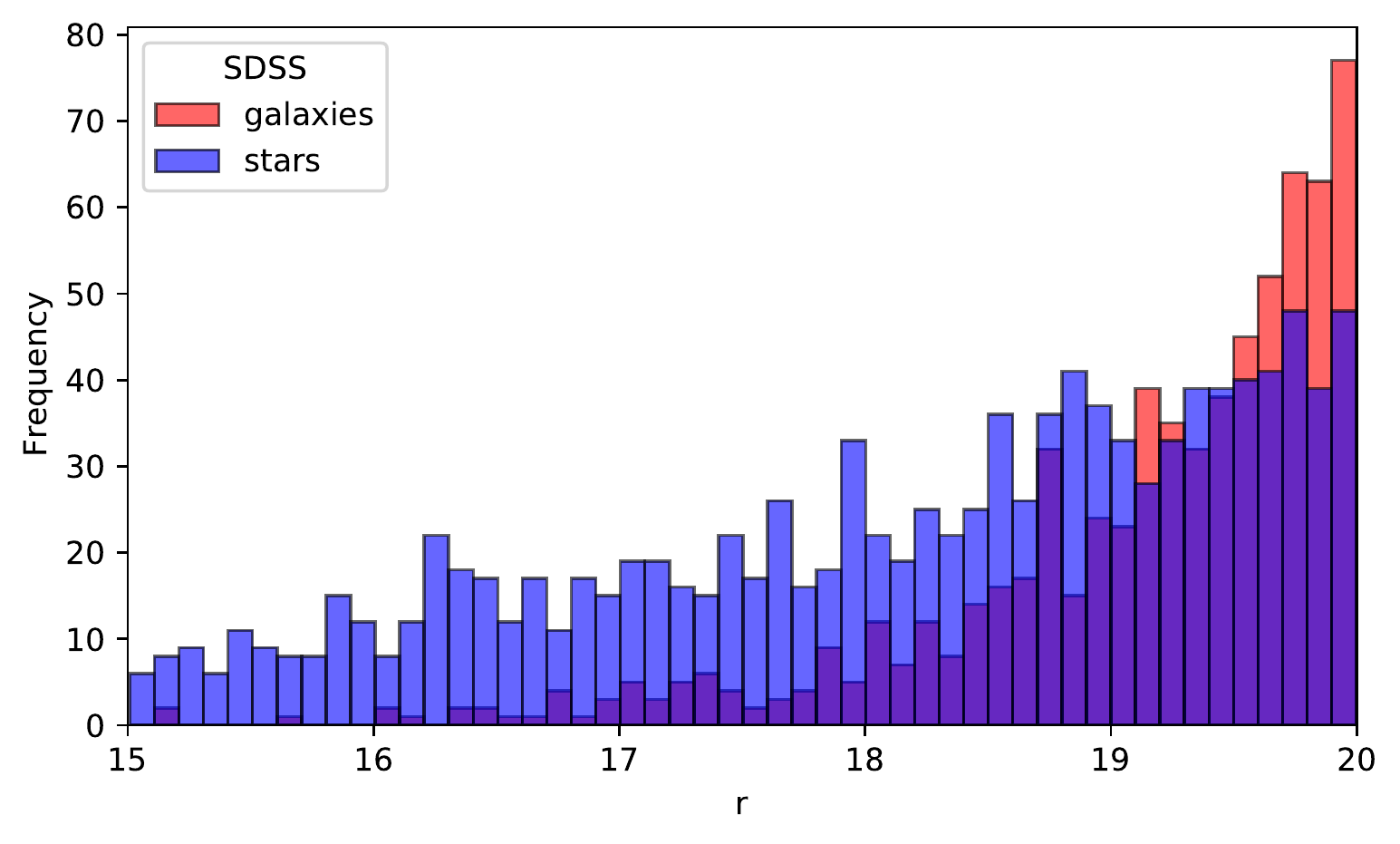}
 \includegraphics[width=.98 \columnwidth]{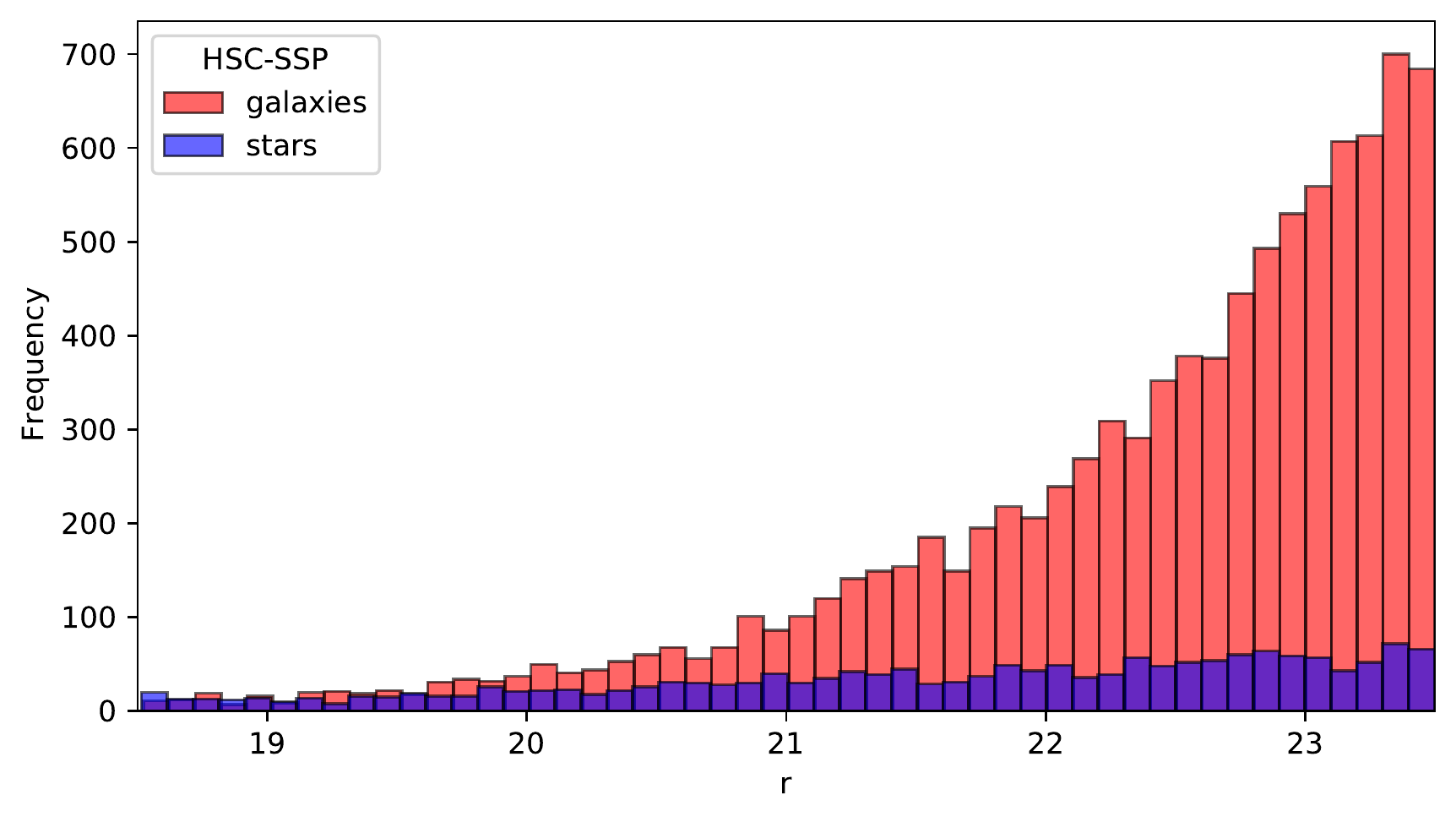}
\caption{
Histograms of the $r$-band magnitudes of the sources of the miniJPAS catalog crossmatched with the SDSS (top) and HSC-SSP (bottom) catalogs. Classification by SDSS and HSC-SSP, respectively. Galaxies are shown in red, stars in semi-transparent blue.}
\label{fig:catalogs}
\end{figure}

\begin{figure}
\centering
 \includegraphics[width=.98 \columnwidth]{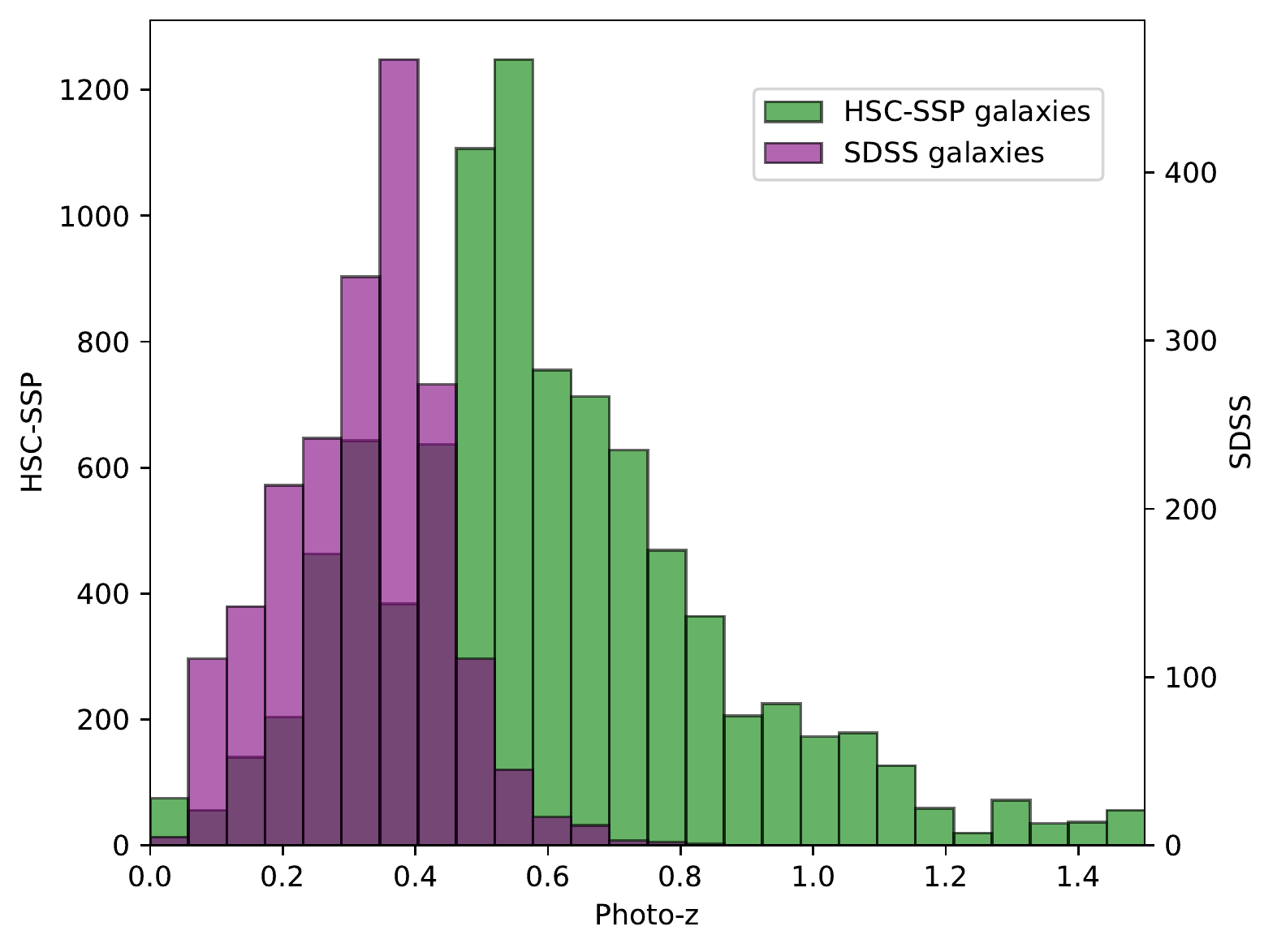}
\caption{Histograms of the photometric redshifts of the galaxies of the miniJPAS catalog crossmatched with the SDSS (semi-transparent purple) and HSC-SSP (green) catalogs. }
\label{fig:z_distrib}
\end{figure}

The goal of this paper is to develop an ML model that can accurately classify the objects detected by  Pathfinder miniJPAS.
As we will consider supervised ML algorithms, we need, for the learning process, a trustworthy classification by some other survey that has a sufficiently high overlap with  miniJPAS.
We use SDSS\footnote{\href{https://www.sdss.org/dr12/}{sdss.org/dr12/}} and HSC-SSP\footnote{\href{https://hsc-release.mtk.nao.ac.jp/doc/}{hsc-release.mtk.nao.ac.jp/doc/}} data, whose classification is expected to be trustworthy within the intervals $15\le r \le20$ and $18.5\le r \le23.5$, respectively.
As said earlier,  by ``stars'' we mean point-like objects that are not galaxies, that is, both stars and quasars. We assume that the classification by SDSS and HSC-SSP is trustworthy within this definition \citep{2015ApJS..219...12A,2019PASJ...71..114A}.

We found 1810 common sources with SDSS, 691 galaxies and 1119 stars, and 11089 common sources with HSC-SSP, 9398 galaxies and 1691 stars. See Fig.~\ref{fig:catalogs} for the $r$-band distributions of stars and galaxies and Fig.~\ref{fig:z_distrib} for the redshift distribution of galaxies.

\subsubsection{SDSS classification}

SDSS is a photometric and spectroscopic survey conducted at the Apache Point Observatory (New Mexico, USA) with a 2.5-m primary mirror. 
We used the SDSS DR12 photometric catalog \texttt{minijpas.xmatch\_sdss\_dr12}\footnote{For details, see \href{http://archive.cefca.es/catalogues/minijpas-pdr201912}{archive.cefca.es/catalogues/minijpas-pdr201912}}.
Stars are defined according to an extendedness (difference between the CModel and PSF magnitudes) less than 0.145.\footnote{\href{https://www.sdss.org/dr12/algorithms/classify}{www.sdss.org/dr12/algorithms/classify/\#photo\_class}}

In order to test the photometric calibration by SDSS we crossmatched the latter with the catalog from the ALHAMBRA (Advance Large Homogeneous Area Medium Band Redshift Astronomical) survey \citep{2008AJ....136.1325M}.%
\footnote{\href{http://svo2.cab.inta-csic.es/vocats/alhambra}{svo2.cab.inta-csic.es/vocats/alhambra}}
We obtained 1055 sources after imposing mask and saturation flags.
As discussed in \citet{Molino:2013oia}, ALHAMBRA provides a trustworthy classification in the magnitude range $15\le r \le 21$.

\begin{figure}
\centering
\includegraphics[width=.98 \columnwidth]{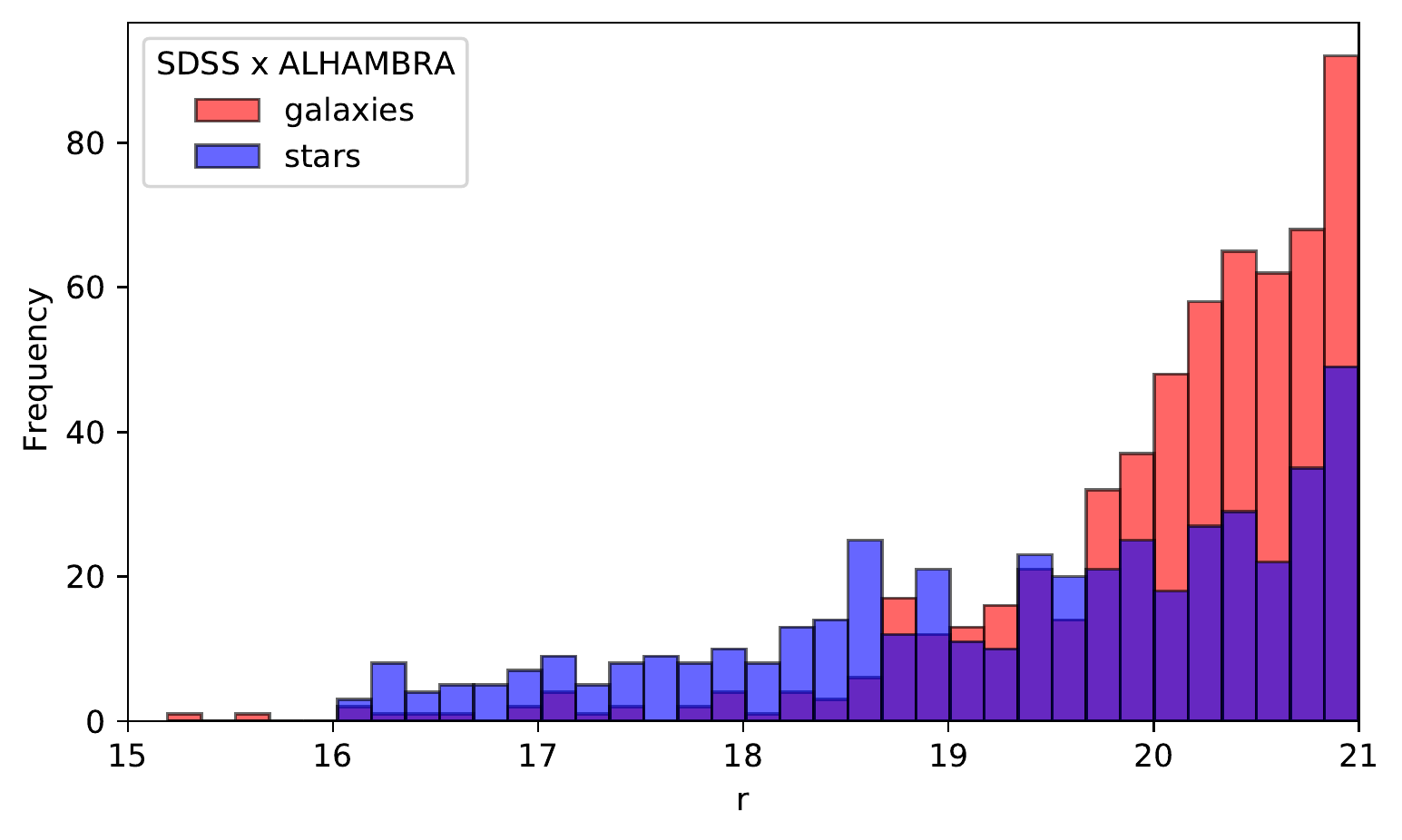}
\includegraphics[width=.98 \columnwidth]{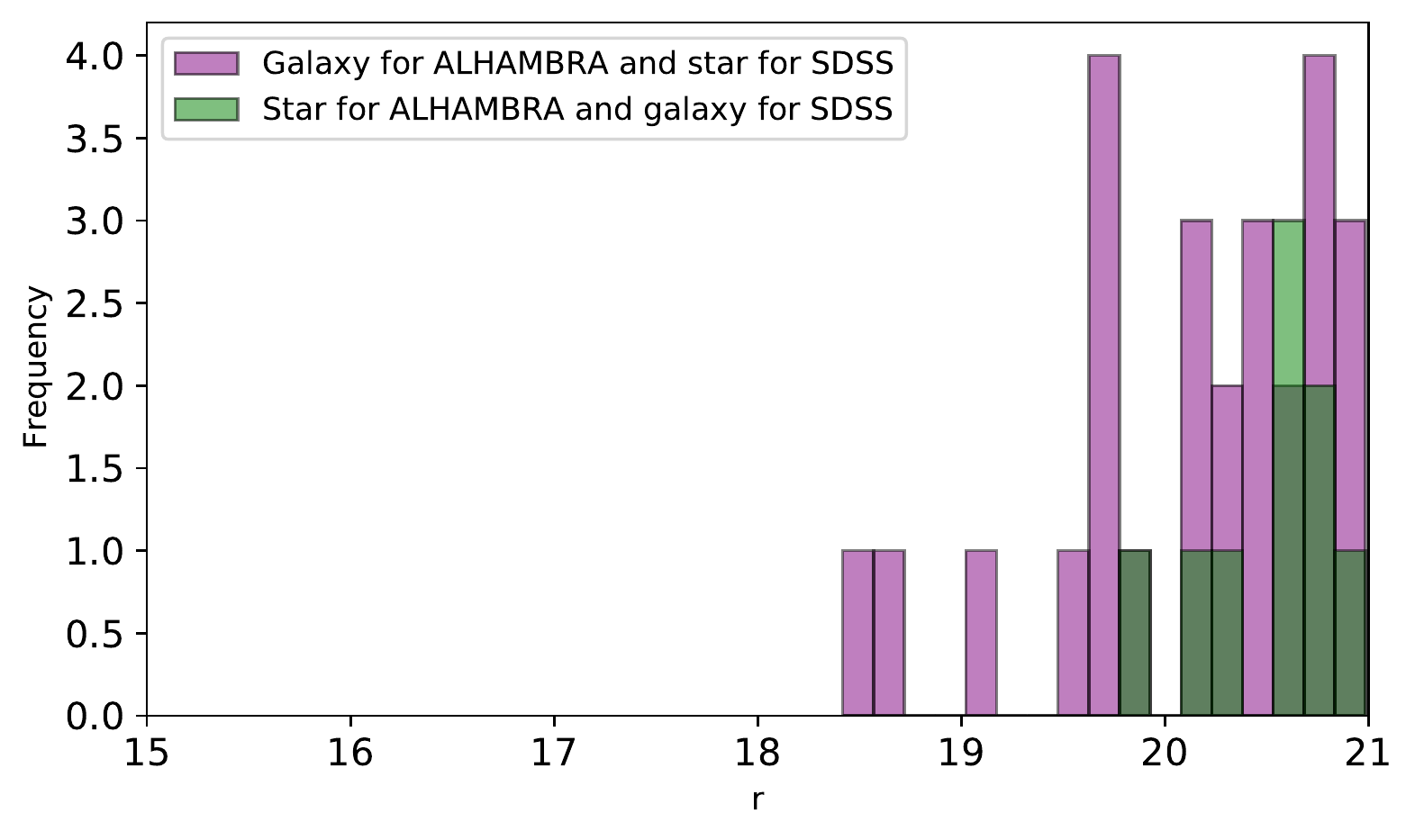}
\caption{
Top: Histograms of the $r$-band magnitudes of the objects resulting
 from the crossmatch between the SDSS catalog used in this paper and ALHAMBRA. Galaxies are shown in red, stars in semi-transparent blue.
Bottom: disagreement between SDSS and ALHAMBRA as a function of $r$ magnitude. Sources classified as galaxies by ALHAMBRA and as stars by SDSS are in purple, while vice versa in semi-transparent green.}
\label{fig:sdssclass}
\end{figure}

As one can see from Fig.~\ref{fig:sdssclass} (top) ALHAMBRA covers the relevant magnitude range and agrees with SDSS well till $r=20$ (bottom).
Indeed, within $15\le r \le 20$, the percentages of false negatives and false positives are 0.2\% and 1.9\%, respectively (positive refers to the object being a galaxy). Note that, for the value added catalog, we will use SDSS in the more limited range $15\le r \le 18.5$ so that the percentages of false negatives and false positives are 0\% and 0.7\%, respectively  (using $p_{\rm cut}=0.5$, see Section~\ref{coma}).



\subsubsection{HSC-SSP classification}

The HSC-SSP is a photometric survey with a 8.2-m primary mirror located in Hawaii, USA. We crossmatched the miniJPAS data with the wide field from the Public Data Release 2.
Stars are defined according to an extendedness less than 0.015.\footnote{\href{https://hsc-release.mtk.nao.ac.jp/doc/index.php/stargalaxy-separation-2/}{hsc-release.mtk.nao.ac.jp/doc/index.php/stargalaxy-separation-2/}}
We used the following data quality constraints:  \texttt{isprimary = True}, \texttt{r\_extendedness\_flag!=1} and \texttt{r\_inputcount\_value>=4} for HSC-SSP, and \texttt{flag=0} and \texttt{mask=0} for miniJPAS.
The crossmatch was performed with the \texttt{TOPCAT}%
\footnote{\href{http://www.star.bris.ac.uk/~mbt/topcat/}{www.star.bris.ac.uk/~mbt/topcat/}}
software with a tolerance of 1 arcsec.

In order to test the photometric calibration by HSC-SSP we crossmatched the latter with the spectroscopic  catalogs from the DEEP2 Galaxy Redshift Survey (1992 sources \cite{2013ApJS..204...21M}). We could not use this spectroscopic catalog to check the photometric SDSS calibration because it does not cover the required magnitude range.

\begin{figure}
\centering
\includegraphics[width=.98 \columnwidth]{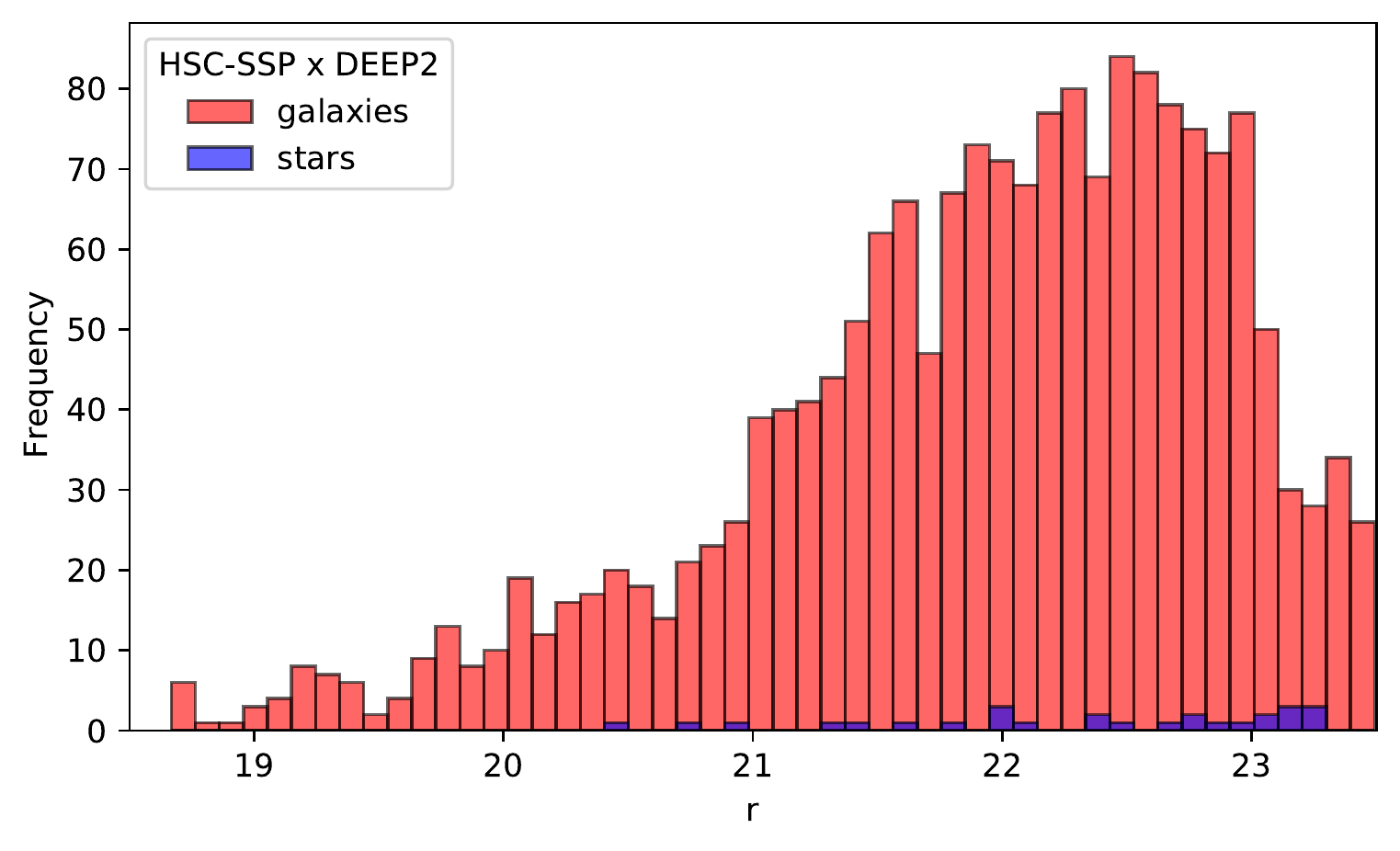}
\includegraphics[width=.98 \columnwidth]{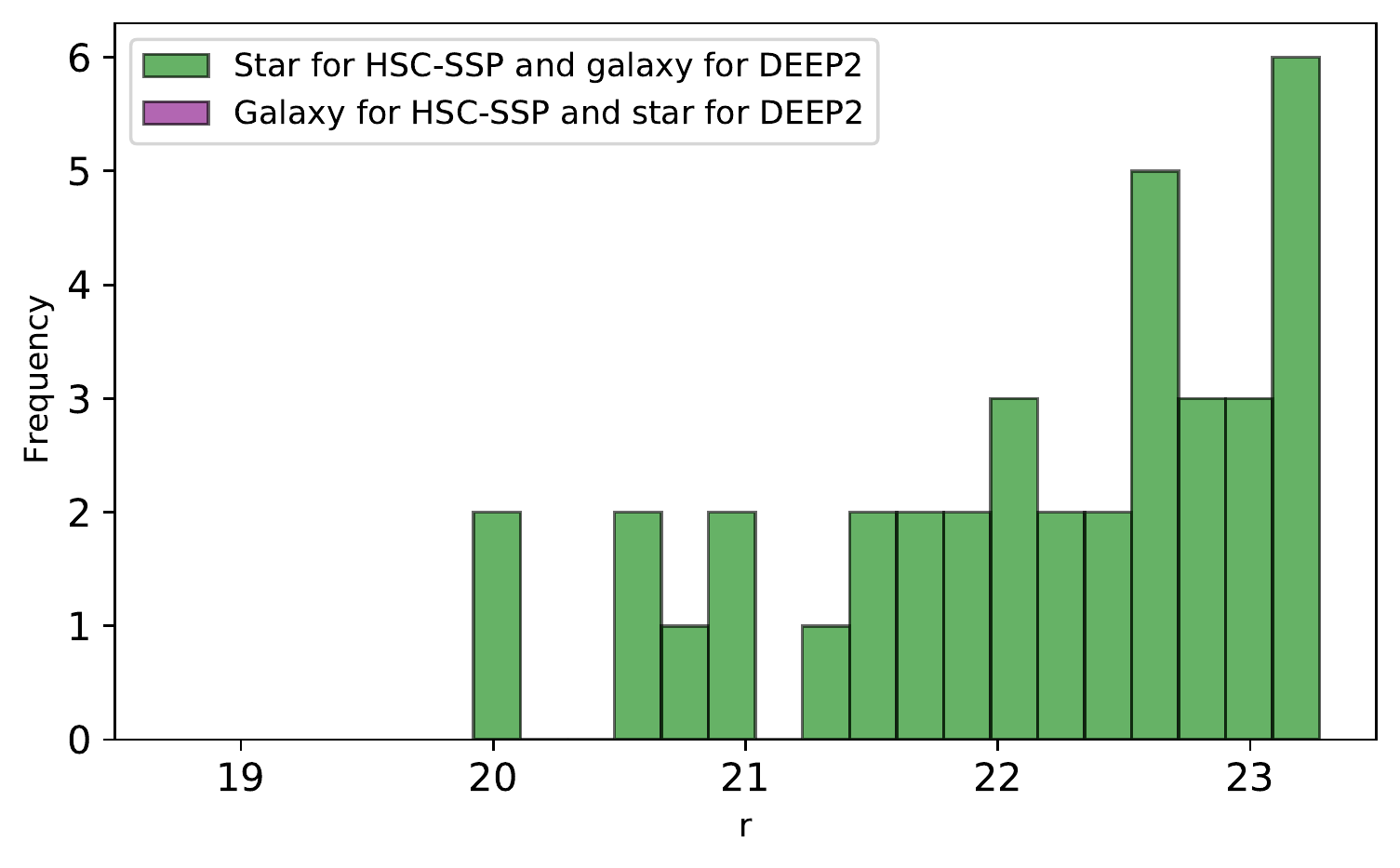}
\caption{
Top: Histograms of the $r$-band magnitudes of the objects resulting
 from the crossmatch between the HSC-SSP catalog used in this paper and DEEP2.
Bottom: disagreement between HSC-SSP and DEEP2 as a function of $r$ magnitude. No object was classified as galaxy by HSC-SSP and star by DEEP2.}
\label{fig:subaruclass}
\end{figure}

As one can see from Fig.~\ref{fig:subaruclass} (top) DEEP2 covers the relevant magnitude range and agrees with HSC-SSP well (bottom).
Indeed, for the range $18.5\le r \le23.5$, the percentages of false negatives and false positives are 1.9\% and 0\%, respectively.


\subsection{Input parameters for the ML algorithms}

The features that are used as input for our algorithms can be grouped into  photometric and  morphological classes. Besides these two sets of features, we also consider the average PSF in the $r$ detection band of the 4 fields of miniJPAS, which is 0.70" for AEGIS1, 0.81" for AEGIS2, 0.68" for AEGIS3 and 0.82" for AEGIS4.
The different PSF values signal different observing conditions: by including the PSF value we let the ML algorithms know that data is not homogeneous.

\subsubsection{Photometric information}

As photometric information we consider the \texttt{MAG\_AUTO} magnitudes associated to the 60 filters together with their errors. The rationale behind including the errors is that, in this way, one can characterize the statistical distribution associated to a magnitude measurement.
Indeed, observations may suffer from inhomogeneity due to varying observing conditions and the measurement errors should be able to account, at least in part, for this potential bias. 
As we will see, how well can one measure the magnitude associated to a filter may be more important than the actual measurement.

As said earlier, sources are detected in the $r$ band so that one may have non-detection in the other filters.
Null or negative fluxes (after background subtraction) are assigned a magnitude value of 99.
The ML algorithms are expected to learn that 99 marks missing values.

\subsubsection{Morphological information}

\begin{figure}
\centering
 \includegraphics[width=.98 \columnwidth]{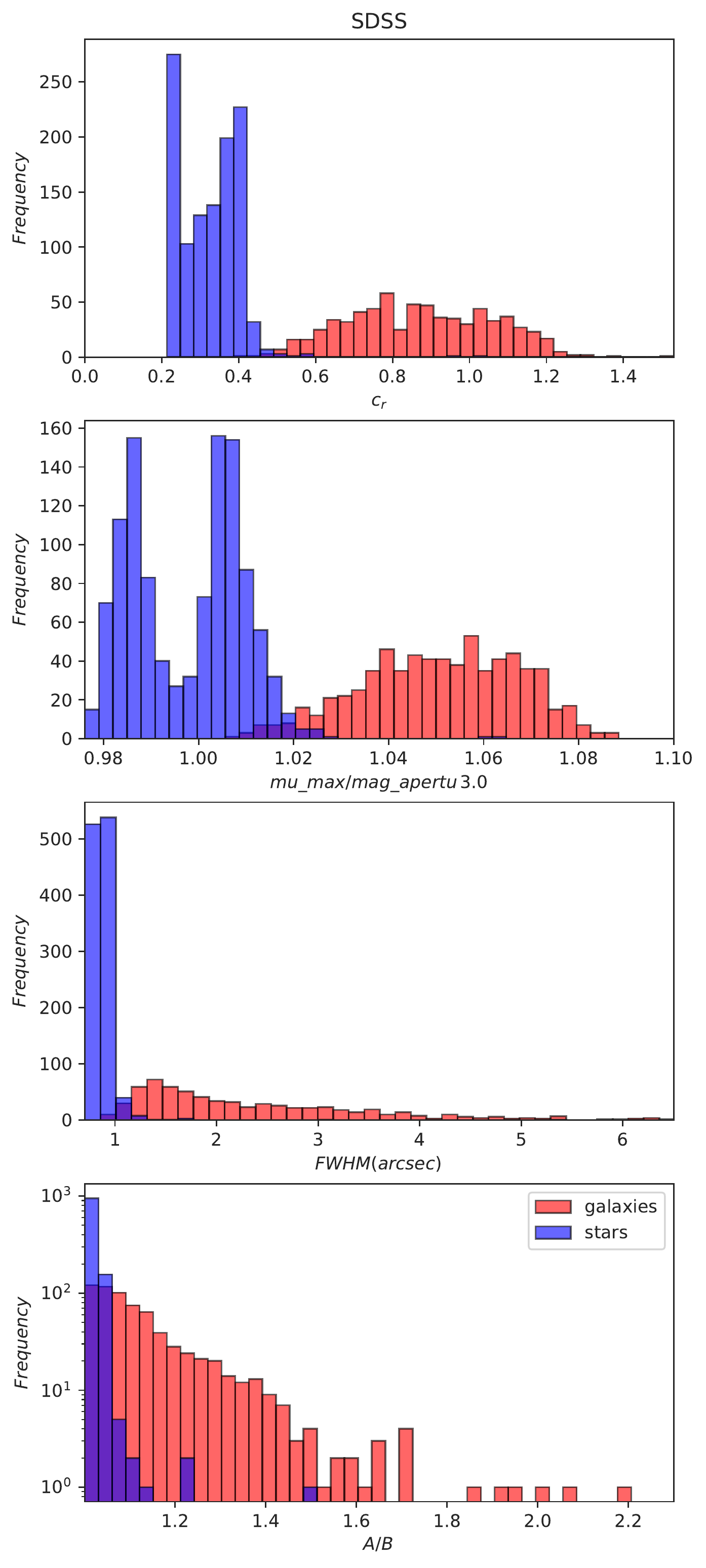}
\caption{
Distributions of the morphological parameters of stars and galaxies for the miniJPAS catalog crossmatched with SDSS. Galaxies are shown in red, stars in semi-transparent blue.}
\label{fig:mor-sdss}
\end{figure}

\begin{figure}
\centering
 \includegraphics[width=.98 \columnwidth]{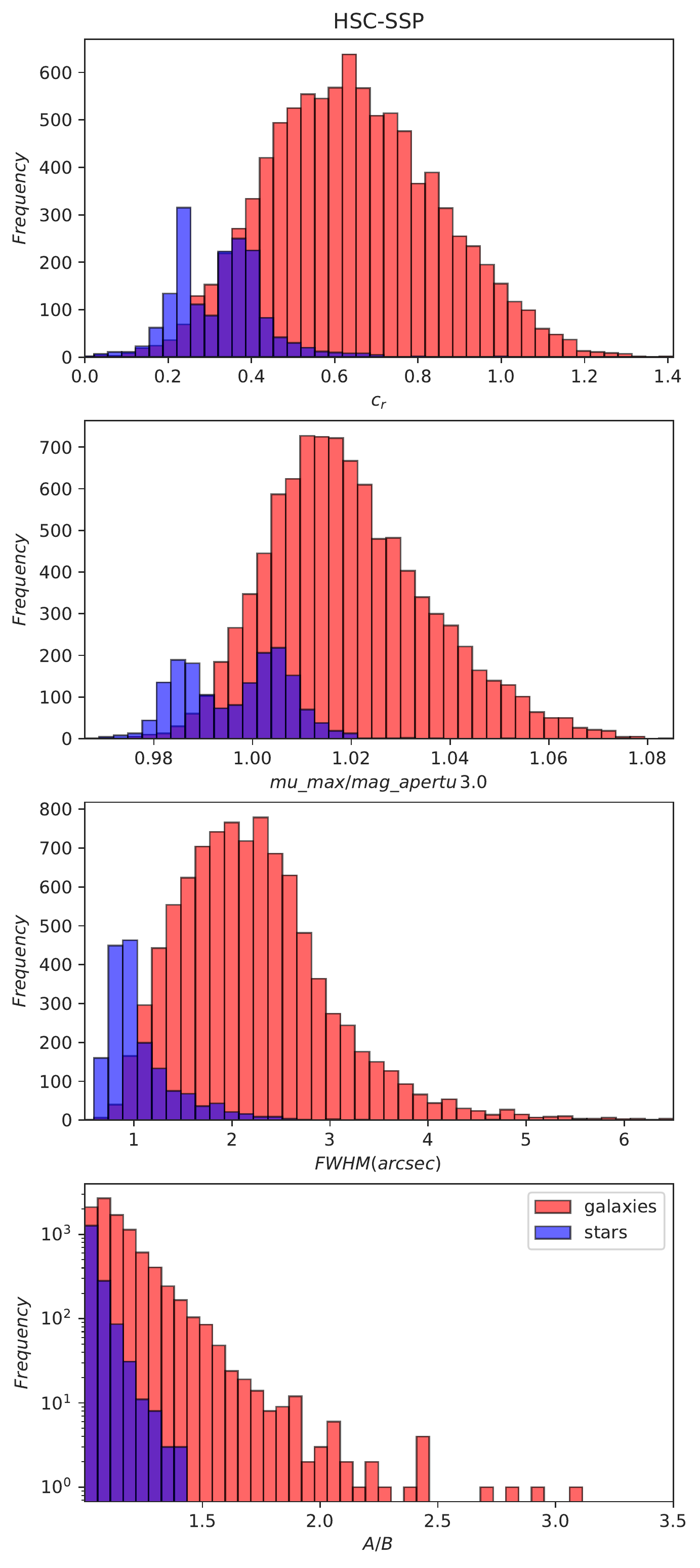}
\caption{
Distributions of the morphological parameters of stars and galaxies for the miniJPAS catalog crossmatched with HSC-SSP. Galaxies are shown in red, stars in semi-transparent blue.}
\label{fig:mor-hsc}
\end{figure}

We consider the following 4 morphological parameters: 
\begin{itemize}

\item  concentration $c_r=r_{1.5''}-r_{3.0''}$, where $r_{1.5''}$ and $r_{3.0''}$ are the $r$-band magnitudes within fixed circular apertures of 1.5'' and  3.0'', respectively,

\item  ellipticity $A/B$, where $A$ and $B$ are the RMS of the light distribution along the maximum and minimum dispersion directions, respectively. 

\item the full width at half maximum $FWHM$ assuming a Gaussian core,

\item \texttt{MU\_MAX/MAG\_APER\_3\_0} ($r$ band), where \texttt{MU\_MAX} and \texttt{MAG\_APER\_3\_0} are the peak surface brightness above background and the magnitude  within 3.0", respectively. Note that here we are taking the ratio in order to have a parameter that is complementary to $c_r$.

\end{itemize}
Figures~\ref{fig:mor-sdss} and~\ref{fig:mor-hsc} show their distributions for stars and galaxies and the two catalogs.
The stellar bimodality in $c_r$ and \texttt{MU\_MAX/MAG\_APER\_3\_0} is due to the fact that the four fields feature a different average PSF.
We discuss these figures when examining feature importance in Section~\ref{featureimpo}.

\subsection{J-PAS star/galaxy classifiers} \label{sgclass}

\begin{figure}
\centering
 \includegraphics[width= \columnwidth]{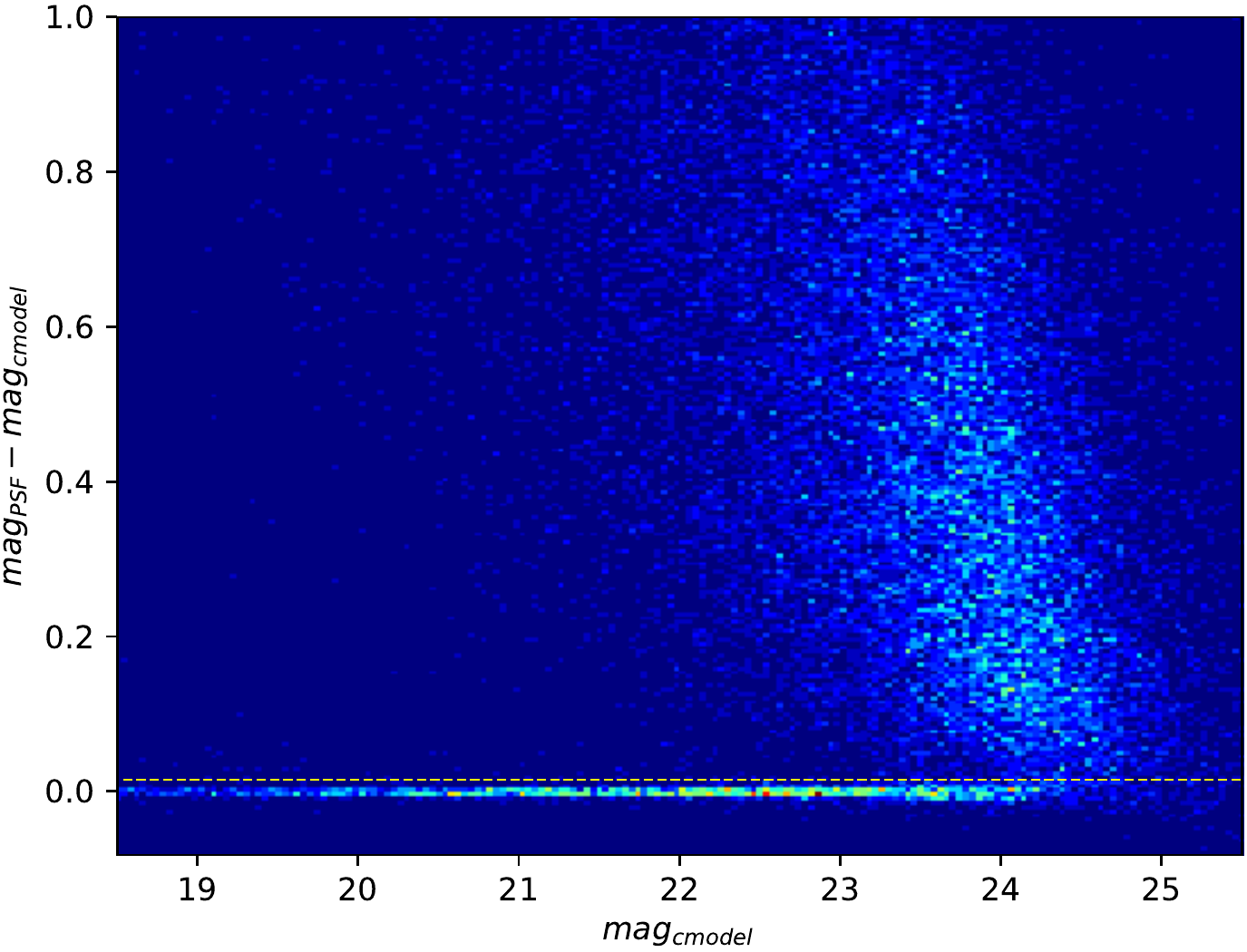}
\caption{Density of objects as a function of extendedness (the difference between $mag_{PSF}$ and $mag_{cmodel}$) and $mag_{cmodel}$ for HSC-SSP data. The yellow line marks an extendedness of 0.015.
According to this morphological classification the sources below the cut are stars and the ones above the cut are galaxies.}
\label{morphology_HSC-SSP}
\end{figure}

Here, we briefly discuss the star/galaxy classifiers available for  miniJPAS.
However, first we show how HSC-SSP classifies objects into stars and galaxies.
This is performed by drawing a ``hard cut'' in the source parameter space.
In Figure~\ref{morphology_HSC-SSP} we plot the difference between $mag_{PSF}$ and $mag_{cmodel}$ as a function of $mag_{cmodel}$ for the HSC-SSP data using their $r$ band \citep[for the definitions see][]{2019PASJ...71..114A}. Stars are expected to have $mag_{PSF} \simeq mag_{cmodel}$ while galaxies, due to their extended structure, should feature $mag_{PSF} > mag_{cmodel}$.
Therefore, one can separate stars from galaxies via a cut in the extendedness parameter $mag_{PSF}-mag_{cmodel}$, which we show with a yellow line in Figure~\ref{morphology_HSC-SSP}.
The disadvantage of this model is that it provides an absolute classification for a scenario in which the uncertainties increase as we move toward weaker magnitudes. Note that for $r_{cmodel} \gtrsim 24$ the separation is not reliable as stars do not cluster anymore around a null extendedness.

\subsubsection{\texttt{CLASS\_STAR}}

SExtractor \citep[Source Extractor,][]{Bertin:1996fj} is a software developed for processing large images (60k $\times$ 60k pixels). It has been widely applied to photometric surveys including miniJPAS. Besides detecting sources, SExtractor also classifies objects into stars and galaxies. The software has two internal classifiers, \texttt{CLASS\_STAR} and \texttt{SPREAD\_MODEL}. miniJPAS includes the classification via \texttt{CLASS\_STAR} which is based on neural networks (see Section~\ref{sec:ann}).\footnote{\href{https://sextractor.readthedocs.io/en/latest/ClassStar.html}{sextractor.readthedocs.io/en/latest/ClassStar.html}}
The network has 10 inputs: 8 isophotal areas, the peak intensity and the ``seeing'' control parameter.
The output is probabilistic and quasars are classified as stars (in agreement with our convention).
\texttt{CLASS\_STAR} is reliable up to $r \sim 21$ \citep[see also][]{Bertin:1996fj}.

\subsubsection{Stellar/galaxy loci classifier} \label{cpdf}

miniJPAS includes the Bayesian classifier (SGLC) developed by \citet{2019A&A...622A.177L} for J-PLUS data.%
\footnote{\href{http://www.j-plus.es/datareleases}{j-plus.es/datareleases}}
The concentration versus magnitude diagram presents  a bimodal distribution, corresponding to compact point-like objects and extended sources. \citet{2019A&A...622A.177L} models both distributions to obtain the probability of each source to be compact or extended.
The  model with suitable priors is then used to estimate the Bayesian probability that a source is a star or a galaxy.
Also in this case quasars are expected to be classified as ``stars.''
This method was updated to miniJPAS data, in particular a different galaxy population model was adopted.
See~\citet{Bonoli:2020ciz} for more details.

\section{Machine learning} \label{mlsec}

Machine learning is a branch of artificial intelligence that includes statistical and computational methods dedicated to providing predictions or taking decisions without being explicitly programmed to perform the task.
Machine learning is employed in a variety of computing tasks, for which the explicit programming of well-performing algorithms is difficult or unfeasible.
ML methods can either be supervised or unsupervised. The former learn from pre-classified data that has known inputs and outputs.
When  classification is unavailable, one relies instead on unsupervised methods, which can group  items that are related in the parameter space, i.e., learn without the need of external information.

In this paper, we focus on binary supervised classification methods.
In this case, the model (the internal parameters of the algorithm) is implicitly adjusted via the ``training set.''
Its performance is then tested with the remaining part of the dataset---the ``test set.''
Specifically, the internal parameters of the prediction function $f: \mathbb{R}^n \to Y$ are trained via the training dataset $\mathbf{x}_i \in \mathbb{R}^n$ ($n$ is the dimensionality of the feature space, $i$ labels the elements of the training set) with classifications $y_i$ $\in \{0,1\}$, where 1 stands for  galaxy and 0 for  star.
Classifiers are divided into non-probabilistic and probabilistic classifiers. The former type of classifier outputs the best class while the latter the probability of the classes (the best class is taken as the one with the highest probability).
Here, we will consider only  binary probabilistic classifiers so that it is $f: \mathbb{R}^n \to [0,1]$, that is, $f$ gives the probability that  an object is a galaxy. The probability of being a star is simply $1-f$.
A value of $f$ close to 1 means that the object is likely a galaxy.

According to the No-Free Lunch theorem there is not an ideal algorithm that performs better than the others in any situation~\citep{6795940}.
As it is impossible to test all the methods available with all the possible choices of hyperparameters, 
we followed the strategy to explore firstly some of the main ML methods, namely, K-Nearest Neighbors (KNN), Decision Trees (DT), Random Forest (RF) and Artificial Neural Networks (ANN).%
\footnote{We also tested Support-Vector Machine (SVM) with the linear, polynomial and Radial Basis Function (RBF) kernels. We found results similar to DT and KNN.}
Subsequently, because of the best response of the RF technique, we decide to focus on decision tree algorithms and ensemble models, so
we added Extremely Randomized Trees (ERT) and Ensemble Classifier (EC) to our analysis.
These algorithms can be used for both regression and classification.%
\footnote{
While classification is used to predict if an object belongs to a class, regression is used to predict real valued outputs that do not belong to a fixed set. For example, regression is used when one uses photometric information in order to predict the source's redshift.}
Here, we will only consider classification.
We implemented these algorithms using the \texttt{scikit-learn}\footnote{\href{https://scikit-learn.org/}{scikit-learn.org}} package written in python \citep{Pedregosa:2012toh}. For more information about supervised learning see \citet{mitchell1997machine,hastie2009elements}.
For the training and test sets we use 80\% and 20\% of the crossmatched catalogs, respectively. The division is performed randomly.
This guarantees a good training and an accurate testing.
A 70\%-30\% split is also a viable alternative.
As mentioned in Section~\ref{xmatch}, the training sets are unbalanced as they feature a different number of galaxies and stars. We will show the purity curves for stars and galaxies in order to estimate the performance for each class.
We now briefly review the six ML algorithms adopted in this paper. The hyperparameters used in the algorithms can be found in appendix \ref{hyperparameter}.


\subsection{K-Nearest-Neighbors}
The KNN algorithm is one of the most simple ML methods \citep{altman1992introduction,hastie2009elements}. It calculates the distance between the element to be classified (within the test set) and the ones belonging to the training set. The predicted class will be calculated using the $k$ nearest neighbors.
Although in this work we use the Euclidean metric, it is possible to choose others metrics to compute the distances. This method is very fast and its computational cost is proportional to the size of training set.

The output of the model is discrete if one uses the majority vote from the $k$ nearest neighbors.\footnote{A vote is a classification by a neighbor.}
Here, we use the probabilistic version which assigns a probability to each class. In this case the classification is given by the average of the nearest $k$ neighbors:
\begin{equation}
f(\mathbf x_q)=\frac{\sum_{i=1}^k w_i f(\mathbf x_i)}{\sum_{i=1}^k w_i} \qquad \textrm{with} \qquad w_i = \frac{1}{d(\mathbf x_q, \mathbf x_i)^2} \,,
\end{equation}
where the sum over the $k$ nearest neighbors is weighted by the weights $w_i$ which are the inverse of the square of the distance $d(\mathbf x_q, \mathbf x_i)$ from the neighbors~($\mathbf x_i$) to the element to be classified~($ \mathbf x_q$, $q$ labels the test set), and $f( \mathbf x_i)=y_i$ are the classifications of the training set.
As discussed in Section~\ref{kfold}, the number $k$ of neighbors is optimized via $k$-fold cross-validation. KNN has the advantage of being simple, intuitive and competitive in many areas. However, its computational complexity increases with the number of data.



\subsection{Decision Trees} \label{sec:dt}

DT methods \citep[see][]{breiman1984classification,hastie2009elements} divide recurrently the parameter space according to a tree structure, following the choice of minimum class impurity of the groups at every split.
To build a Decision Tree we  first define an Information Gain (IG) function:
\begin{equation}
   IG(D_p,x_t)=I(D_p)-\frac{N_{\rm left}}{N_p}I(D_{\rm left})-\frac{N_{\rm right}}{N_p}I(D_{\rm right}) \,,
\end{equation}
where $D_p$ is the parent dataset of size $N_p$, $D_{\rm left}$ and $D_{\rm rigth}$ are the child datasets of sizes $N_{\rm left}$ and $N_{\rm right}$, respectively, and $I$ is a function called impurity.  At every step the dataset is divided according to the feature and threshold $x_t$%
\footnote{Within our notation, $x_t$ is the threshold for the feature that maximizes~$IG$ (there are $n$ features).}
that maximize the $IG$ function, or, equivalently, that minimize the impurity in the children dataset.
We considered several impurity functions, such as entropy, classification error and Gini. For example, the latter is:
\begin{equation} \label{inpu}
    I_G(m)= 1-\sum_{i=0,1} p(i|m)^2 \,,
\end{equation}
where $p(i|m)$ is the fraction of data belonging to the class $i$ (0 or 1) for a particular node $m$ that splits the parent dataset into the child datasets.
After the growth of the tree is completed, the feature space is divided with probabilities associated to each class, and the probability for a test element is that of the region it belongs to.

During the branching process described above, some features appear more often than others. Using this frequency we can measure how important each feature is in the prediction process. We define the importance of each feature as:
\begin{equation} \label{impo}
      Imp(x)=\sum_t \frac{N_p}{N_{\rm tot}}IG(D_p,x_t) \,,
\end{equation}
where $N_{\rm tot}$ is the size of the dataset.
The higher the number of times a feature branches a tree, higher  its importance.
Note that the first features that divide the tree tend to be of greater importance because the factor $N_p/N_{\rm tot}$ in Eq.~\eqref{impo} decreases as the tree grows ($N_p$ decreases).
DT is characterized by an easy interpretability and handling but it is sensitive to small changes in the training set, thus suffering from potential biases.


\subsection{Random Forest}
Random Forest \citep{breiman2001random,hastie2009elements} is an ensemble algorithm built from a set of decision trees (the forest). Each tree generates a particular classification and the RF prediction is the combination of the different outputs.
Each tree is different because of the stochastic method used to find the features when maximizing the IG function.
Moreover, using the bootstrap statistical method, different datasets are built from the original one in order to grow more trees. 
For the discrete case the output is built from the majority vote, as seen with the KNN algorithm. For the probabilistic case we calculate the RF output as the average of the probabilities of each class for each tree.
Finally, one computes the feature importances $Imp(x)$ for each tree of the ensemble and then averages them to obtain the RF feature importance. 
The diversity of trees decreases the bias as compared to DT, generating  globally better models. On the other hand, RF requires greater memory and time as compared to DT.


\subsection{Extremely Randomized Trees}

Extremely Randomized Trees \citep{geurts2006extremely} is an ensemble method similar to RF. There are only two differences between RF and ERT. The first is that ERT originally does not use bootstrap, although the implementation in \texttt{scikit-learn} allows one to insert it in the analysis. The second is that, while RF tries to find the best threshold for a features via the $IG$ function, in ERT the division is done randomly.
Then, of all the randomly generated splits, the split that yields the highest score is chosen to split the node.
For large datasets ERT algorithms are faster than RF ones and yield a similar performance \citep{geurts2006extremely}.


\subsection{Artificial Neural Networks} \label{sec:ann}

Artificial Neural Networks mimic the functioning of the nervous system, being able to recognize patterns from a representative dataset \citep[for an introduction see][]{mitchell1997machine,hastie2009elements}.
The model we will use in our analysis consists of a simple supervised model called Multilayer Perceptron (MLP).

MLP consists of a set of perceptrons arranged in different layers. A perceptron, or artificial neuron, is a binary classifier.
The data features are inserted in the input layer, the learning process occurs in the hidden layers, and the object classification is performed by the output layer. The information in the hidden layers is passed through each perceptron several times until convergence. In this algorithm, we can have several layers containing hundreds of perceptrons.
To train the neural network, one uses a Cost Function that should be minimized. As learning method we use backpropagation \citep{1986Natur.323..533R}.
We use \texttt{LBFGS}, \texttt{Stochastic Gradient Descent} and \texttt{Adam} cost functions, besides various activation functions. The values of the hyperparameters that give the best performance are given in Appendix \ref{hyperparameter}. In particular, we adopted 1 hidden layer with 200 neurons.
ANN algorithms are very competitive  and have the ability to deal with complex non-linear problems, but possess low interpretability and require powerful processors.

Finally, we briefly discuss the differences between \texttt{CLASS\_STAR} and the ANN classifier used in this work.
First, our ANN classifier is trained on real miniJPAS data, while \texttt{CLASS\_STAR} was trained on simulated images. Second, although they both feature one hidden layer, \texttt{CLASS\_STAR} has an input layer with 10 parameters and a hidden layer with 10 neurons while our classifier uses an input layer with 64 parameters (4 morphological features plus 60 photometric bands) and has a hidden layer with 200 neurons.


\subsection{Ensemble Classifiers}

The Ensemble method aims to construct a meta classifier from the union of different algorithms. Generally, when efficiently combined, these classifiers can perform  better  than the single best algorithm.
In order to combine the classifiers we adopt the weighted sum rule with equal weights.
The probability prediction function $f$ can be written as:
\begin{equation}
     f(\textbf x_q)=  \frac{\sum^m_{j=1}w_j f_{j}(\textbf x_q)}{\sum^m_{j=1}w_j} \,,
\end{equation}
where $f_{j}(\textbf x_q)$ is the probabilistic binary classification from the classifier $j$ and $m$ is the number of classifiers considered. 
We implemented this algorithm using the \texttt{VotingClassifier} function from \texttt{scikit-learn}. In the following, the ensemble classifier (EC) comprises ANN, RF and SGLC methods with equal weight ($w_j =1/3$).
Note that EC is  not a pure ML classifier as it uses SGLC, see Section~\ref{cpdf}.
These algorithms generally will inherit the advantages and disadvantages of the methods which are based on.


\section{Performance metrics} \label{metrics}

We will now introduce the metrics that we adopt in order to assess the performance of the classifiers.
See \citet{mitchell1997machine,hastie2009elements} for more details.

\subsection{Confusion Matrix} \label{coma}


As we are considering probabilistic classifiers, the classification of sources into stars or galaxies depends on a probability threshold $p_{\rm cut}$ to be specified. In our case, all objects with $f>p_{\rm cut}$ will be classified as galaxies.
The choice of $p_{\rm cut}$ depends on completeness and purity requirements.

Once $p_{\rm cut}$ is specified, one can summarize the classification performance using the confusion matrix, which thoroughly  compares predicted and true values.
For a binary classifier the confusion matrix has four entries: True Positives (TP), True Negatives (TN), False Positives (FP) and False Negatives (FN). 
TP are sources correctly classified as galaxies by the model. TN are sources correctly classified as stars. FN are sources classified as stars by the model when, actually, they are galaxies. FP are sources classified as galaxies when they are stars.

\subsection{Metrics}

The receiver operating characteristic (ROC) curve represents a comprehensive way to summarize the performance of a classifier. It is a parametric plot of the true positive rate (TPR) and false positive rate (FPR) as a function of $p_{\rm cut}$:
\begin{equation}
TPR(p_{\rm cut})=\frac{TP}{TP+FN} \qquad FPR(p_{\rm cut})=\frac{FP}{FP+TN}
\end{equation}
with $0\le p_{\rm cut} \le 1$.
TPR is also called ``recall'' and, in astronomy, is the completeness.
The performance of a classifier can  then be summarized with the area under the curve (AUC). The AUC can assume values between 0 and 1. A perfect classifier has a value of 1, while a random classifier, on average, a value of 1/2.

The purity curve is a useful method to assess the performance of an unbalanced classifier (as the training set does not feature the same number of stars and galaxies).
It is a parametric plot of the completeness (or recall) and the purity (or precision) as a function of $p_{\rm cut}$:
\begin{equation}
\text{Purity} = \frac{TP}{TP+FP} \,.
\end{equation}
In order to summarize the purity curve, we consider the average precision (AP) which is the area under the purity curve and takes values between 0 and 1.


Finally, one can measure the algorithm performance with the mean squared error ($MS\!E$) defined as:
\begin{equation}
MS\!E= \frac{1}{N_{\rm test}} \sum_{q=1}^{N_{\rm test}} \left (y_q- f(\mathbf x_q) \right)^2 \,,
\end{equation}
where $y_q$ are the test-set classifications and $N_{\rm test}$ is the test-set size. $MS\!E = 0$ characterizes a perfect performance. In the present case of a binary classifier it is $MS\!E=(FP + FN)/N_{\rm test}$.

\subsection{$k$-fold cross-validation}\label{kfold}

We use the $k$-fold cross-validation method in order to optimize the algorithm's hyperparameters, test for overfitting and underfitting and estimate the errors on AUC and AP.
$k$-fold cross-validation separates the training data in $k$ equal and mutually exclusive parts (we adopt $k=10$). The model is trained in $k-1$ parts and validated in the remaining one, called validation. This process is repeated cyclically $k$ times.
The final result is the mean and standard deviation of the metric.

The ML methods described in Section~\ref{mlsec} depend on several internal hyperparameters (for example, the number $k$ of neighbors in KNN). In order to optimize them we performed $k$-fold cross-validation for several hyperparameter configurations. The results of the next Section are relative to the best configuration according to the AUC. See Appendix \ref{hyperparameter} for details.

We also tested the ML algorithms against overfitting and underfitting. The former happens when the training is successful (high AUC)  but not the testing (low AUC). The latter when training and testing are not successful (both AUC's are low). We checked that the average AUC from the $k$-fold cross-validation agrees with the AUC from the test set; all the methods pass this test.

Finally, we can use $k$-fold cross-validation in order to estimate the error in the determination of the AUC and AP.
This will help us understand if the differences between two estimators are significative and also how sensitive a classifier is with respect to the division of the dataset into training and test sets.


\section{Results} \label{results}

\setlength{\tabcolsep}{8pt}
\renewcommand{\arraystretch}{1.3}
\begin{table*}
\caption{
Performance of the classifiers considered in this paper for the miniJPAS catalog crossmatched with the SDSS catalog ($15\le r \le20$, top) and with the HSC-SSP catalog ($18.5\le r \le23.5$, bottom).
The best performance is marked in bold (EC is not considered).
$P$ stands for the analysis that uses only photometric bands while $M\!+\!P$ stands for the analysis that uses  photometric bands together with morphological parameters.
\label{table_foda}}
\centering
\begin{tabular}{@{}|l|llllll|l@{}}
\hline
\textbf{miniJPAS-SDSS}  & $AUC_{M+P}$  &  $AUC_{P}$  & $AP_{M+P}^{\rm gal}$ &  $AP_{P}^{\rm gal}$ & $MS\!E_{M+P}$ &   $MS\!E_{P}$ \\
\hline
\hline 
SGLC           &  0.994                 & --                    &   0.989		         &  --                     & 0.006   &    --     \\
\texttt{CLASS\_STAR}     &  0.997                 & --                    &   0.993 		         &  --                     & 0.032   &  	--	    \\
KNN            &  $0.996\!\pm\!0.003$  & $0.991\!\pm\!0.007$ &   $0.990\!\pm\!0.008$ &  $0.984\!\pm\!0.009$  & 0.015   &  0.027   \\
DT             &  $0.992\!\pm\!0.006$  & $0.984\!\pm\!0.012$ &   $0.983\!\pm\!0.011$ &  $0.974\!\pm\!0.018$  & 0.011   &  0.032   \\
RF             &  $\mathbf{0.997\!\pm\!0.006}$   & $0.996\!\pm\!0.004$ &   $0.992\!\pm\!0.009$ &  $0.995\!\pm\!0.010$  & 0.006   &  0.019   \\ 
EC             &  0.997                 & 0.997                &   0.995                &  0.996                 & 0.006   &  0.014   \\ 
ANN            &  $0.997\!\pm\!0.004$  & ${0.988\!\pm\!0.009}$ &   $\mathbf{0.994\!\pm\!0.017}$ &  $0.983\!\pm\!0.015$  & 0.012   &  0.043   \\ 
ERT            &  $0.997\!\pm\!0.002$  & $\mathbf{0.997\!\pm\!0.003}$ &   $0.993\!\pm\!0.006$ &  $\mathbf{0.996\!\pm\!0.004}$  & \textbf{0.005} &  \textbf{0.019}   \\
\hline
\hline
\textbf{miniJPAS-HSC-SSP}  & $AUC_{M+P}$  &  $AUC_{P}$  & $AP_{M+P}^{\rm gal}$ &  $AP_{P}^{\rm gal}$ & $MS\!E_{M+P}$ &   $MS\!E_{P}$ \\
\hline
\hline
SGLC           &  0.970                &  --                    &  0.992                 &  --                      & 0.040 &  --    \\
\texttt{CLASS\_STAR}     &  0.956                &  --                    &  0.991                 &  --                      & 0.053 &  --    \\
KNN            &  $0.950\!\pm\!0.010$ &  $0.824\!\pm\!0.023$ &  $0.989\!\pm\!0.003$  &  $0.959 \!\pm \!0.006$ & 0.053 & 0.098 \\
DT             &  $0.961\!\pm\!0.009$ &  $0.855	\!\pm\!0.017$ &  $0.990\!\pm\!0.003$  &  $0.959 \!\pm \!0.007$ & 0.061 & 0.132 \\
RF             &  $0.978\!\pm\!0.005$ &  $\mathbf{0.938\!\pm\!0.007}$ &  $0.995\!\pm\!0.002$  &  $\mathbf{0.986 \!\pm \!0.002}$ & 0.032 & 0.054 \\ 
EC             &  0.979                &  0.967                &  0.996                 &  0.993                  & 0.031 & 0.040 \\ 
ANN            &  $0.970\!\pm\!0.007$ &  $0.885\!\pm\!0.014$ &  $0.993\!\pm\!0.003$  &  $0.969 \!\pm \!0.005$ & 0.036 & 0.070 \\ 
ERT            &  $\mathbf{0.979\!\pm\!0.006}$ &  $0.931\!\pm\!0.006$ &  $\mathbf{0.995\!\pm\!0.002}$  &  $0.982 \!\pm \!0.002$ & \textbf{0.032} & \textbf{0.053} \\ 
\hline
\end{tabular}
\end{table*}


We now present our results for the algorithms introduced in Sections~\ref{mlsec} applied to the crossmatched catalogs described in Section~\ref{xmatch}.
Regarding stars and galaxy number counts we refer the reader to the miniJPAS presentation paper~\citep{Bonoli:2020ciz}.


\subsection{miniJPAS-SDSS catalog}

The performance of the star/galaxy classifiers considered in this paper for the miniJPAS catalog crossmatched with the SDSS catalog in the magnitude interval $15\le r \le20$ is excellent. The results are summarized in Table~\ref{table_foda}, where the best result are marked in bold (EC is not considered as it is not a pure ML classifier).%
\footnote{We omit  the corresponding figures as they are not informative given the excellent performance.}
The errors on the pure-ML classifiers are estimated via $k$-fold cross-validation.
In order to assess the importance of photometric bands and morphological parameters, the analysis considers two cases: only photometric bands ($P$ subscript in the table) and photometric bands together with morphological parameters ($M+P$  subscript in the table).
Note that this distinction does not apply to SGLC and \texttt{CLASS\_STAR} as they always include the use of morphological parameters.

Regarding the analysis with photometric bands only, the best ML methods are RF and ERT, showing the power of combining several trees when making a prediction.
Remarkably, using only photometric information, RF and ERT outperform SGLC and \texttt{CLASS\_STAR}.
If now we add  morphological information, the almost perfect performance of RF and ERT does not improve, showing again that, in this magnitude range, photometric information is sufficient.
In Table~\ref{table_foda} we also show the $MS\!E$, whose results agree with the ones from the ROC and purity curves.

\begin{figure}
\centering
 \includegraphics[width=.98 \columnwidth]{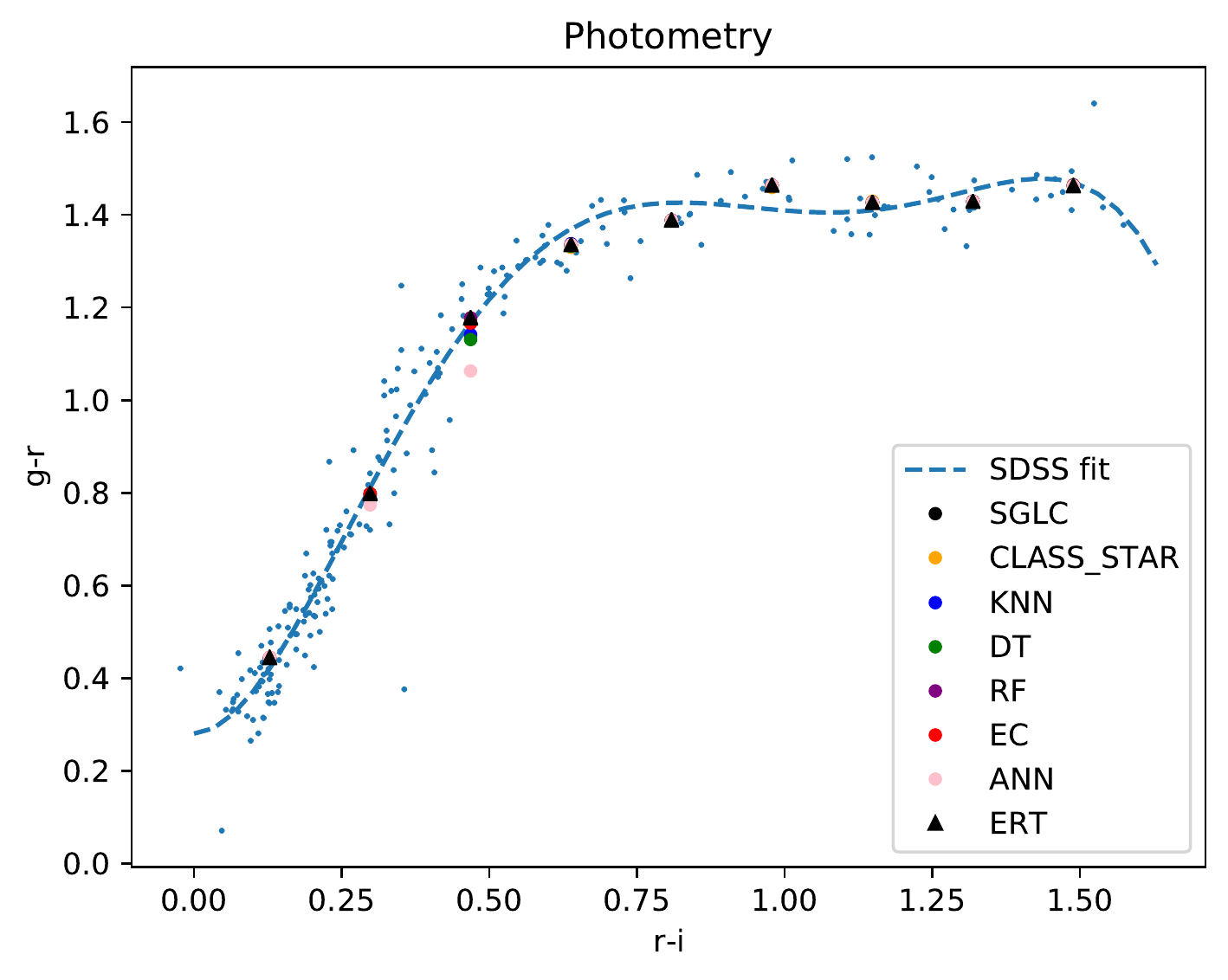}
 \includegraphics[width=.98 \columnwidth]{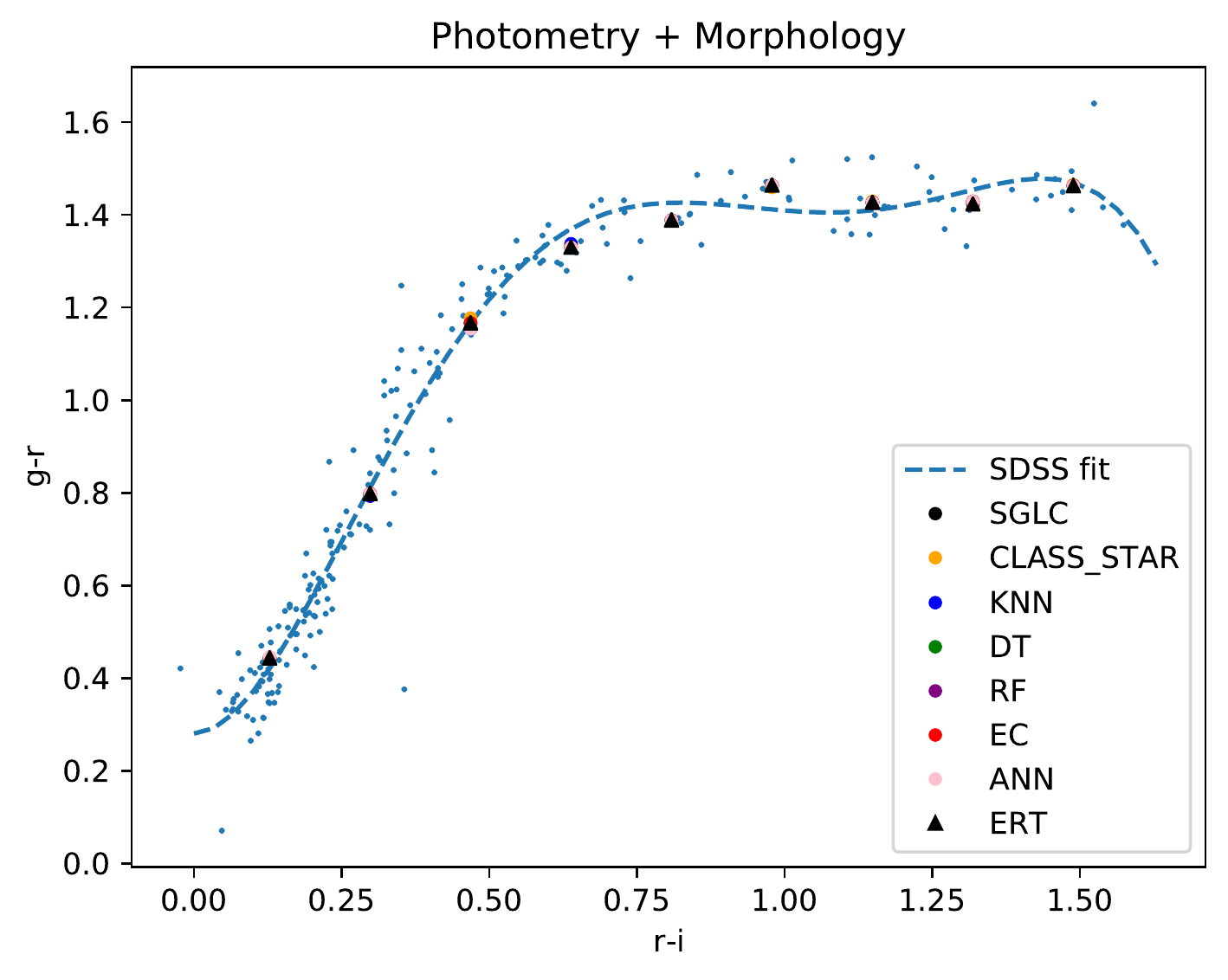}
\caption{The blue small dots represent the stellar locus for the objects classified as stars ($p\le p_{\rm cut}=0.5$) of the miniJPAS catalog crossmatched with the SDSS catalog in the magnitude interval $15\le r \le20$. The dashed line represent a polynomial fit to the stellar locus. The top panel is relative to the analysis that uses only photometric bands, while the bottom panel is relative to the analysis that also uses morphological information. The colored larger symbols represent the mean stellar locus provided by the different ML models.
For comparison it is shown also the classification by \texttt{CLASS\_STAR} and SGLC that always use morphological parameters.\label{locus_photo_SDSS}}
\end{figure}

Another way to analyze qualitatively the performance of a classifier is via a color-color diagram for objects classified as stars ($p \le p_{\rm cut}=0.5$).
Figure \ref{locus_photo_SDSS} shows the stellar locus in the $g-r$ versus $r-i$ color space.
The blue line is a fifth-degree polynomial fit, based on miniJPAS data that were classified as stars by SDSS.
The various markers represent the averages of each classifier for different bins.
We observe a small dispersion around the curve, which decreases when morphological parameters are included.
This indicates that the classifiers and the classification from SDSS are in good agreement.

\subsection{miniJPAS-HSC-SSP catalog} \label{hscresu}

\begin{figure*}
\centering
 \includegraphics[width=.98 \columnwidth]{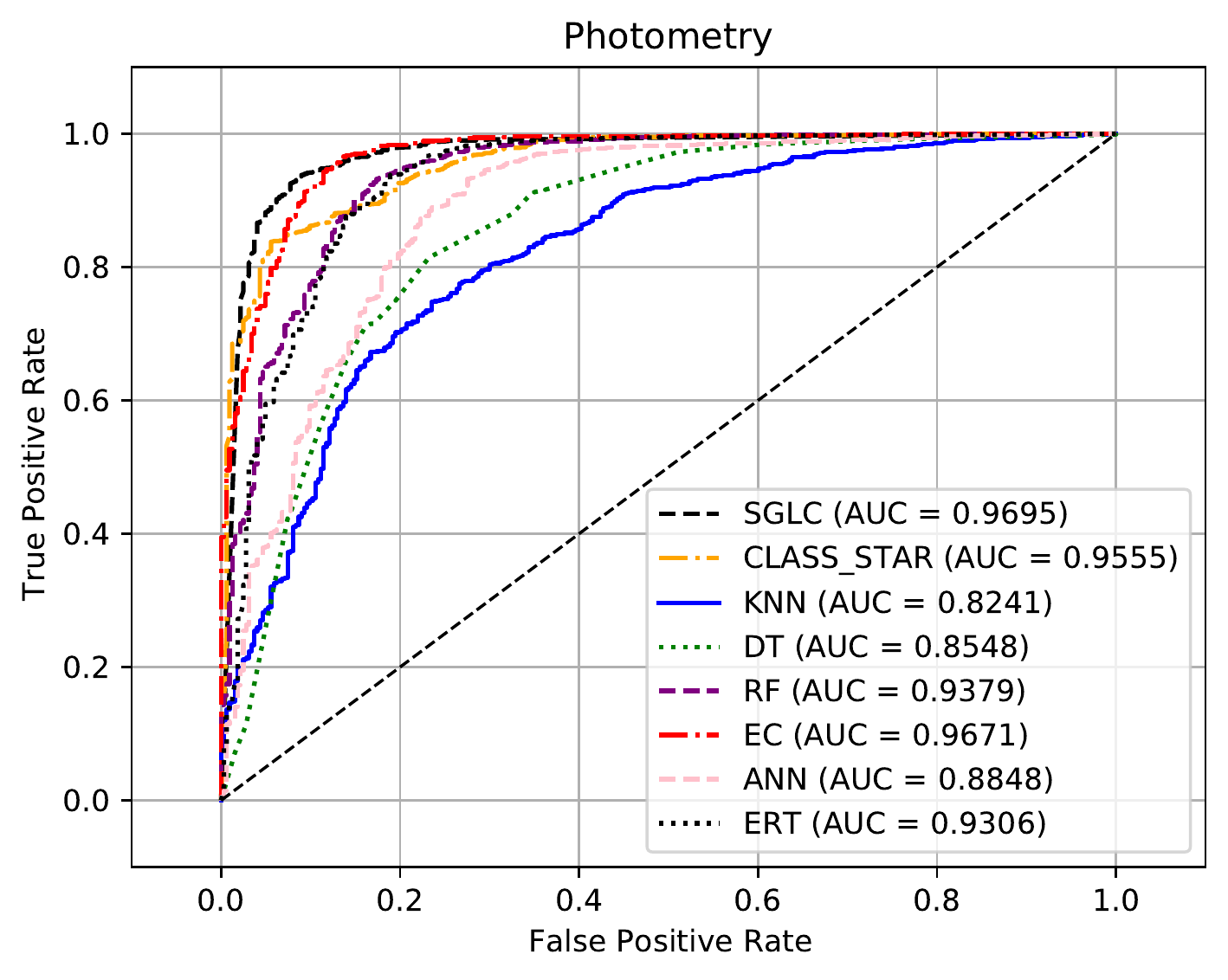}
 \includegraphics[width=.98 \columnwidth]{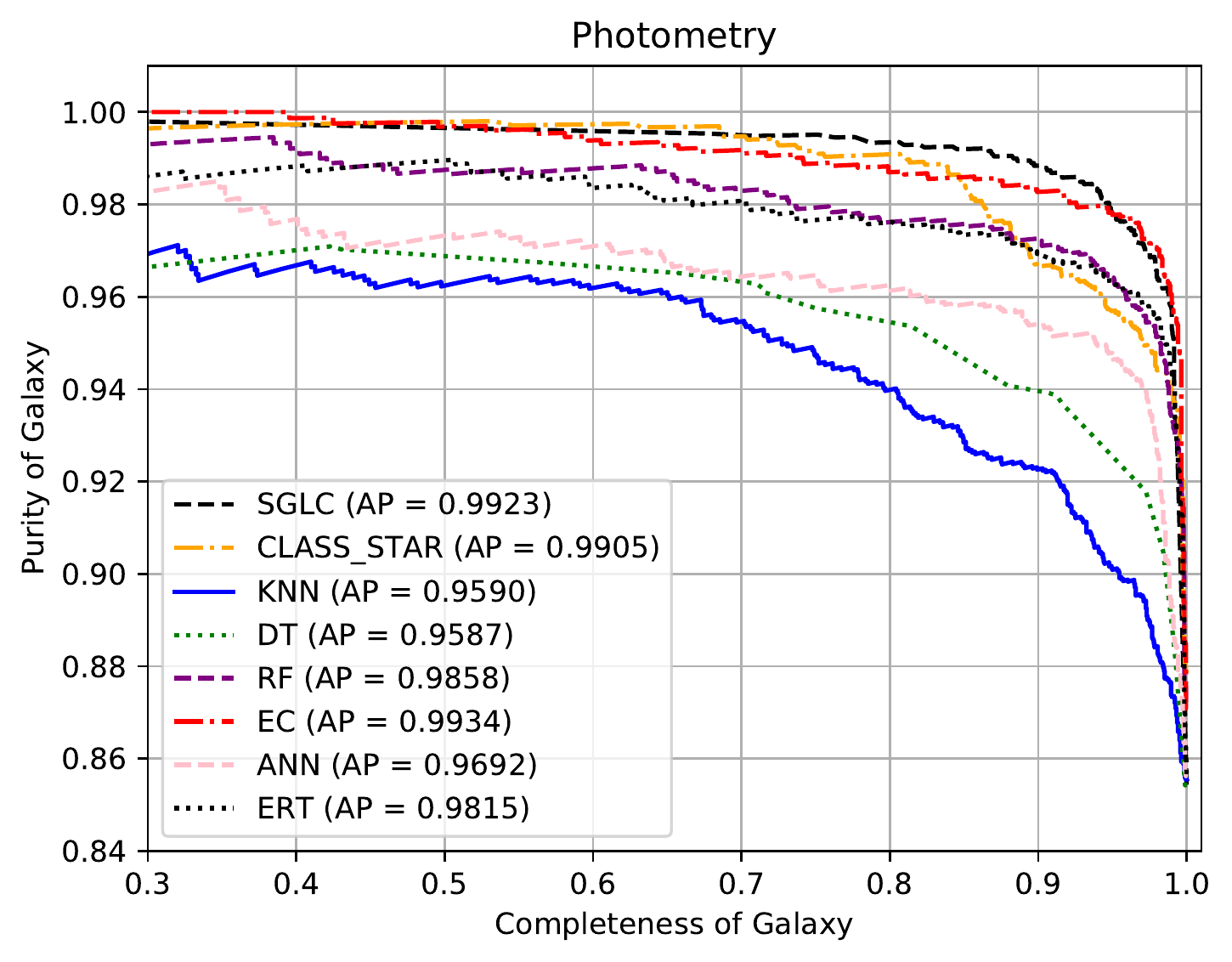}
 \includegraphics[width=.98 \columnwidth]{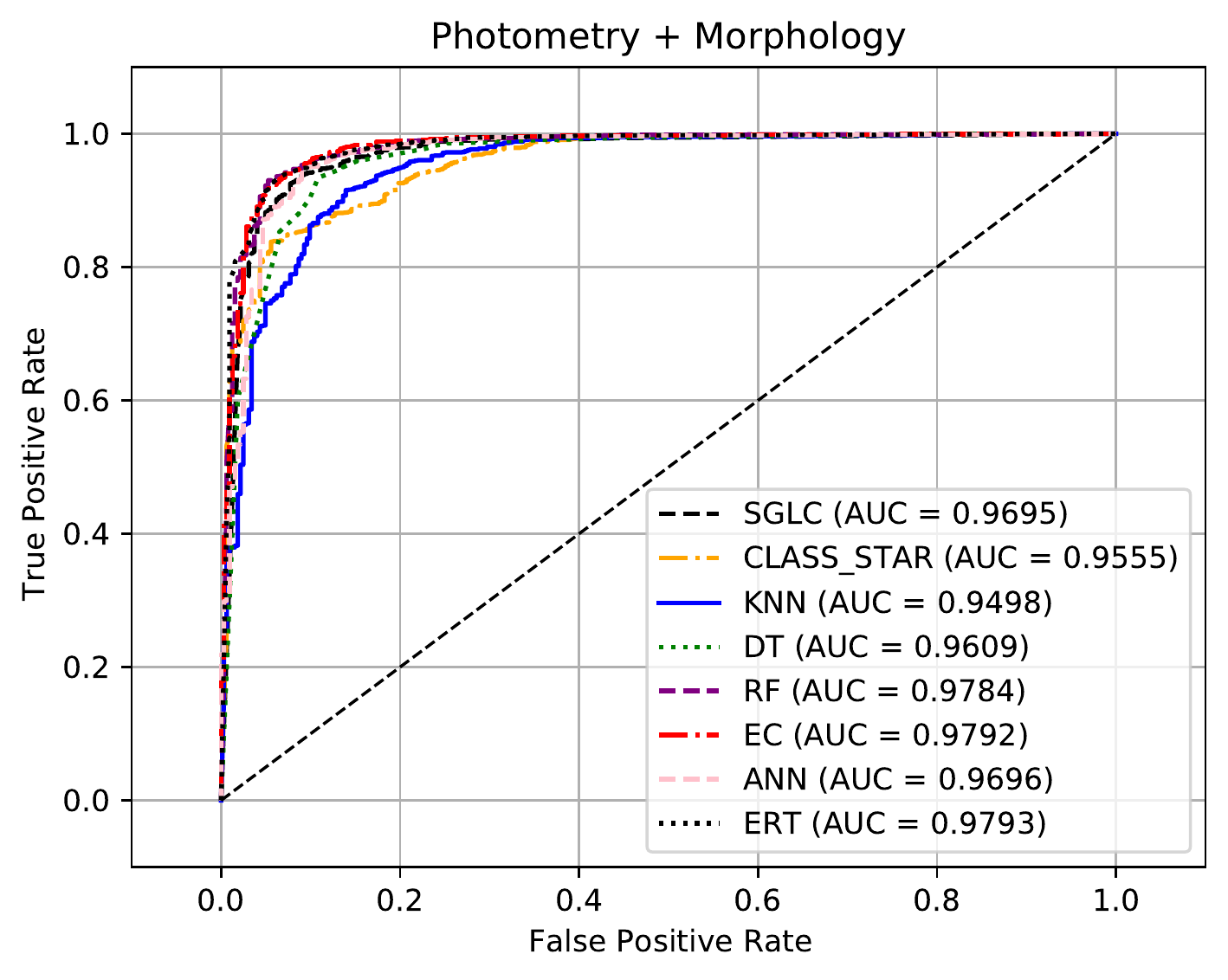}
 \includegraphics[width=.98 \columnwidth]{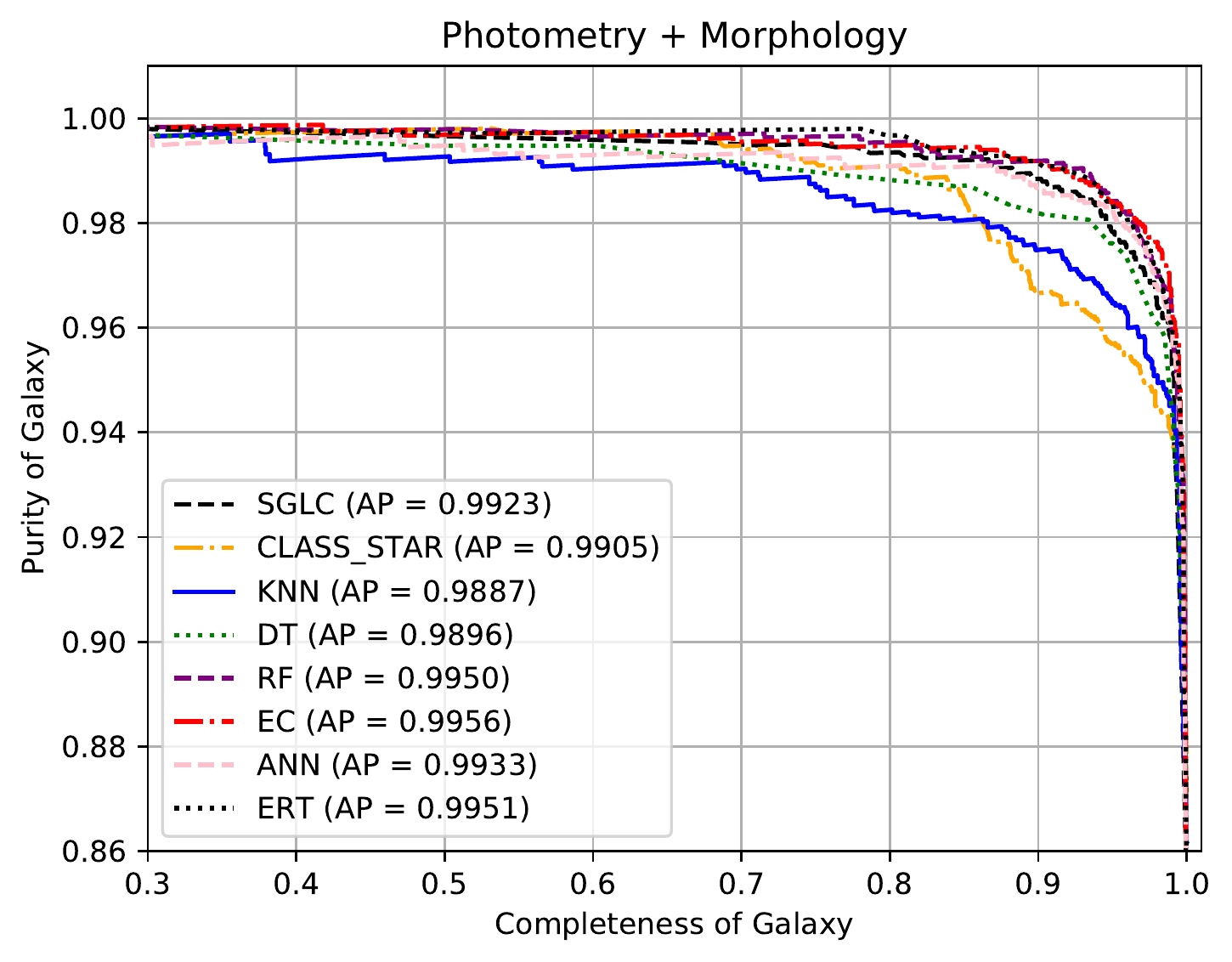}
\caption{
ROC curves (left panels) and purity curves for galaxies (right panels) for the classifiers considered in this paper for the miniJPAS catalog crossmatched with the HSC-SSP catalog in the magnitude interval $18.5\le r \le23.5$.
The top panels are relative to the analysis that uses only photometric bands while the bottom panels to the analysis that uses  photometric bands and morphological parameters. For comparison it is shown also the classification by \texttt{CLASS\_STAR} and SGLC that always use morphological parameters. Note that the axes ranges are varied in order to better show the curves.
The results are summarized in Table~\ref{table_foda} (bottom).\label{roc_hsc_morpho}}
\end{figure*}

As shown in the previous Section, star/galaxy classification in the range $15\le r \le20$ is not problematic.
However, the scenario changes when one moves to fainter magnitudes. As the amount of light decreases, with less information reaching the telescope, the performance of the algorithms decreases to the point that it is important to look for alternative solutions such as ML.
Here, we  present the analysis of the previous Section applied to the miniJPAS catalog crossmatched with the HSC-SSP catalog in the magnitude interval $18.5\le r \le23.5$.

Figure~\ref{roc_hsc_morpho} and Table~\ref{table_foda} show the results. Using photometric information only, the RF algorithm achieves the remarkable score of $AUC=0.938$. Although it is less performant than SGLC and \texttt{CLASS\_STAR} (that use morphology), this result shows that ML has the potential of identifying compact galaxies, which share the same morphology of stars.
Also, it has been argued that models that use just photometry can classify QSO's as extragalactic objects better than models that use morphological parameters \citep{2019arXiv190908626C}.
The use of the morphological parameters improves the performance of the ML methods to the point that ERT and RF perform better than \texttt{CLASS\_STAR} and SGLC.
In Appendix~\ref{aegis1} we repeat the analysis of Figure~\ref{roc_hsc_morpho} for the mJP-AEGIS1 field, which is the miniJPAS pointing with the best point spread function (PSF).

It is interesting to note that, although the classifiers feature lower $AUC$'s and higher $MS\!E$'s as compared to the analyses of the previous Section, the $AP$'s reach similar values, even when we use only photometric bands. This is due to this dataset having many more galaxies and only 15.3\% of stars. Therefore, even if there are contaminations by stars, the impact is lower.

\begin{figure}
\centering
 \includegraphics[width= \columnwidth]{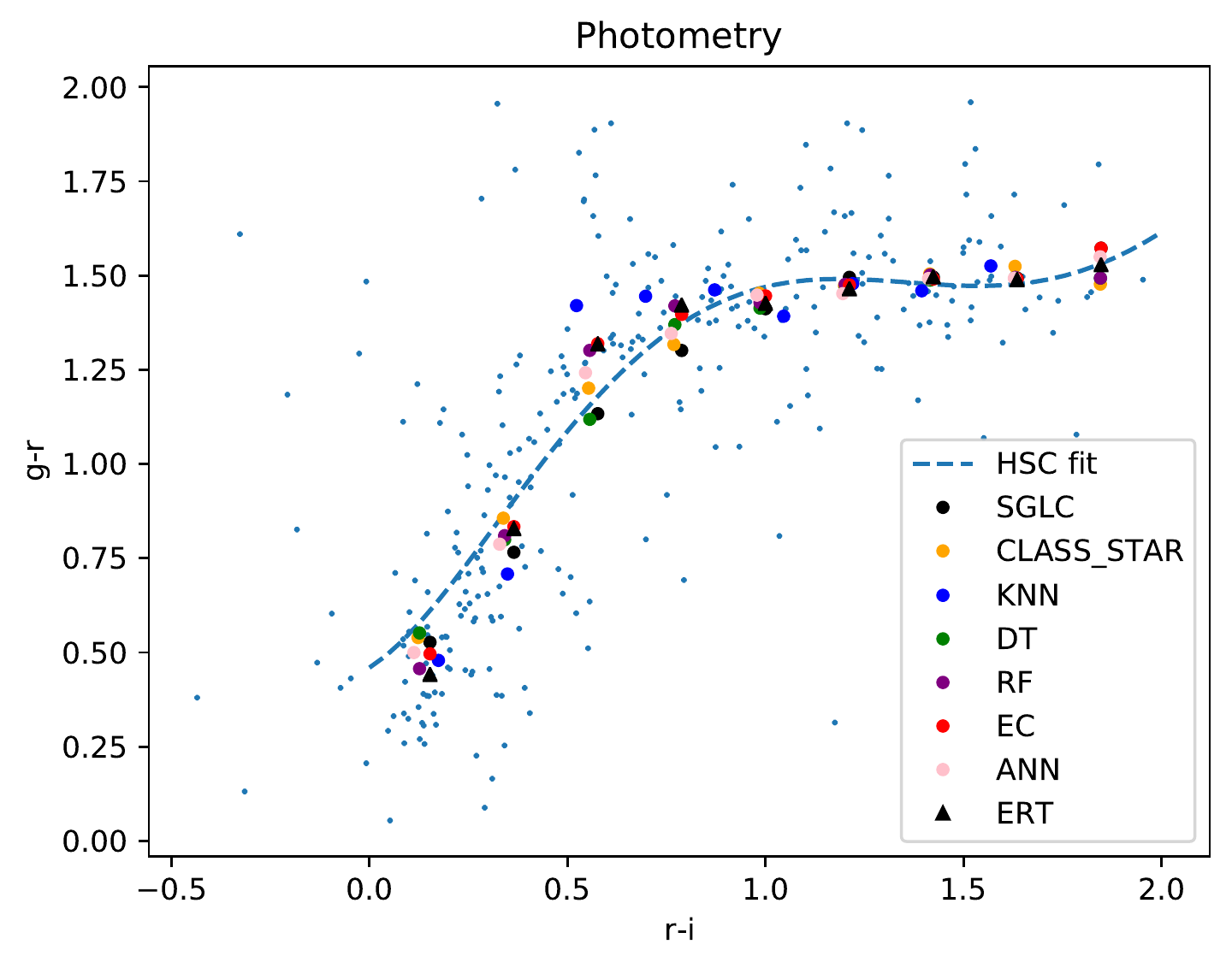}
 \includegraphics[width= \columnwidth]{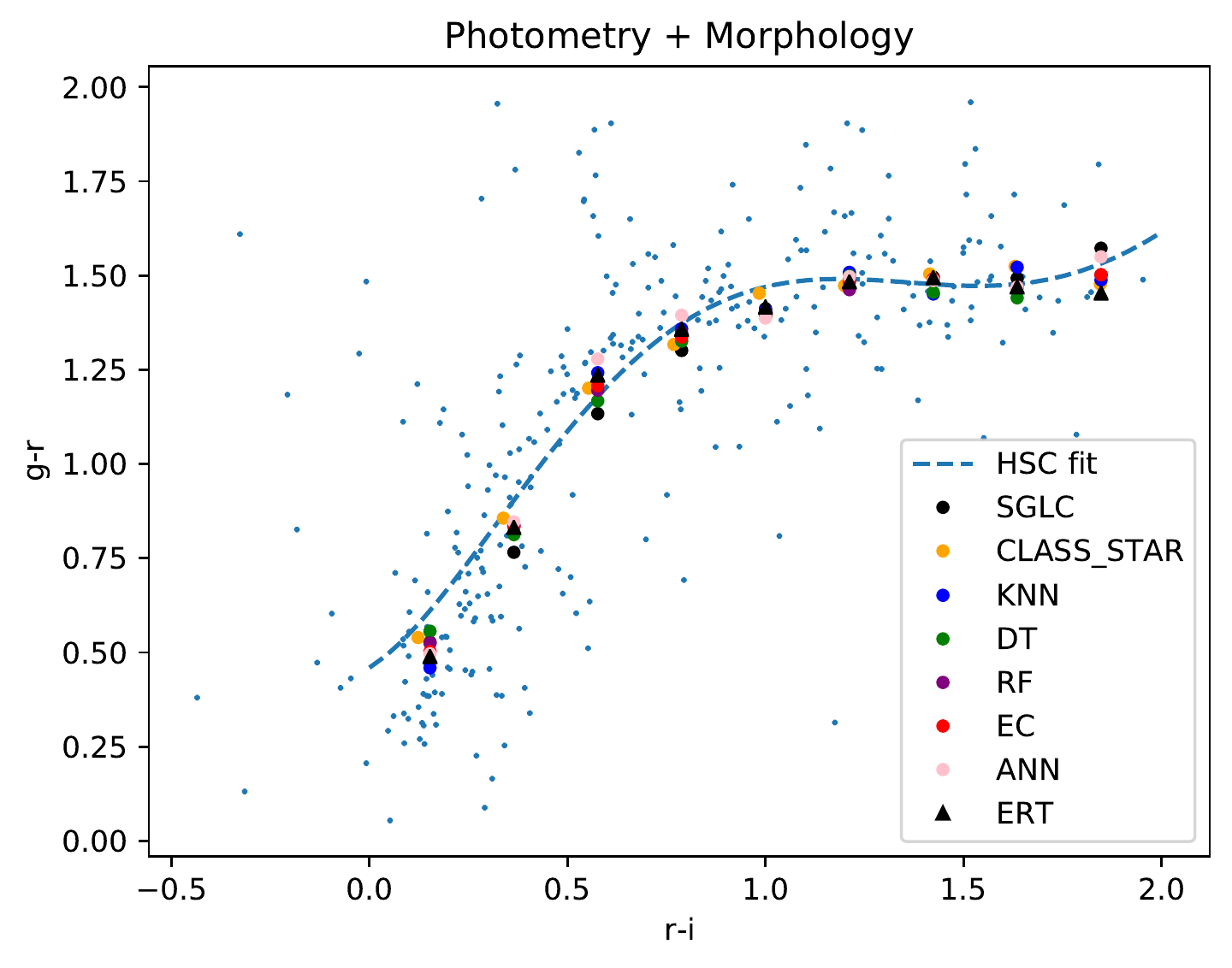}
\caption{
The blue small dots represent the stellar locus for the objects classified as stars ($p\le p_{\rm cut}=0.5$) of the miniJPAS catalog crossmatched with the HSC-SSP catalog in the magnitude interval $18.5\le r \le23.5$. The dashed line represent a polynomial fit to the stellar locus. The top panel is relative to the analysis that uses only photometric bands, while the bottom panel is relative to the analysis that also uses morphological information. The colored larger symbols represent the mean stellar locus provided by the different ML models.
For comparison it is shown also the classification by \texttt{CLASS\_STAR} and SGLC that always use morphological parameters.}
\label{locus_photo_HSC-SSP}
\end{figure}

Finally, in Figure~\ref{locus_photo_HSC-SSP} we show the stellar locus. We can observe a greater dispersion as compared with Figure~\ref{locus_photo_SDSS}, especially when we use only photometric bands in the analysis. Nevertheless,
the ML methods return the correct shape of the stellar locus and their performance is similar to the one by SGLC.

\subsection{Value added catalog}
\label{vac}

The ultimate goal of this work is to release a value added catalog with our best alternative classification.
In the previous Section we studied star/galaxy classification in the (partially overlapping) magnitude ranges $15\le r \le20$ and $18.5\le r \le23.5$. Here, in order to have a uniform dependence on $p_{\rm cut}$, we wish to produce a catalog that is obtained using a single classifier.
As seen in Section~\ref{xmatch}, in the magnitude range $18.5\le r \le20$, the classification by HSC-SSP is more reliable than the one by SDSS. Therefore, we consider the classification by SDSS in the range $15\le r <18.5$ and the one by HSC-SSP in the range $18.5\le r \le23.5$.
This catalog spans the magnitude range $15\le r \le 23.5$ and features a total of 11763 sources, 9517 galaxies and 2246 stars. We call it XMATCH catalog.

\begin{figure}
\centering
 \includegraphics[width= \columnwidth]{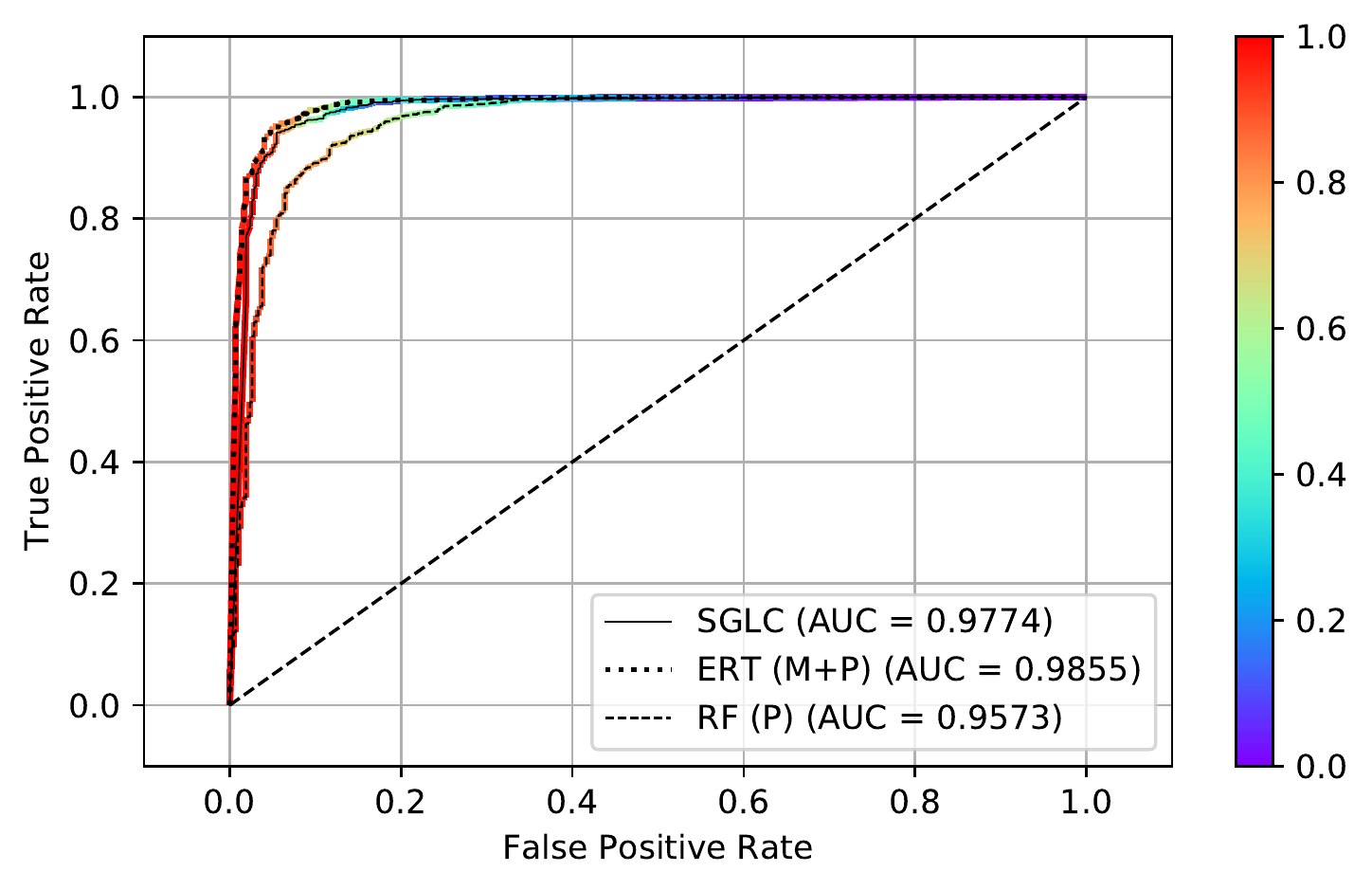}
 \includegraphics[width= \columnwidth]{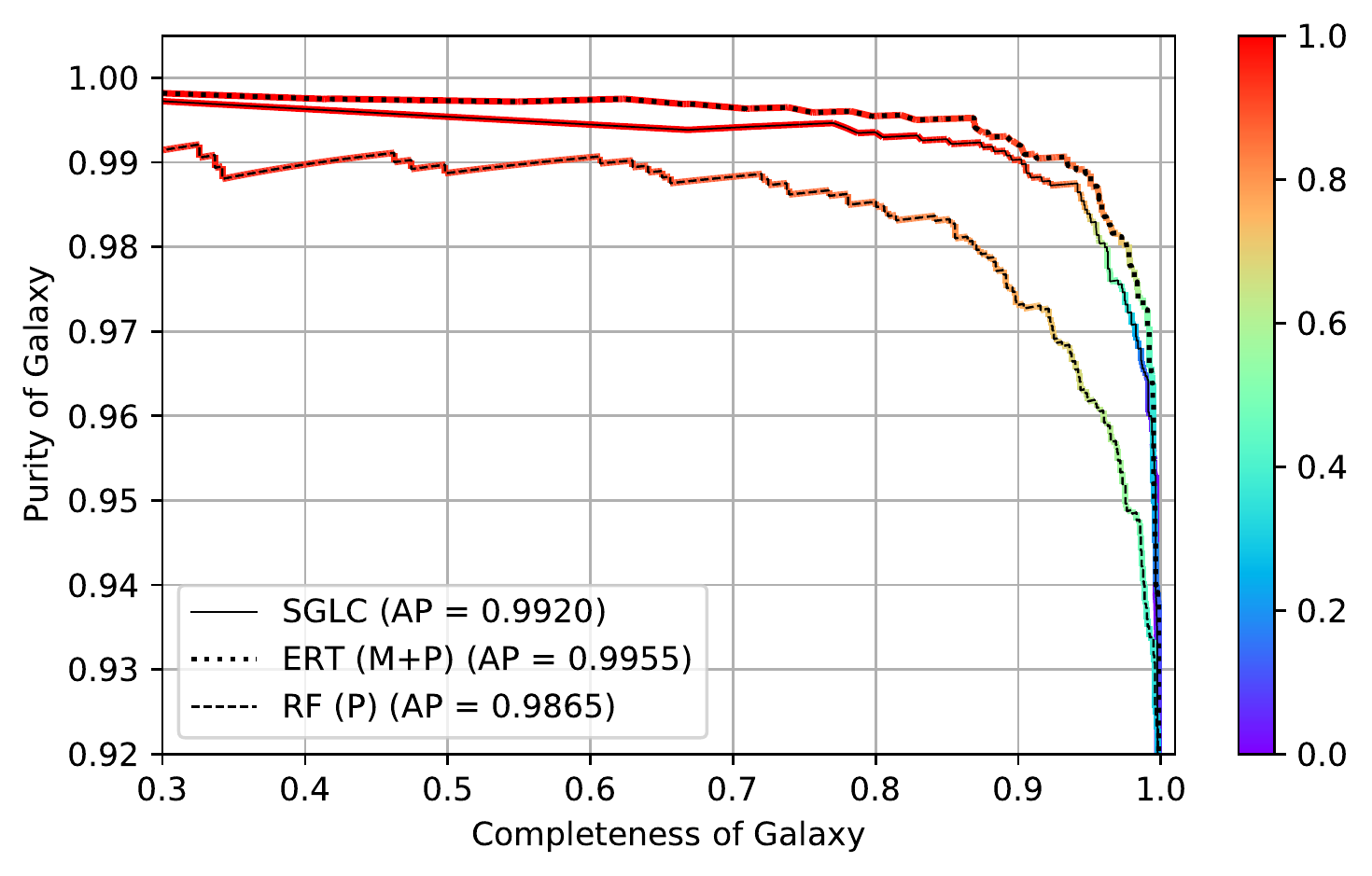}
 \includegraphics[width= \columnwidth]{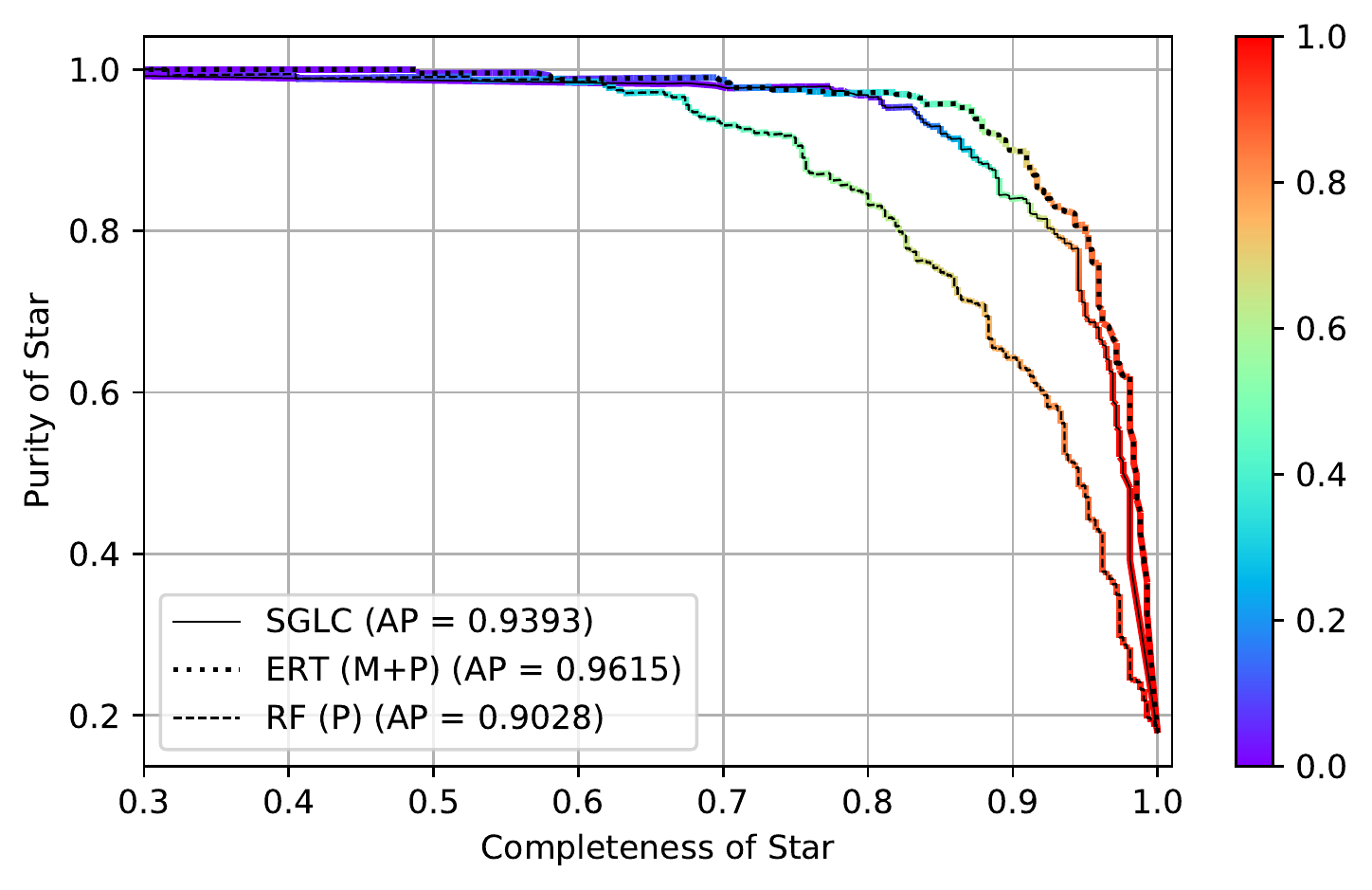}
\caption{ROC curve (top panel) and purity curve for galaxies (middle panel) and stars (bottom panel) for  RF (no morphology), ERT (with morphology) and SGLC for sources in the magnitude range $15 \le r \le 23.5$.
The color coding indicates the probability threshold $p_{\rm cut}$. Note that the axes ranges are varied in order to better show the curves.}
\label{cROCK}
\end{figure}

Next, we train and test  all the models on this catalog.
Using only photometric information the best classifier is RF, which reaches $AUC=0.957 \pm 0.008$, close to the performance of SGLC that uses morphological information.
Using photometric and morphological information  the best classifier is ERT, which, with $AUC=0.986\pm 0.005$, outperforms SGLC.%
\footnote{In Appendix~\ref{morphology_analysis} we show the analysis that considers only morphological information. We find that RF and ANN yield $AUC=0.970 \pm 0.006$.}
Figure~\ref{cROCK} shows the ROC curve and the purity curve for galaxies and stars for the three classifiers above, with the addition of the probability threshold $p_{\rm cut}$ via color coding. 
These plots are meant to help choosing the probability threshold that best satisfies one's needs of completeness and purity (see also Appendix~\ref{histo}).
These plots were made with the code available at
\href{https://github.com/PedroBaqui/minijpas-astroclass}{github.com/PedroBaqui/minijpas-astroclass}.
As shown in the bottom panel of Figure~\ref{cROCK}, the AP of stars is quite good (and significantly better than SGLC), showing that the fact that we used an unbalanced set did not affect the results regarding the least represented class.

\begin{figure}
\centering
 \includegraphics[width= \columnwidth]{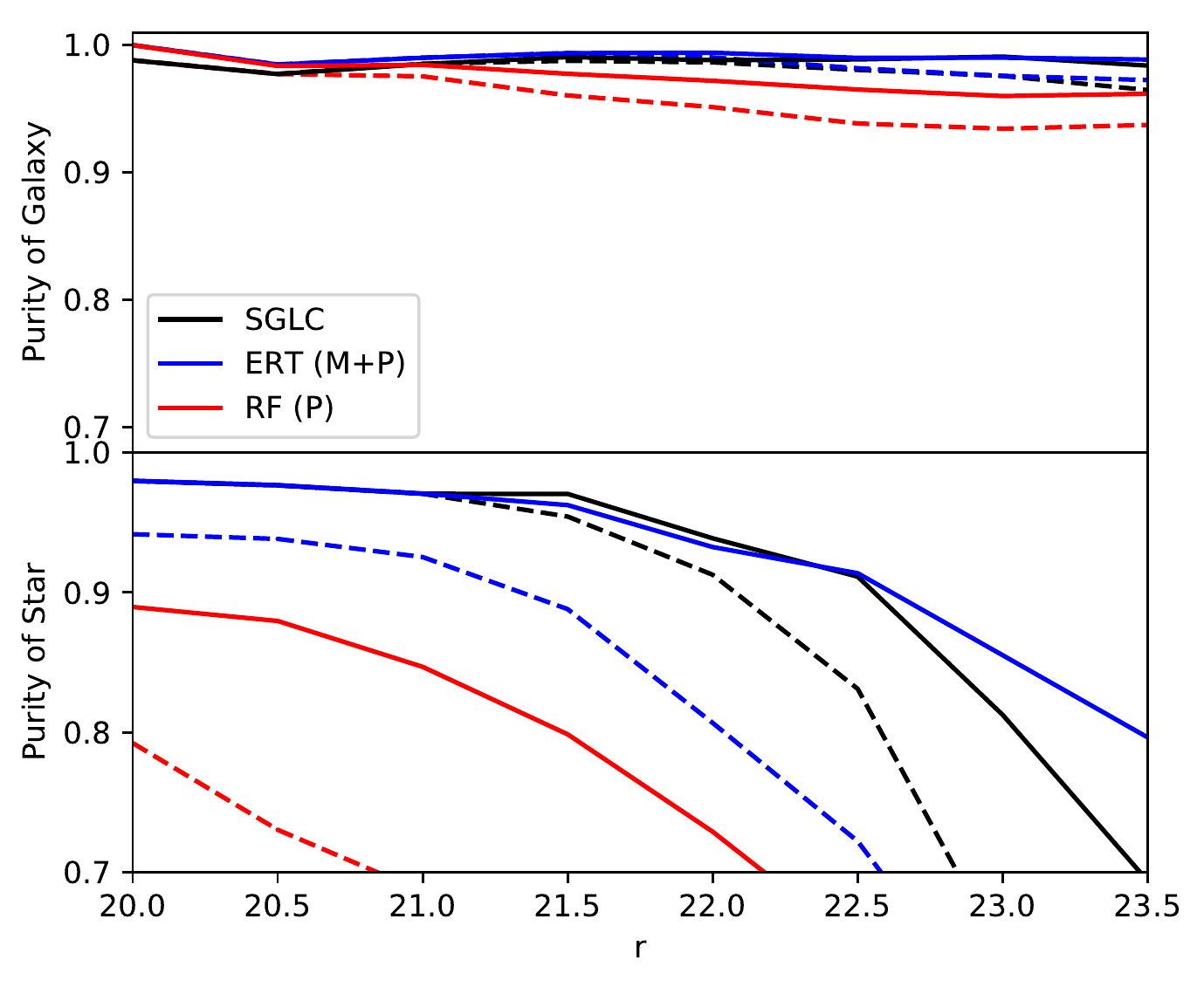}
\caption{Cumulative purity of the galaxy (top) and star (bottom) samples as a function of magnitude for the ML classifiers of Fig.~\ref{cROCK}, for a fixed completeness of 95\% (solid line) and 99\% (dashed line).}
\label{fixcomp}
\end{figure}

Finally, we show in Figure~\ref{fixcomp} the cumulative purity of the galaxy and star samples as a function of $r$ magnitude for a fixed completeness of 95\% and 99\%, which are achieved by choosing a suitable $p_{\rm cut}$.
For a completeness of 95\% and the ERT classifier, the purity of the galaxy sample remains higher than 99\% throughout the magnitude range, better than SGLC.
Regarding stars, for a completeness of 95\% and ERT, purity remains higher that 90\% for $r<22.5$. For fainter stars, ERT outperforms SGLC.

In order to build our catalog, we applied our two best classifiers (RF without morphology and ERT with morphology) to the 29551 miniJPAS sources in the magnitude range $15 \le r \le 23.5$.
It is important to note that, given the completeness of miniJPAS \citep[see][]{Bonoli:2020ciz},  sources outside this magnitude interval are less likely to enter scientific studies.
The catalog is publicly available at \href{https://j-pas.org/datareleases}{j-pas.org/datareleases} via the ADQL table \texttt{minijpas.StarGalClass}.
See Appendix~\ref{adql} for more informations and an ADQL query example.

%
%

\subsection{Feature importance}
\label{featureimpo}

We use the RF algorithm (see Eq.~\ref{impo}) to assess feature importance which can give us insights on the way objects are classified.
The 15 most important features are listed in Table~\ref{tab:f-sdss}. The full tables are provided as machine readable supplementary material.

When including morphological parameters, FWHM is the most important feature.
This agrees with the distributions of FWHM in Figs.~\ref{fig:mor-sdss} and \ref{fig:mor-hsc} which show a good separation between stars and galaxies.
Although this separation is less evident for the other parameters, they also contribute to classification.
In particular, the mean PSF is the fourth most importante feature, while the least important morphological feature is  the ellipticity parameter $A/B$.
To some extent, these results could depend on the choice of the impurity function (see Eq.~\eqref{inpu}). We tested different impurity functions and  confirmed that morphological parameters are generally more important than photometric bands.

When using photometric information only, the importance of the features is more evenly distributed as more features work together towards object classification.
In particular, broad bands are not necessarily more important than narrow bands and errors (the width of the distribution) are as important as the measurements (central value of the distribution).
In other words, the full characterization of the measurement seems to be important.

\begin{figure}
\centering
 \includegraphics[width= \columnwidth]{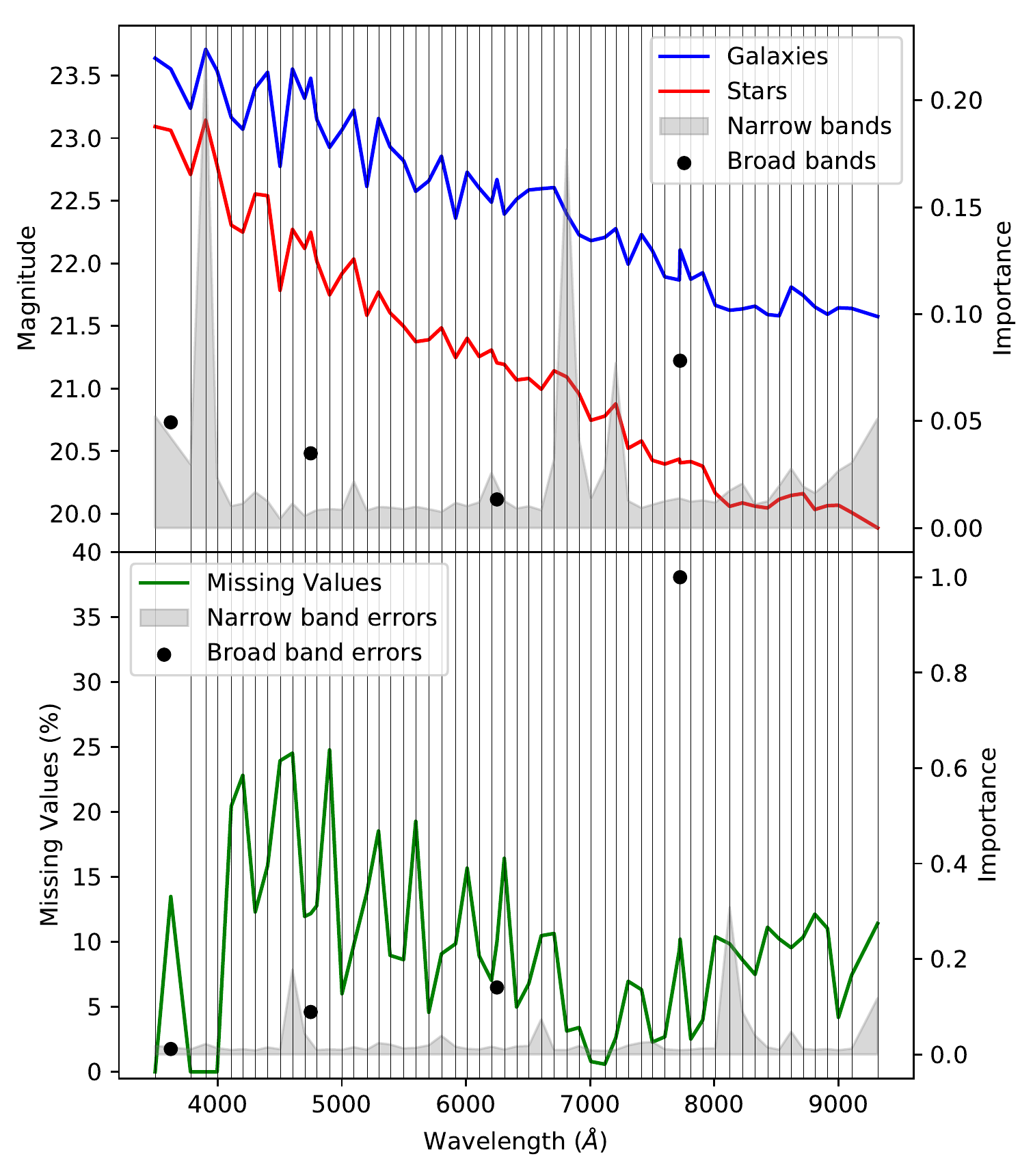}
\caption{
Top: The shaded area represents the relative importance (see Eq.~\ref{impo}) of the narrow-band filters as function of the filters' wavelength for the analysis relative to the full magnitude range $15\le r \le 23.5$ (see Section~\ref{vac}). The importance of the 4 broad-band filters is shown using black circles.
The red and blue lines show the average photo-spectrum of stars and galaxies, respectively.
Bottom: as the top panels but for the relative importance of the magnitude errors. The green line shows the percentage of missing values (magnitude of 99) for the narrow band filters.
\label{features}}
\end{figure}

In order to get a physical insight on the regions of the spectrum that matter most for classification, we show in Figure~\ref{features} (top)
the relative importance of the filters's magnitudes as function of the filters' wavelength together with the median star and galaxy photo-spectrum.
It is clear that there are regions systematically more important than others (neighboring filters with higher importance) and that there is correlation between the most important regions and the average features in the spectra.
In the bottom panel of Figure~\ref{features} we show the importance of the magnitude errors, which also show regions that are systematically more important than others. Particularly important is the error on the $i$ band.
In the same panel we also show the fraction of missing values (magnitude of 99) for each narrow band filter.
We can see that this fraction anti-correlates with the filter importance (top panel).

\setlength{\tabcolsep}{10pt}
\renewcommand{\arraystretch}{1.2}
\begin{table}
\caption{Feature importance with ($M+P$) and without ($P$) morphological parameters for the analysis relative to the full crossmatched catalog XMATCH ($15\le r \le 23.5$, see Section~\ref{vac}).
The importance is normalized relative to the best feature. The quantity max/ap3 is \texttt{MU\_MAX/MAG\_APER\_3\_0}.
The full tables are provided as machine readable supplementary material. See also Figure~\ref{features}.}
\centering
\begin{tabular}{p{1cm}l|p{1cm}l}
\hline
\hline
\multicolumn{2}{c}{XMATCH ($P$)} & \multicolumn{2}{c}{XMATCH ($P+M$)} \\
\hline
Feature            & Importance & Feature         & Importance  \\
\hline
iSDSSerr           &  1.00 &  FWHM           &  1.00  \\
J0810err           &  0.31  &  $c_r$          &  0.30	 \\
J0390              &  0.22  &  max/ap3  	   &  0.18	 \\
J0460err           &  0.18  &  \textrm{PSF}&  0.10\\
J0680              &  0.18  &  iSDSSerr       &  0.08	 \\
rSDSSerr           &  0.14  &  J0820err       &  0.02	 \\
J1007err           &  0.12  &  J0390err       &  0.02	 \\
J0820err           &  0.09  &  A/B            &  0.01	 \\
gSDSSerr           &  0.09  &  J1007err       &  0.01	 \\
iSDSS              &  0.08  &  J0810err       &  0.01	 \\
J0720              &  0.08  &  J0390          &  0.01  \\
J0660err           &  0.07  &  gSDSS          &  0.009  \\
uJAVA              &  0.05 &  uJAVAerr       &  0.008  \\
J1007              &  0.05  &  J0790err       &  0.008  \\
uJPAS              &  0.05  &  J0680          &  0.007  \\
...	               &...         &...              &...         \\
\hline \hline
\end{tabular}
\label{tab:f-sdss}
\end{table}

\subsection{Transmission curve variability}
\label{trans}

The transmission curves of the narrow band filters vary according to the relative position in the filters. 
In particular, the transmission curve variability depends on the SED of each object so that the map of relative variation in flux for a given filter is different for objects with different SEDs.
This effect should affect classifications that depend strongly on particular narrow spectral features (even more if they fall in one of the edges of the narrow band transmission curve) and would have almost no effect when considering mainly the continuum.
As we use photometric data, our results could be impacted by this effect.

miniJPAS data, in particular the size of the XMATCH catalog, does not allow us to perform a thorough investigation of this effect. Therefore, we explore this issue by dividing the test set into the 4 quadrants of the filter area and compute the $AUC$ for each quadrant. The filter coordinates are given in pixels via the \texttt{X\_IMAGE} and \texttt{Y\_IMAGE} variables ($9000 \times 9000$ pixels).
As can be seen from Table~\ref{tab:trans}, the $AUC$ variation is compatible with the overall performance of $AUC=0.957 \pm 0.008$ (RF) and $AUC=0.986\pm 0.005$ (ERT), showing that the effect should not strongly bias our results.

\setlength{\tabcolsep}{10pt}
\renewcommand{\arraystretch}{1.2}
\begin{table}
\caption{Area under the curve ($AUC$) for the 4 filter quadrants relative to the best classifiers shown in Figure~\ref{cROCK}.}
\centering
\begin{tabular}{c|cc}
\hline
\hline
  RF ($P$)          & $\texttt{X} <4500$ & $4500\le \texttt{X} \le 9000$           \\
\hline
$\texttt{Y} <4500$           &  0.9633  &  0.9592             \\
$4500\le \texttt{Y} \le9000$           &  0.9449  &  0.9588         	 \\
\hline
\hline
  ERT ($P+M$)          & $\texttt{X} <4500$ & $4500\le \texttt{X} \le 9000$           \\
\hline
$\texttt{Y} <4500$           &  0.9917  &  0.9775             \\
$4500\le \texttt{Y} \le9000$           &  0.9822  &  0.9938         	 \\
\hline
\hline
\end{tabular}
\label{tab:trans}
\end{table}

\section{Conclusions} \label{conclusions}

In this work we applied different machine learning methods for the classification of sources of miniJPAS.
The goal was to build models that are competitive with and complementary to those adopted in the miniJPAS 2019 public data release and to offer to the astronomical community a value added catalog with an alternative classification.
As we considered supervised ML algorithms, we classified the miniJPAS objects that are in common with SDSS and HSC-SSP, whose classifications are trustworthy within the magnitude intervals $15\le r \le20$ and $18.5\le r \le23.5$, respectively. We used as input the magnitudes associated to the 60 filters along with their errors, 4 morphological parameters and the mean PSF of the pointings. The output of the algorithms is probabilistic.
We tested K-Nearest Neighbors, Decision Trees, Random Forest, Artificial Neural Networks, Extremely Randomized Trees and  Ensemble Classifier.

Our results show that ML is able to classify objects into stars and galaxies without the use of morphological parameters.
This makes ML classifiers quite valuable as they can distinguish compact galaxies from stars, differently from methods that necessarily use morphological parameters in the classification process.
Of course, the inclusion of morphological parameters improves the results to the point that ERT can outperform \texttt{CLASS\_STAR} and SGLC (the default classifier in J-PAS).

We used the RF algorithm to assess feature importance.
When using morphological parameters, FWHM is the most important feature.
When using photometric information only, we observe that broad bands are not necessarily more important than narrow bands and errors (the width of the distribution) are as important as the measurements (central value of the distribution).
In other words, the full characterization of the measurement seems to be important.
We have also shown that ML  can give meaningful insights on the regions of the spectrum that matter most for classification.

After having validated our methods, we applied our best classifiers, with and without morphology, to the full dataset.
This classification is available as a value added catalog at \href{https://j-pas.org/datareleases}{j-pas.org/datareleases} via the ADQL table \texttt{minijpas.StarGalClass}.
The ML models are available at \href{https://github.com/J-PAS-collaboration/StarGalClass-MachineLearning}{github.com/J-PAS-collaboration/StarGalClass-MachineLearning}.
Our catalog both validates the quality of SGLC and produces an independent classification that can be useful to test the robustness of subsequent scientific analyses. In particular, our classification uses the full photometric information, with and without morphology, which is important for faint galaxies whose morphology is similar to the one of stars.

We conclude stressing that our methodology can be further improved both at the algorithmic and at the data input level.
A promising avenue is the direct use of the object images with convolutional neural networks.
This approach has the potential of outperforming presently available classifiers.

\begin{acknowledgements}

POB thanks, for financial support, the Coordenação de Aperfeiçoamento de Pessoal de Nível Superior - Brasil (CAPES) - Finance Code 001.
VM thanks CNPq (Brazil) and FAPES (Brazil) for partial financial support. This project has received funding from the European Union’s Horizon 2020 research and innovation programme under the Marie Skłodowska-Curie grant agreement No 888258.
LADG is supported by the Ministry of Science and Technology of Taiwan (grant MOST 106-2628-M-001-003-MY3), and by the Academia Sinica (grant AS-IA-107-M01).
ES has been partly supported by the Spanish State Research Agency (AEI) Projects AYA2017-84089 and MDM-2017-0737 at Centro de Astrobiología (CSIC-INTA), Unidad de Excelencia María de Maeztu.
MQ is supported by the Brazilian research agencies CNPq and FAPERJ.
RGD acknowledges financial support from the State Agency for Research of the Spanish MCIU through the ``Center of Excellence Severo Ochoa'' award to the Instituto de Astrofísica de Andalucía (SEV-2017-0709) and through the projects AyA2016-77846-P and  PID2019-109067GB-100.
LS acknowledges support from Brazilian agencies CNPq (grant 304819/2017-4) and FAPESP (grant 2012/00800-4).\\
This work made use of the Virgo Cluster at Cosmo-ufes/UFES, which is funded by FAPES and administrated by Renan Alves de Oliveira.\\
Based on observations made with the JST/T250 telescope and PathFinder camera for the miniJPAS project at the Observatorio Astrof\'{\i}sico de Javalambre (OAJ), in Teruel, owned, managed, and operated by the Centro de Estudios de F\'{\i}sica del  Cosmos de Arag\'on (CEFCA). We acknowledge the OAJ Data Processing and Archiving Unit (UPAD) for reducing and calibrating the OAJ data used in this work.\\
Funding for OAJ, UPAD, and CEFCA has been provided by the Governments of Spain and Arag\'on through the Fondo de Inversiones de Teruel; the Arag\'on Government through the Research Groups E96, E103, and E16\_17R; the Spanish Ministry of Science, Innovation and Universities (MCIU/AEI/FEDER, UE) with grant PGC2018-097585-B-C21; the Spanish Ministry of Economy and Competitiveness (MINECO/FEDER, UE) under AYA2015-66211-C2-1-P, AYA2015-66211-C2-2, AYA2012-30789, and ICTS-2009-14; and European FEDER funding (FCDD10-4E-867, FCDD13-4E-2685).\\
Based on data from ALHAMBRA Data Access Service the at CAB (CSIC-INTA).\\
Funding for the DEEP2 Galaxy Redshift Survey has been provided by NSF grants AST-95-09298, AST-0071048, AST-0507428, and AST-0507483 as well as NASA LTSA grant NNG04GC89G.\\
The Hyper Suprime-Cam (HSC) collaboration includes the astronomical communities of Japan and Taiwan, and Princeton University. The HSC instrumentation and software were developed by the National Astronomical Observatory of Japan (NAOJ), the Kavli Institute for the Physics and Mathematics of the Universe (Kavli IPMU), the University of Tokyo, the High Energy Accelerator Research Organization (KEK), the Academia Sinica Institute for Astronomy and Astrophysics in Taiwan (ASIAA), and Princeton University. Funding was contributed by the FIRST program from the Japanese Cabinet Office, the Ministry of Education, Culture, Sports, Science and Technology (MEXT), the Japan Society for the Promotion of Science (JSPS), Japan Science and Technology Agency (JST), the Toray Science Foundation, NAOJ, Kavli IPMU, KEK, ASIAA, and Princeton University.\\
This paper makes use of software developed for the Large Synoptic Survey Telescope. We thank the LSST Project for making their code available as free software at  \href{http://dm.lsst.org}{dm.lsst.org}.\\
This paper is based [in part] on data collected at the Subaru Telescope and retrieved from the HSC data archive system, which is operated by Subaru Telescope and Astronomy Data Center (ADC) at National Astronomical Observatory of Japan. Data analysis was in part carried out with the cooperation of Center for Computational Astrophysics (CfCA), National Astronomical Observatory of Japan.\\
Funding for SDSS-III has been provided by the Alfred P. Sloan Foundation, the Participating Institutions, the National Science Foundation, and the U.S. Department of Energy Office of Science. The SDSS-III website is sdss3.org. SDSS-III is managed by the Astrophysical Research Consortium for the Participating Institutions of the SDSS-III Collaboration including the University of Arizona, the Brazilian Participation Group, Brookhaven National Laboratory, Carnegie Mellon University, University of Florida, the French Participation Group, the German Participation Group, Harvard University, the Instituto de Astrofisica de Canarias, the Michigan State/Notre Dame/JINA Participation Group, Johns Hopkins University, Lawrence Berkeley National Laboratory, Max Planck Institute for Astrophysics, Max Planck Institute for Extraterrestrial Physics, New Mexico State University, New York University, Ohio State University, Pennsylvania State University, University of Portsmouth, Princeton University, the Spanish Participation Group, University of Tokyo, University of Utah, Vanderbilt University, University of Virginia, University of Washington, and Yale University.

\end{acknowledgements}

\bibliographystyle{aaArxivDoi} 
\bibliography{biblio.bib} 

\begin{appendix}

\section{Purity curves for stars}
\label{CPstars}

\begin{figure}
\centering
 \includegraphics[width=.98 \columnwidth]{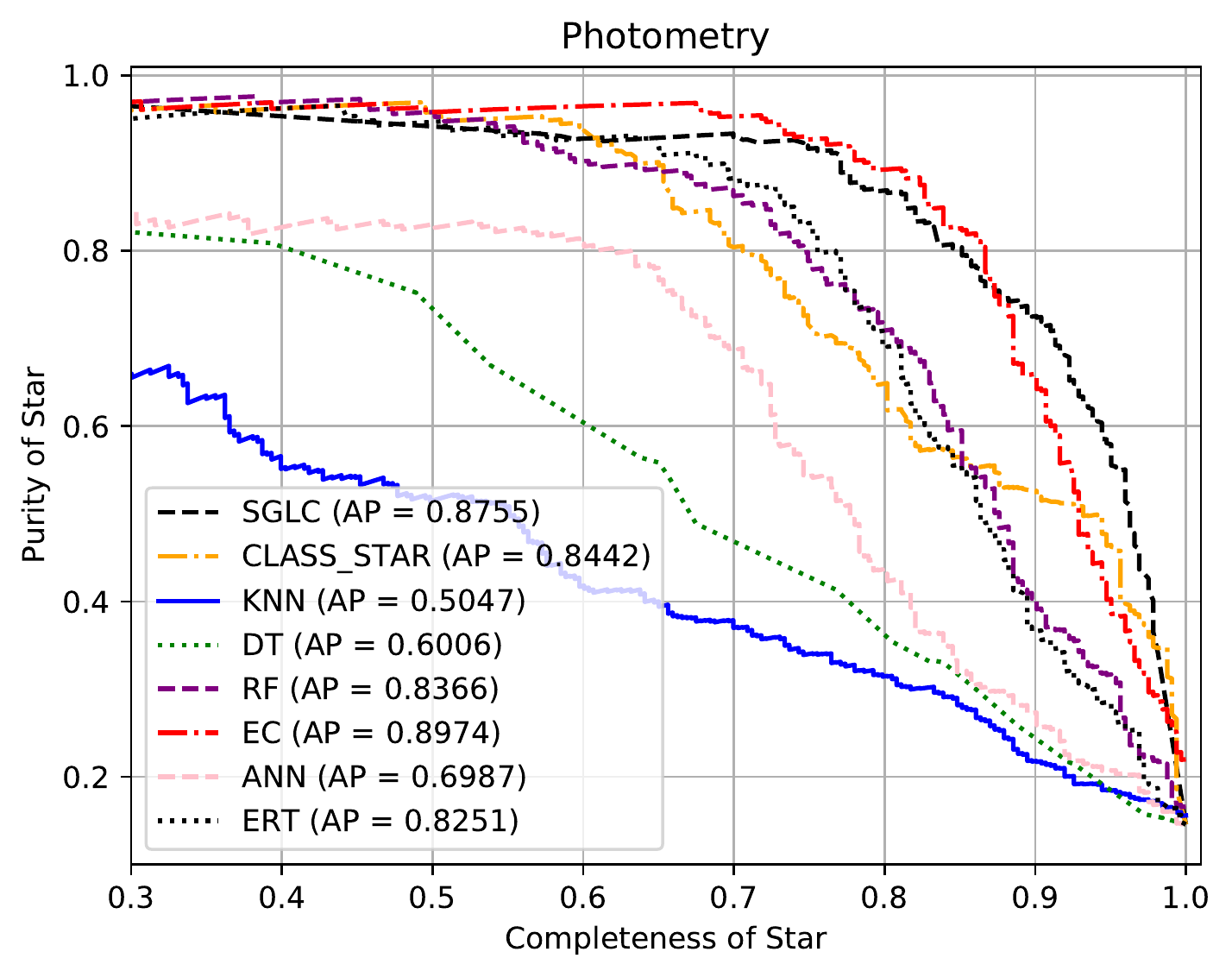}
 \includegraphics[width=.98 \columnwidth]{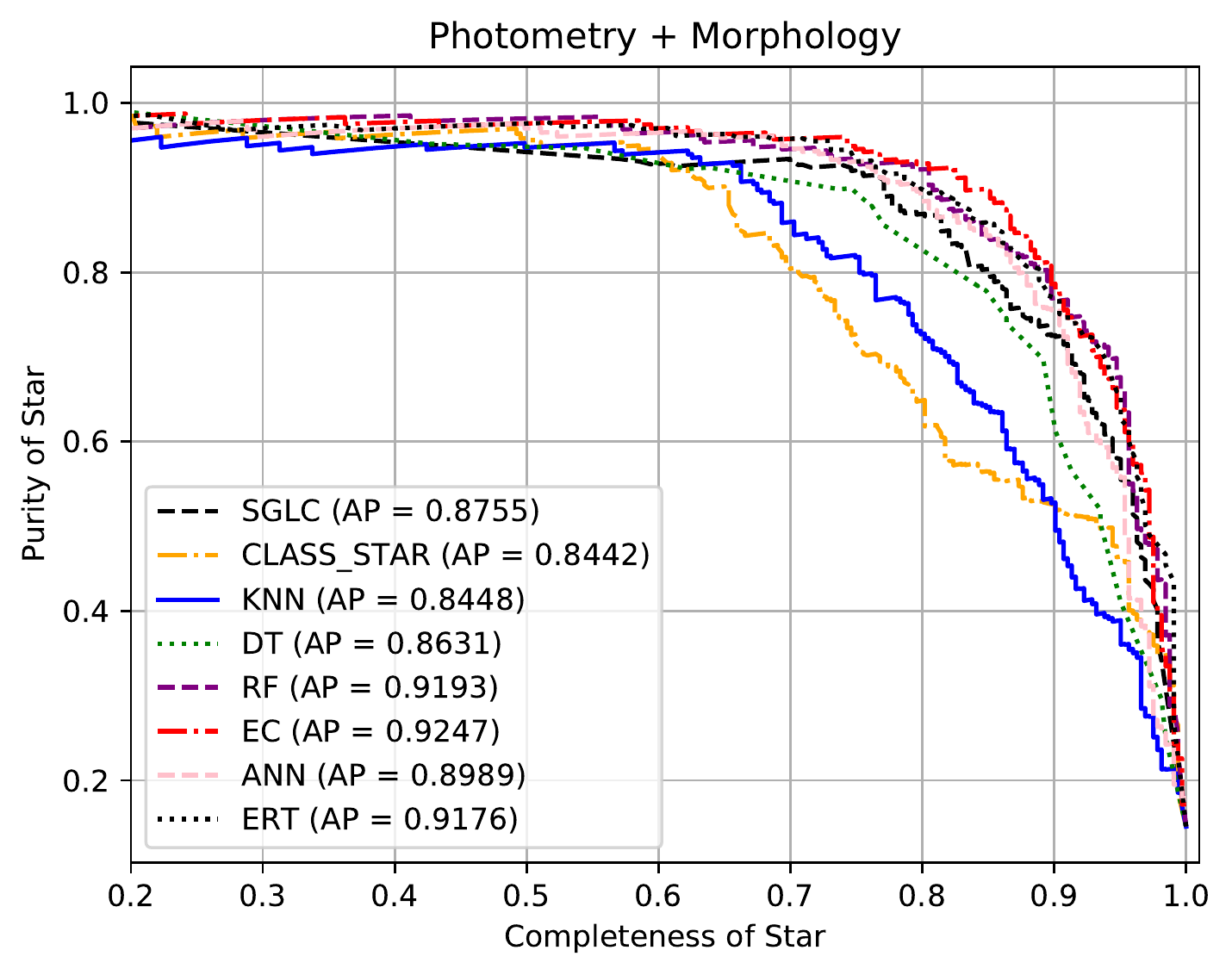}
\caption{Purity curves for stars using J-PAS data with HSC-SSP classification.
The top panel uses only photometric information while the bottom one uses also morphology.
For comparison it is  shown the classification by \texttt{CLASS\_STAR} and SGLC that always use morphological parameters. Note that the axes ranges are varied in order to better show the curves.}
\label{precision_recall_star_SDSS_photo}
\end{figure}

For completeness we report in Figure~\ref{precision_recall_star_SDSS_photo} the purity curves relative to the stars. For a comparison see, for example, \citet{2018MNRAS.481.5451S,2012ApJ...760...15F,2019MNRAS.483..529C}.

\section{Classification vs.~probability threshold}
\label{histo}

\begin{figure*}
\centering
 \includegraphics[width=.98 \columnwidth]{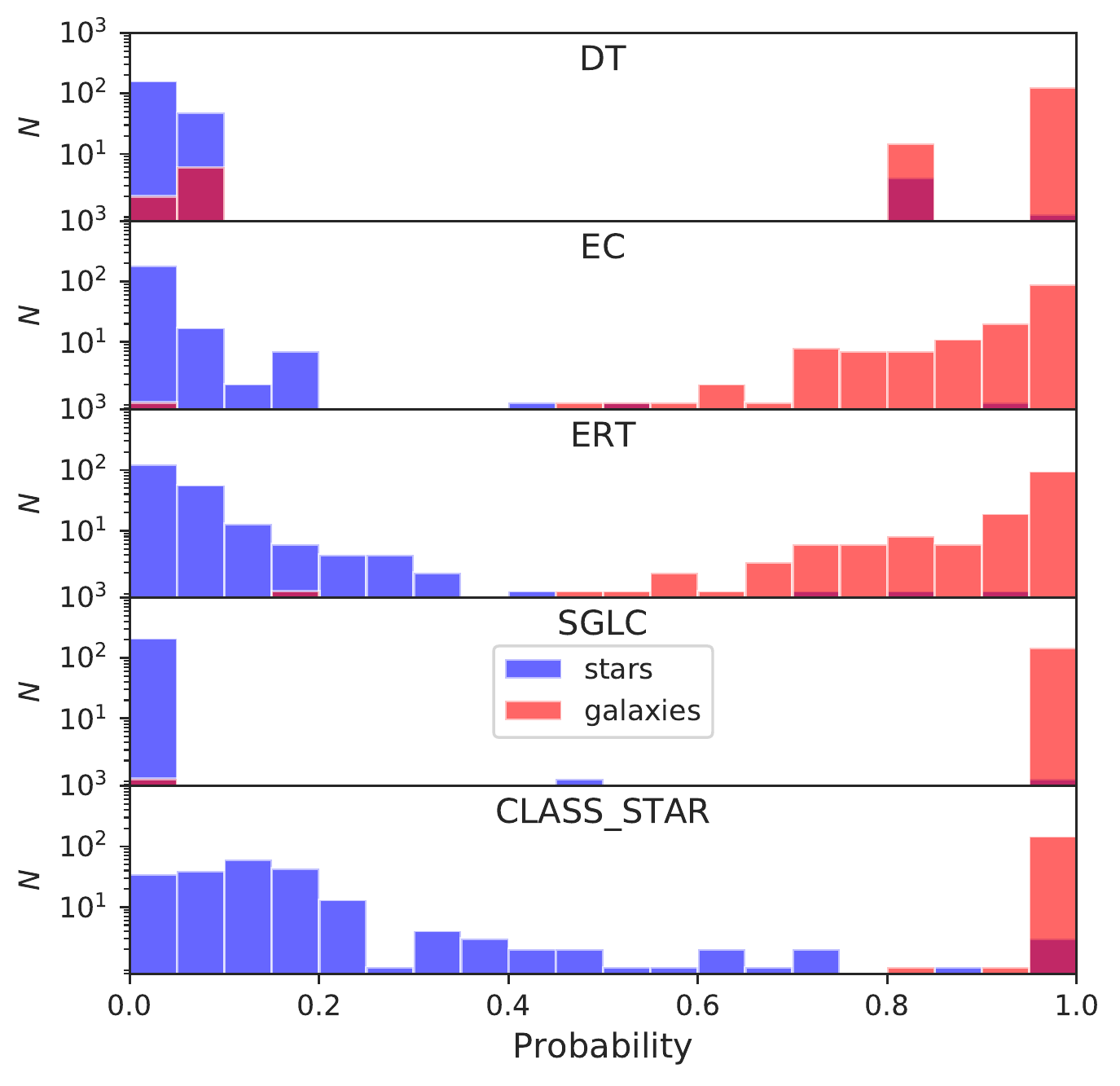}
 \includegraphics[width=.98 \columnwidth]{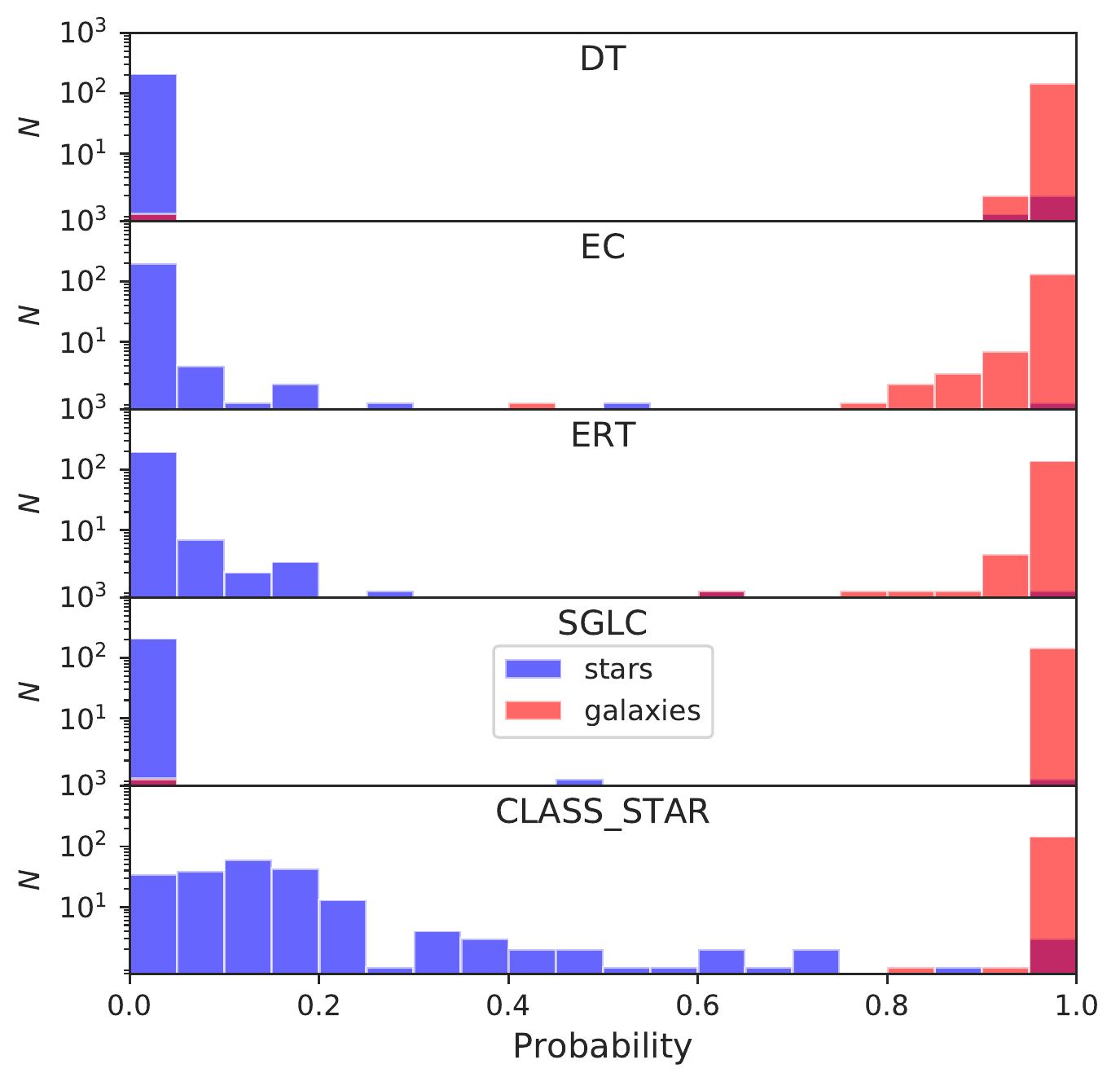}
 \includegraphics[width=.98 \columnwidth]{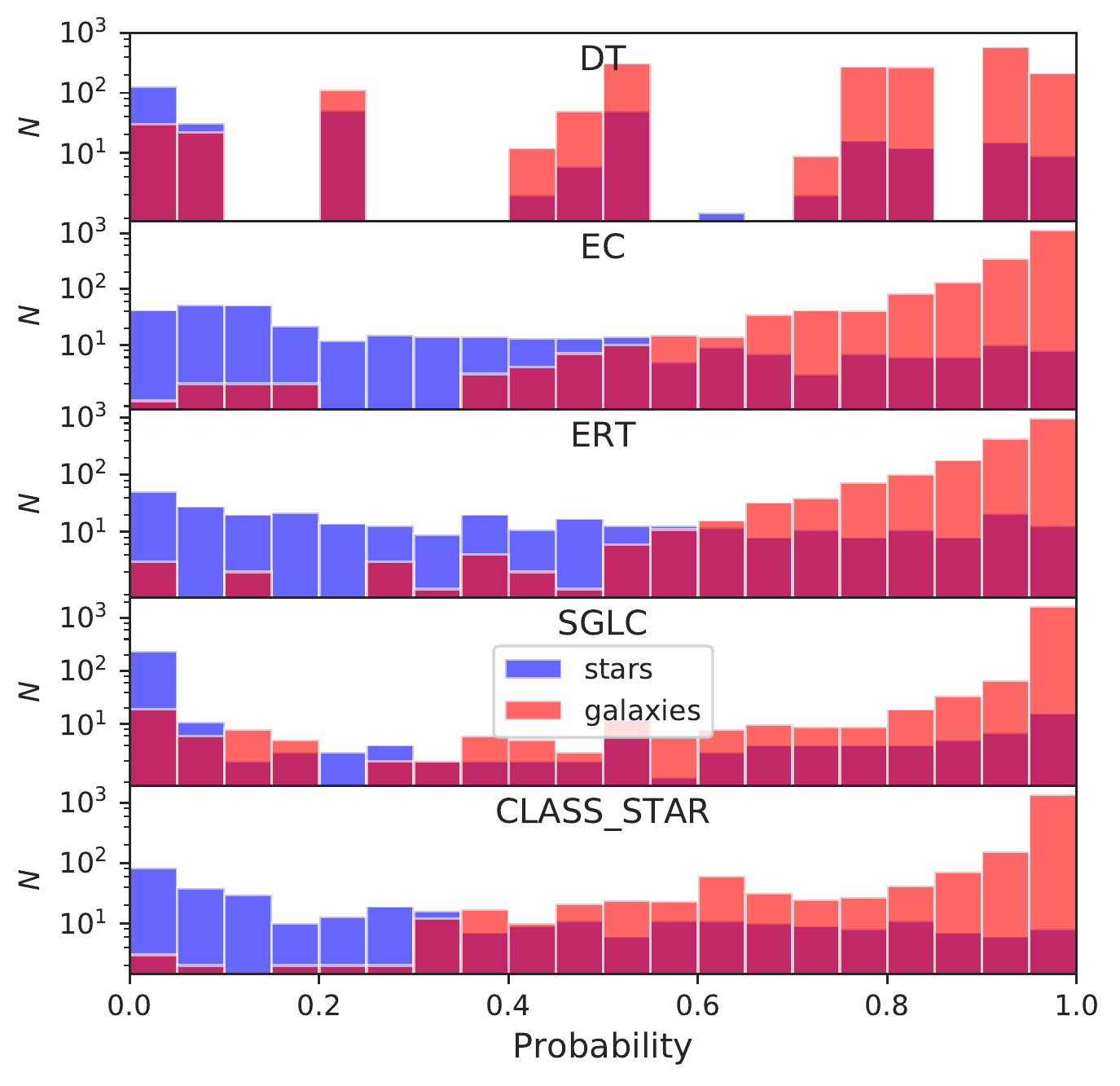}
 \includegraphics[width=.98 \columnwidth]{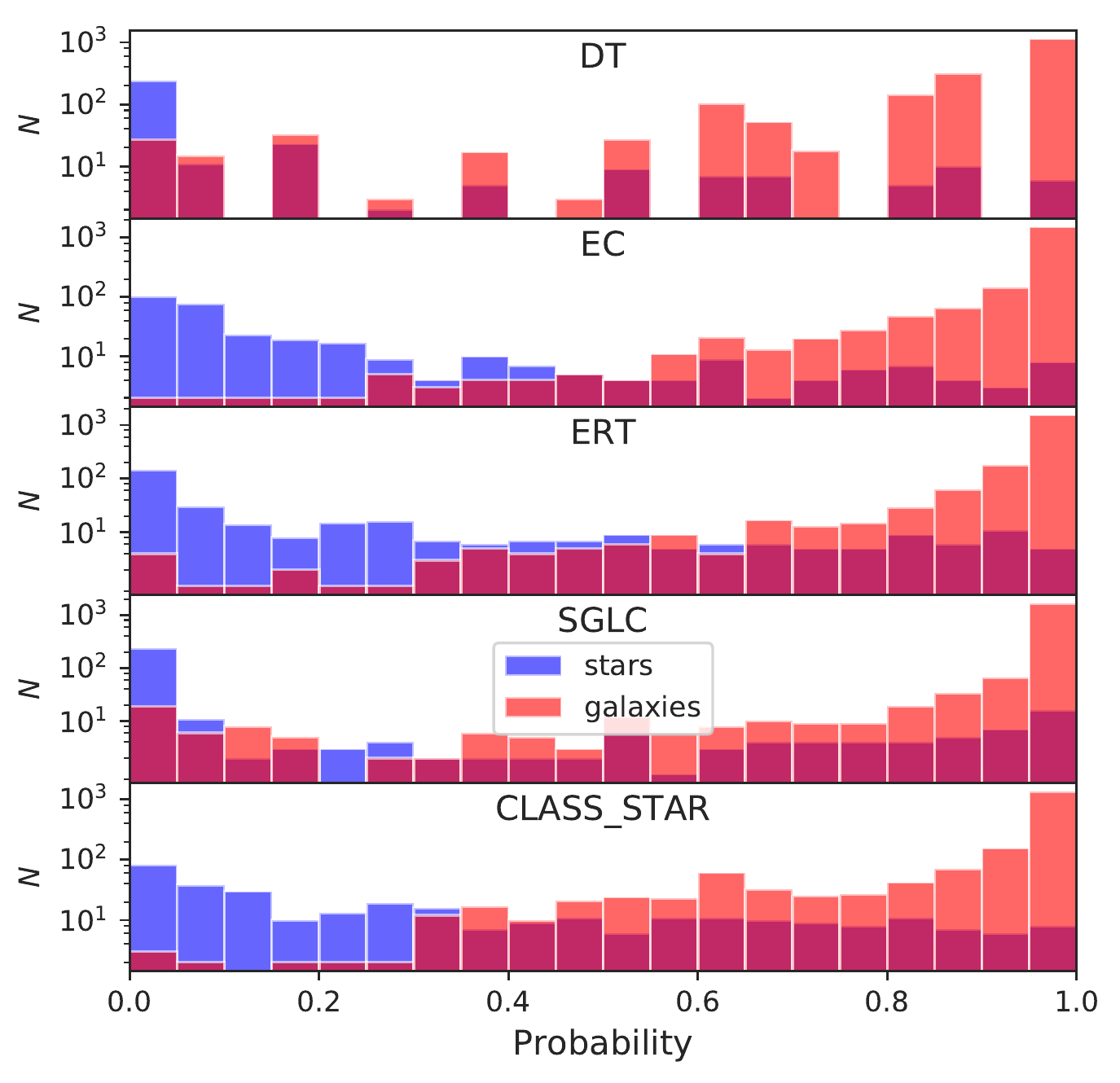}
\caption{Histograms of the probability that a source belongs to the class of galaxy.
The histograms relative to actual stars and galaxies, as classified by SDSS (top) and HSC-SSP (bottom), are in blue and semi-transparent red, respectively.
The panels on the left use only photometric information while the ones on the right use also morphology.
For comparison it is shown also the classification by \texttt{CLASS\_STAR} and SGLC that always use morphological parameters.}
\label{hist1d_photo_SDSS}
\end{figure*}

We show in Figure~\ref{hist1d_photo_SDSS} the histograms of the probabilities that the objects received from the classifiers.
In red we have objects classified as galaxies and in blue as stars.
These plots allow us to assess the performance of the algorithms from a different point of view.
When one choses a value for $p_{\rm cut}$, all the objects to the right of this value will be classified as galaxies while the ones with lower probability will be classified as stars.
It is then clear that an ideal algorithm should have well separated and non-intersecting probability distributions for stars and galaxies.

A first remark is that the addition of morphology makes the distributions tighter and with less intersections.
Similar results were obtained with CFHTLenS data \citep{2015MNRAS.453..507K} .
We observe instead that the probability distribution of galaxies for \texttt{CLASS\_STAR} is more concentrated than the probability distribution for stars. This leads us to the conclusion that \texttt{CLASS\_STAR} has a  tendency to classify galaxies better than stars.
It is also clear that by varying $p_{\rm cut}$ one can sacrifice the completeness of the dataset in favor of a higher purity of galaxies.

\section{mJP-AEGIS1 field}
\label{aegis1}

As said earlier, miniJPAS consists of 4 fields, each of approximately 0.25 deg$^2$ field-of-view \citep[for details see][]{Bonoli:2020ciz}.
The mJP-AEGIS1 has 20016 objects and features an $r$-band PSF which is similar to mJP-AEGIS3 ($\sim$0.7") and better than mJP-AEGIS2 and mJP-AEGIS4 ($\sim$0.8"). It is then interesting to repeat  for  mJP-AEGIS1 the analysis relative to HSC-SSP (see Section~\ref{hscresu}). We do not consider the analysis relative to SDSS as the  crossmatched catalog would be too small.

\begin{figure*}
\centering
 \includegraphics[width=.98 \columnwidth]{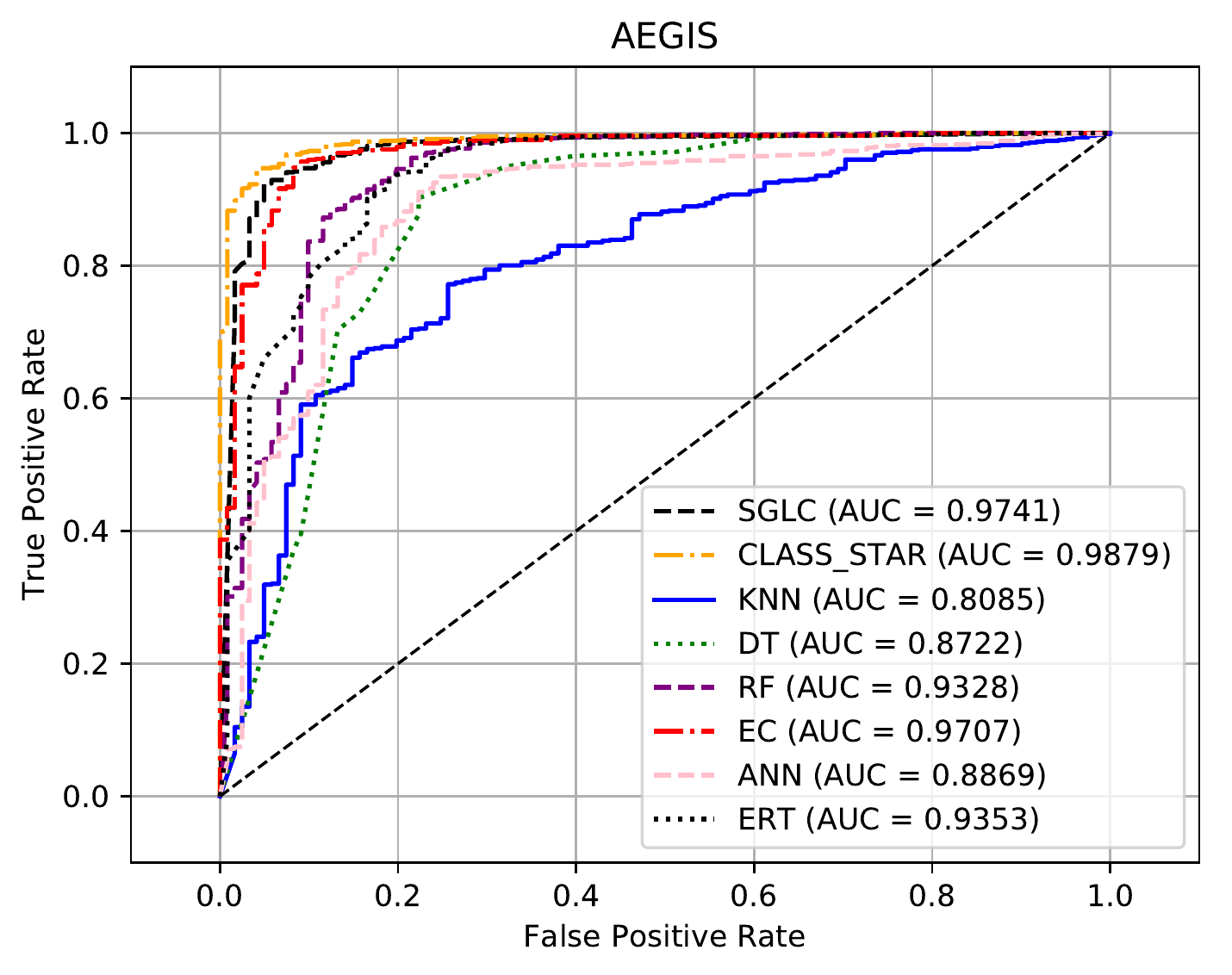}
 \includegraphics[width=.98 \columnwidth]{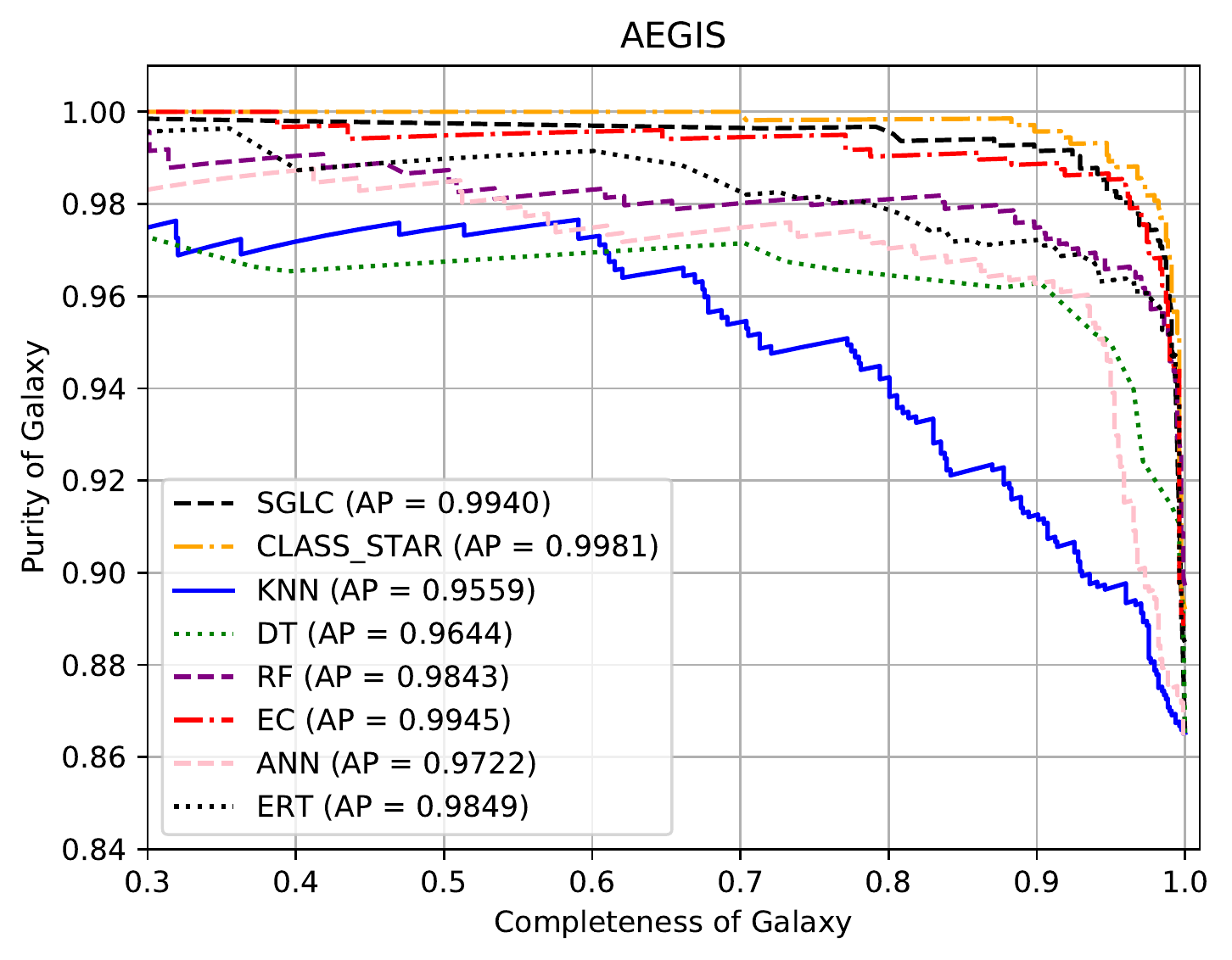}
 \includegraphics[width=.98 \columnwidth]{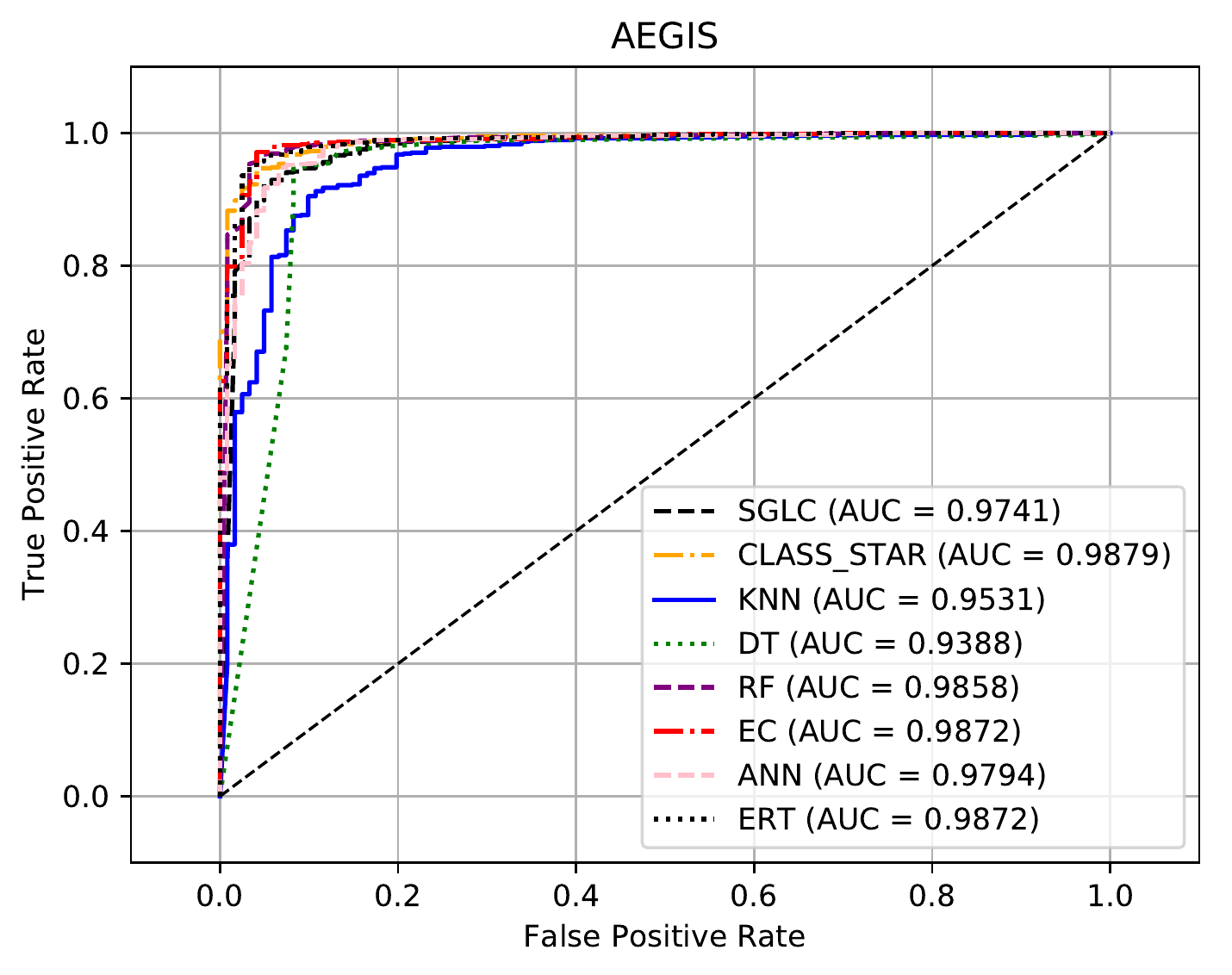}
 \includegraphics[width=.98 \columnwidth]{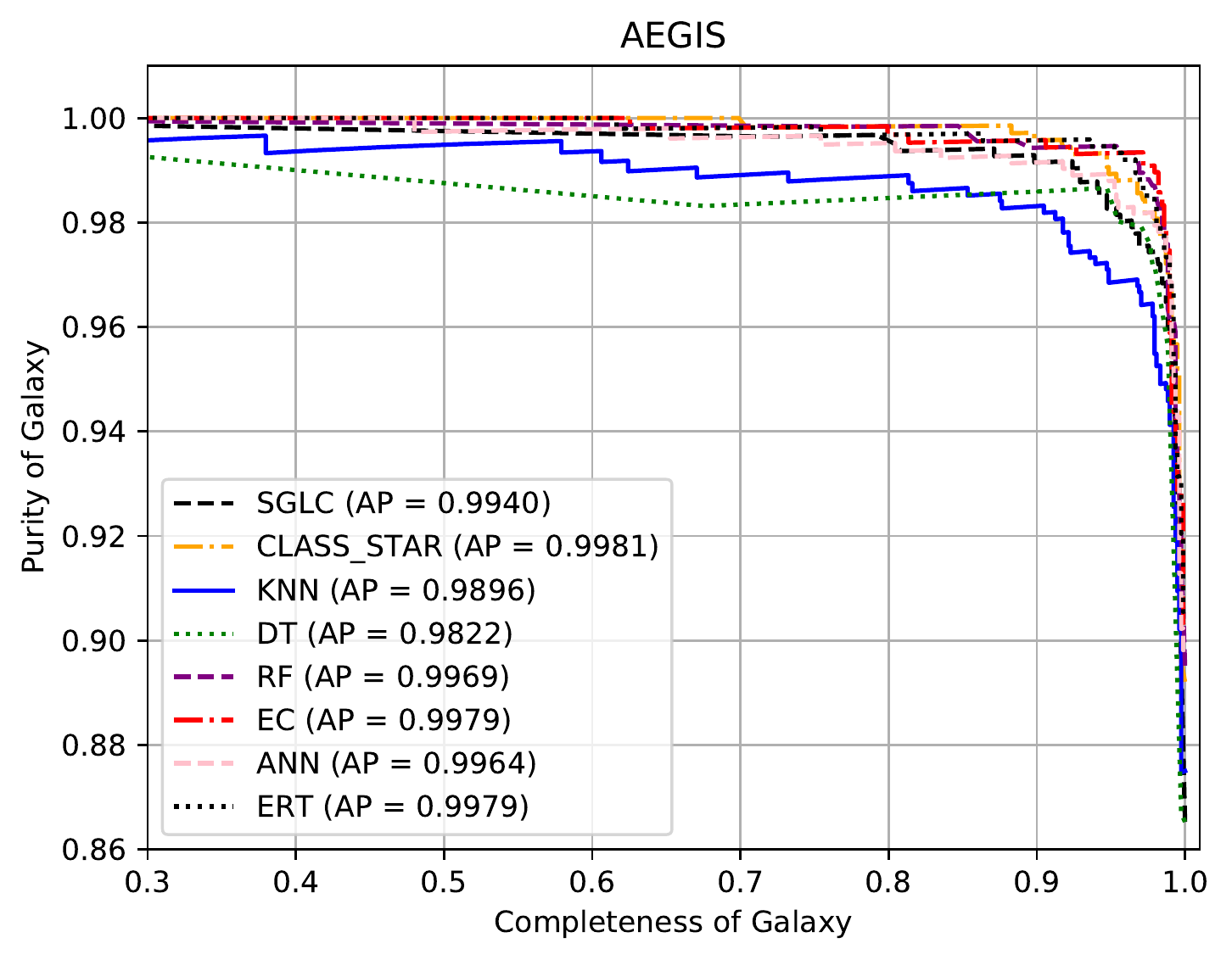}
\caption{
ROC curves (left panels) and purity curves for galaxies (right panels) for the classifiers considered in this paper for the AEGIS1 field  crossmatched with the HSC-SSP catalog in the magnitude interval $18.5\le r \le23.5$.
The top panels are relative to the analysis that uses only photometric bands while the bottom panels to the analysis that uses  photometric bands and morphological parameters. For comparison it is shown also the classification by \texttt{CLASS\_STAR} and SGLC that always use morphological parameters. Note that the axes ranges are varied in order to better show the curves.\label{roc_hsc_morpho-1}}
\end{figure*}

The crossmatch of mJP-AEGIS1 with HSC-SSP in the range $18.5 \le r \le 23.5$ has 4486 objects, 3809 galaxies and 677 stars.
We show the results in Figure~\ref{roc_hsc_morpho-1}, which should be compared with the analysis that considers the full miniJPAS catalog in Figure~\ref{roc_hsc_morpho}.
It is clear that the results relative to the various classifiers improve as expected. In particular, when considering both morphological and photometric features, ERT goes from $AUC=0.979$ (Fig.~\ref{roc_hsc_morpho}) to $AUC=0.987$ (Fig.~\ref{roc_hsc_morpho-1}).

\section{ADQL query}
\label{adql}

The value added catalog with the ERT and RF classifications is publicly available at \href{https://j-pas.org/datareleases}{j-pas.org/datareleases} via the ADQL table \texttt{minijpas.StarGalClass}.
The column \texttt{ert\_prob\_star} gives the probability $1-f$ of being a star provided by the  ERT classifier, using both morphological and photometric information.
The column \texttt{rf\_prob\_star} gives the probability $1-f$ of being a star provided by the RF classifier, using only photometric information. Note that here, in order to follow the convention of the \texttt{minijpas.StarGalClass} table, we are using the probability $1-f$ of being a star and not, as in the rest of this work, the probability $f$ of being a galaxy.

In order to facilitate access to our results we now report a simple query example that allows one to access the classifications generated by ML along with the miniJPAS photometric bands with flag and mask quality cuts:
\begin{verbatim}
SELECT

t1.MAG_AUTO[minijpas::uJAVA] as uJAVA,
t1.MAG_AUTO[minijpas::J0378] as J0378,
t1.MAG_AUTO[minijpas::J0390] as J0390,
t1.MAG_AUTO[minijpas::J0400] as J0400,
t1.MAG_AUTO[minijpas::J0410] as J0410,
t2.ert_prob_star,
t2.rf_prob_star

FROM

minijpas.MagABDualObj t1

JOIN

minijpas.StarGalClass t2

ON

t1.tile_id = t2.tile_id AND
t1.number=t2.number

WHERE

t1.flags[minijpas::rSDSS]=0 AND
t1.mask_flags[minijpas::rSDSS]=0

\end{verbatim}

\section{Analysis using only morphology}
\label{morphology_analysis}

\begin{figure}
\centering
 \includegraphics[width=.98 \columnwidth]{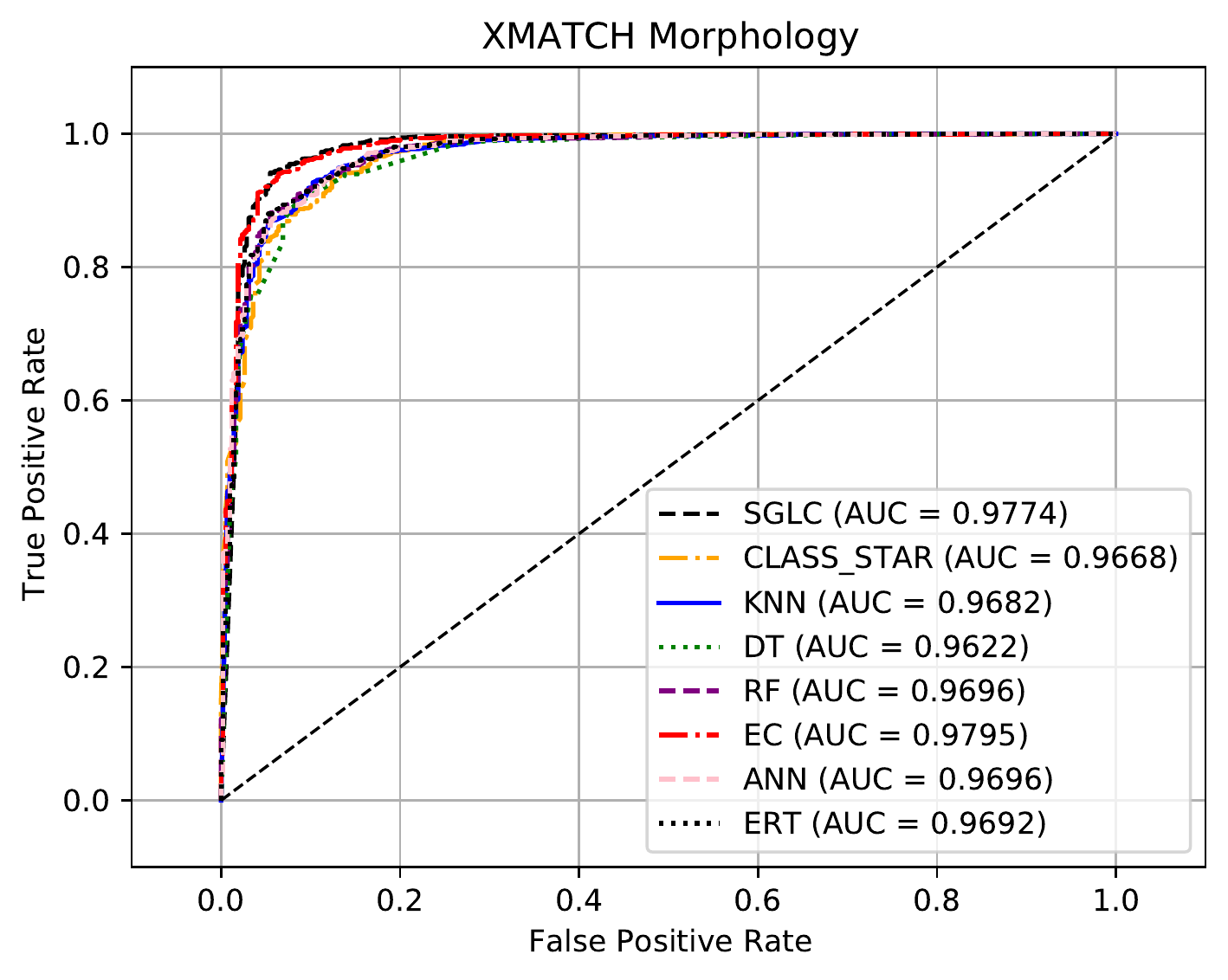}
 \includegraphics[width=.98 \columnwidth]{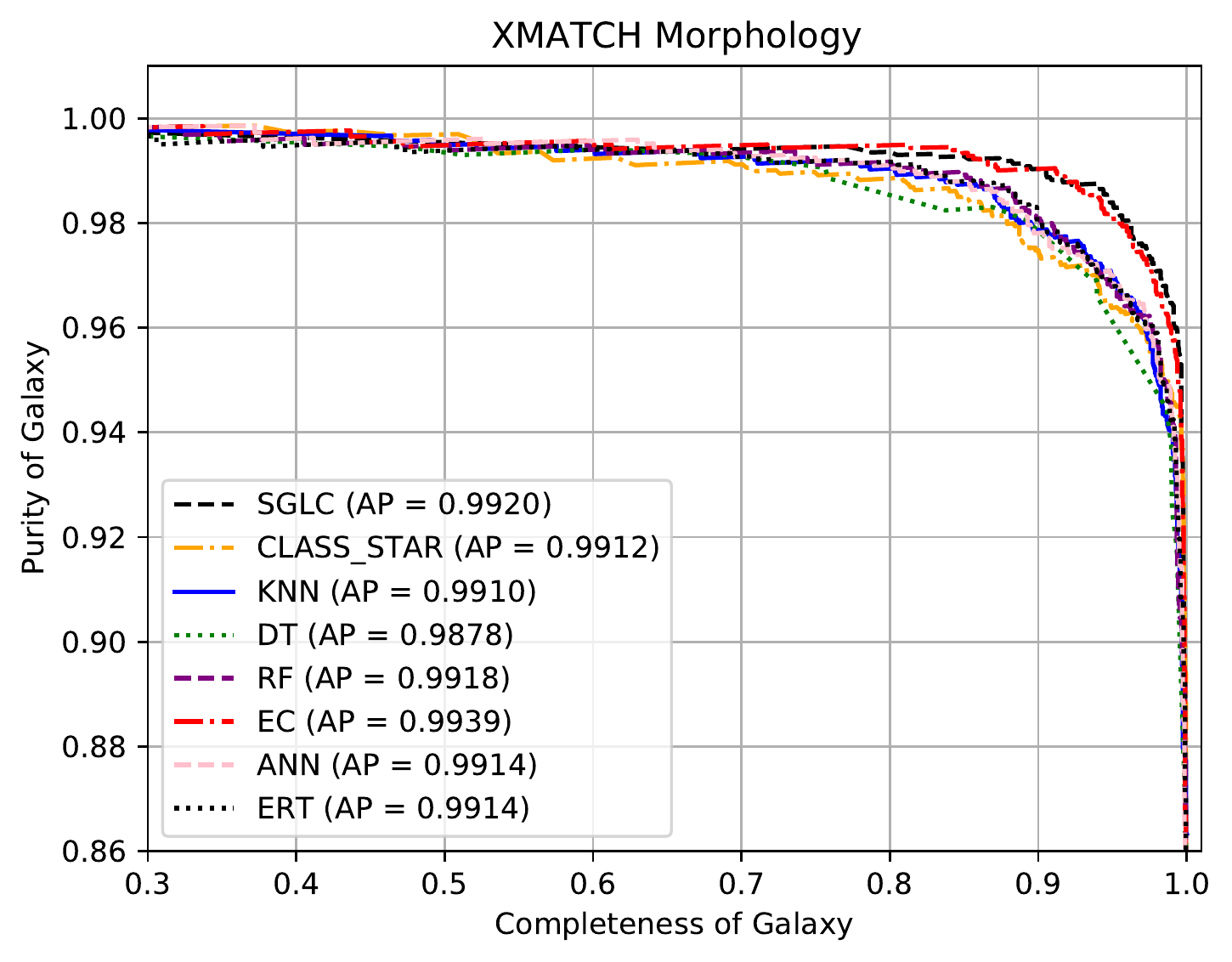}
 \includegraphics[width=.98 \columnwidth]{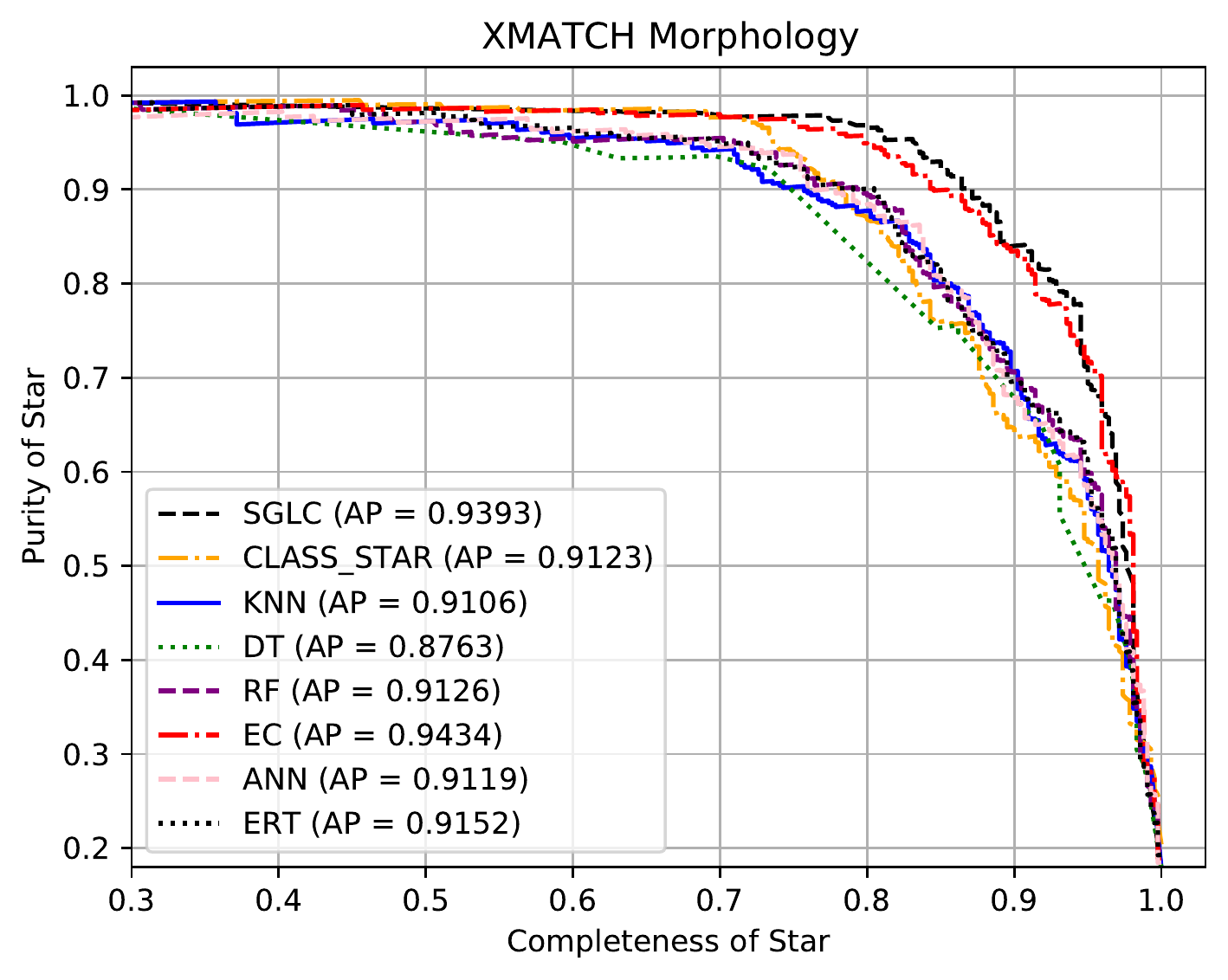}
\caption{ROC curve (top panel) and purity curve for galaxies (middle panel) and stars (bottom panel) when using only morphological parameters in the magnitude range $15 \le r \le 23.5$ of the XMATCH catalog.}
\label{fig:only_morpho}
\end{figure}

Here, we apply the ML methods discussed in the main text to the XMATCH catalog using only morphological parameters as input. In the Figure~\ref{fig:only_morpho} one can see the performances of the different algorithms. Comparing these results with the ones of Figure~\ref{cROCK} (morphology $+$ photometry), we observe that the inclusion of the photometric bands in the analysis increases the performance of the models.
When using only morphological parameters the AUC, galaxy AP ad star AP of the best pure ML classifier are 0.9696, 0.9918 and 0.9152, respectively. Instead, when using morphological parameters and photometric bands the AUC, galaxy AP ad star AP of the best pure ML classifier are 0.9855, 0.9955 and 0.9615, respectively.

\section{Hyperparameter optimization}
\label{hyperparameter}

The best values of the hyperparameters were found  via the \texttt{GridSearchCV} method.
They were selected according to the best mean from $k=10$ folds using the AUC-ROC metric.
Below are the hyperparameters relative to the XMATCH catalog; the unspecified hyperparameters take the default values by \texttt{scikit-learn} version 0.23.1.
We fixed \texttt{random seed = 5}.

\subsection{Using photometric bands only}
\begin{verbatim}
KNN

n_neighbors: 50, weights: distance

DT

class_weight: balanced, criterion: gini, 
max_depth: 5, max_features: None, random_state: 5

RF

n_estimators: 100, bootstrap: True, class_weight:
balanced_subsample, criterion: entropy, 
max_depth: 10, max_features: None

ANN

activation: logistic, hidden_layer_sizes: 200,
learning_rate: constant, solver: adam, 
max_iter: 200, tol=0.0001

ERT

n_estimators: 200, bootstrap: False, 
class_weight: balanced_subsample, criterion:
entropy, max_depth: 20, max_features: None
\end{verbatim}

\subsection{Using  morphological parameters only}
\begin{verbatim}
KNN

n_neighbors: 100, weights: distance

DT

class_weight: balanced, criterion: entropy,
max_depth: 5, max_features: None


RF

bootstrap: True, class_weight: balanced_
subsample, criterion: entropy, max_depth: 5, 
max_features: None, 'n_estimators': 100


ANN

activation: relu, hidden_layer_sizes: 200, 
learning_rate: constant, solver: adam,
max_iter: 200, tol=0.0001

ERT

bootstrap: False, class_weight: balanced_
subsample, criterion: entropy, max_depth: 10,
max_features: None, n_estimators: 200

\end{verbatim}

\subsection{Using photometric bands together morphology parameters}
\begin{verbatim}
KNN

n_neighbors: 100, weights: distance

DT

class_weight: balanced, criterion: entropy,
max_depth: 5, max_features: None

RF

n_estimators: 100, bootstrap: True,
class_weight: balanced_subsample, 
criterion: entropy, max_depth: 20,
max_features: None

ANN

activation: logistic, hidden_layer_sizes: 
200, learning_rate: constant, solver: sgd,
max_iter: 200, tol=0.0001

ERT

n_estimators: 200, bootstrap: False, 
class_weight: balanced_subsample, criterion:
entropy, max_depth: 20, max_features: None
\end{verbatim}

\end{appendix}

\end{document}